\documentclass[twocolumn]{aastex631}

\shorttitle{AASTeX v6.3.1 Sample article}
\shortauthors{Gowanlock et al.}
\graphicspath{{./}{figures/}}

\usepackage{pifont}
\newcommand{\xmark}{\ding{55}}%

\usepackage{amssymb,amsmath}
\usepackage{todonotes}
\usepackage{xspace}
\usepackage{subfigure}
\usepackage{tabularx}
\usepackage{makecell}
\usepackage{multirow}
\usepackage{longtable}
\usepackage{booktabs}
\usepackage{comment}

\newcolumntype{R}{>{\raggedleft\arraybackslash}X}

\usepackage{forest, tikz}
\usetikzlibrary{decorations.pathreplacing}

\def\snaps{\textsc{SNAPS}\xspace}
\def\snapshotdata{\texttt{SS1}\xspace}
\def\tessdata{\texttt{TESSC1}\xspace}
\def\lcdbdata{\texttt{LCDB}\xspace}
\def\intersectsnapstess{\texttt{SS1$\cap$TESSC1}\xspace}
\def\intersectlcdbdata{\texttt{LCDB$\cap$SS1$\cap$TESSC1}\xspace}

\def\lsp{\textsc{LSP}\xspace}
\def\ss{\textsc{SuperSmoother}\xspace}

\begin{document}

\title{Asteroid Period Solutions from Combined Dense and Sparse Photometry}

\correspondingauthor{Michael Gowanlock}
\email{michael.gowanlock@nau.edu}

\author[0000-0002-0826-6204]{Michael Gowanlock}
\affiliation{School of Informatics, Computing, and Cyber Systems \\
Northern Arizona University \\
P.O. Box 5693 \\
Flagstaff, AZ 86011, USA}
\affiliation{Department of Astronomy and Planetary Science \\
Northern Arizona University \\
P.O. Box 6010 \\
Flagstaff, AZ 86011, USA}

\author[0000-0003-4580-3790]{David E. Trilling}
\affiliation{Department of Astronomy and Planetary Science \\
Northern Arizona University \\
P.O. Box 6010 \\
Flagstaff, AZ 86011, USA}
\affiliation{School of Informatics, Computing, and Cyber Systems \\
Northern Arizona University \\
P.O. Box 5693 \\
Flagstaff, AZ 86011, USA}

\author[0009-0005-9955-1500]{Andrew McNeill}
\affiliation{Department of Physics\\Lehigh University\\ 16 Memorial Drive East\\ Bethlehem, PA 18015, USA}

\author[0000-0002-6676-1713]{Daniel Kramer}
\affiliation{School of Informatics, Computing, and Cyber Systems \\
Northern Arizona University \\
P.O. Box 5693 \\
Flagstaff, AZ 86011, USA}
\affiliation{Department of Astronomy and Planetary Science \\
Northern Arizona University \\
P.O. Box 6010 \\
Flagstaff, AZ 86011, USA}

\author[0000-0002-6292-9056]{Maria Chernyavskaya}
\affiliation{Department of Astronomy and Planetary Science \\
Northern Arizona University \\
P.O. Box 6010 \\
Flagstaff, AZ 86011, USA}



\begin{abstract}
Deriving high quality lightcuves for asteroids and other periodic sources from survey data is challenging due to many factors, including the sparsely sampled observational record and diurnal aliasing which is a signature imparted into the periodic signal of a source that is a function of the observing schedule of ground-based telescopes. In this paper, we examine the utility of combining asteroid observational records from the Zwicky Transient Facility (ZTF) and the Transiting Exoplanet Survey Satellite (TESS) which are the ground- and space-based facilities, respectively, to determine to what degree the data from the space-based facility can suppress diurnal aliases. Furthermore, we examine several optimizations that are used to derive the rotation periods of asteroids which we then compare to the reported rotation periods in the literature. Through this analysis we find that we can reliably derive the rotation periods for $\sim$85\% of our sample of 222 objects that are also reported in the literature and that the remaining $\sim$15\% are difficult to reliably derive as many are asteroids that are insufficiently elongated which produces a lightcurve with an insufficient amplitude and consequently, an incorrect rotation period. We also investigate a binary classification method that biases against reporting incorrect rotation periods. We conclude the paper by assessing the utility of using other ground- or space-based facilities as companion telescopes to the forthcoming Rubin Observatory.
\end{abstract}



\section{Introduction}\label{sec:motivation}

There are several benefits of using either ground- or space-based telescopic surveys to derive the physical properties of astrophysical phenomena, and in this paper, we examine the physical properties of asteroids. 

It is well-known that when deriving the period for periodic sources, ground-based observatories will impart a periodic signal as a function of the diurnal cycle~\citep{VanderPlas2018}. These aliases typically appear in a periodogram at frequencies that are multiples of 1~day$^{-1}$. In contrast, space-based observatories do not suffer from diurnal aliases because they are not prevented from observing at any time during the day.

The Transiting Exoplanet Survey Satellite (TESS) is a space-based observatory with a primary mission to detect transiting exoplanets~\citep{ricker2015transiting}, where the telescope stares at a field for a long duration. This telescope has observed intervening asteroids within its field of view. TESS rarely encounters the same asteroid between two pointings, which limits the temporal extent for which a given object is observed. This time window directly impacts the range of viable rotation periods for an asteroid.  In contrast, ground-based surveys are able to derive the rotation periods of asteroids with long periods because they may observe the asteroid numerous times over several years. 

In summary, in the context of surveys, ground-based telescopes are limited by diurnal observing schedules and space-based telescopes that stare at a field for long durations are only sensitive to detecting asteroids with short rotation periods. By combining datasets from ground- and space-based observatories, there are several potential benefits, including: $(i)$ decreasing the fraction of objects that have a period solution at aliases; $(ii)$ improving the sensitivity to long rotation periods; and, $(iii)$ utilizing partial lightcurve data that is of little value on its own, but could be used to augment the observational records of other catalogs.

Based on the above motivation, in this paper, we examine two applications for combining data from ground- and space-based observatories, which are summarized as follows:
\begin{enumerate}
\item To improve the overall fidelity of derived asteroid rotation period solutions by comparing to the rotation periods reported in the literature. 
\item To examine the use of a companion telescope (either ground- or space-spaced) to the Rubin Observatory to assess how this companion may improve rotation period fidelity. 
\end{enumerate}

In the context of surveys, periodic sources often have lightcurve periods that are difficult to constrain because surveys produce sparse photometry. The first application above is similar to other papers in the literature that develop new methods that either maximize the fraction of correctly assigned lightcurve periods in survey data or examine which methods should be applied to a particular application~\citep{graham2013comparison,drake2013probing,suveges2015comparative,oelkers2017variability,Coughlin2021,KRAMER2023}. The second application is motivated by assessing the impact that a companion telescope to the Rubin Observatory would have on improving Solar System science.

The paper is organized as follows. Section~\ref{sec:data} outlines the datasets used in the paper. Section~\ref{sec:lomb_scargle} describes the period finding algorithm used in the paper and associated parameters. It also illustrates the baseline distribution of rotation periods across each dataset. Section~\ref{sec:comparison_to_LCDB} presents the proposed optimizations used to maximize derived period fidelity as compared to the solutions in the literature that are reported by the Light Curve Data Base (\lcdbdata). With a selection of optimizations established, Section~\ref{sec:comparison_full_intersection} applies them to the full sample of objects that are found in both ZTF and TESS datasets. Section~\ref{sec:rubin_companion} assesses the utility of using other facilities to augment the capabilities of the Rubin Observatory. Finally, Section~\ref{sec:conclusion} concludes the paper.

\begin{deluxetable*}{l|l|l|l}[!t]
\tablecaption{Summary of datasets and properties.}\label{tab:datasets}
\tablewidth{\textwidth}
\tabletypesize{\footnotesize}
\tablehead{
\colhead{Dataset Name}&\colhead{Description}&\colhead{Number of Objects}&\colhead{Number of Observations}
}
\startdata
\snapshotdata&\makecell[l]{The first \snaps alert broker data release containing lightcurve properties\\of asteroids derived from ZTF~\citep{trilling2023solar}.}&28,638&2,145,478\\
\tessdata&Lightcurve properties of asteroids derived from TESS~\citep{mcneill2023tess}.&28,878&5,637,892\\
\intersectsnapstess&The intersection of objects in \snapshotdata and \tessdata.&3168&1,453,748\\
\enddata
\end{deluxetable*}

\section{Data}\label{sec:data}
Table~\ref{tab:datasets} outlines the datasets used in this paper. \snapshotdata and \tessdata are from ZTF~\citep{2019PASP..131a8002B} and TESS~\citep{ricker2015transiting}, respectively. \intersectsnapstess contains only those objects that are found in both \snapshotdata and \tessdata.

As will be shown, we compare our period solutions to \lcdbdata~\citep{warner2009}. However, there are only 222 objects that are in \intersectsnapstess and \lcdbdata (hereafter denoted as \intersectlcdbdata). While there are more objects in \intersectsnapstess (3168), we have no baseline for comparison for all of these objects, so we use the \intersectlcdbdata dataset for several of our comparisons. 

We use a version of \lcdbdata from 2020 that excludes TESS data deposited by~\citet{pal2020}~and~\citet{mcneill2023tess} such that when comparing our derived period solutions to \lcdbdata we do not accidentally compare to some of our own period solutions as derived by our prior work in~\citet{mcneill2023tess}. Otherwise, we may inadvertently increase the total fraction of correctly derived period solutions.



\subsection{Datasets: \snapshotdata and \tessdata}\label{sec:datasets_SS1_TESSC1}

\begin{figure*}[!t]
\centering

\subfigure[\snapshotdata ($\mu=76.1$)]{
\includegraphics[width=0.31\textwidth]{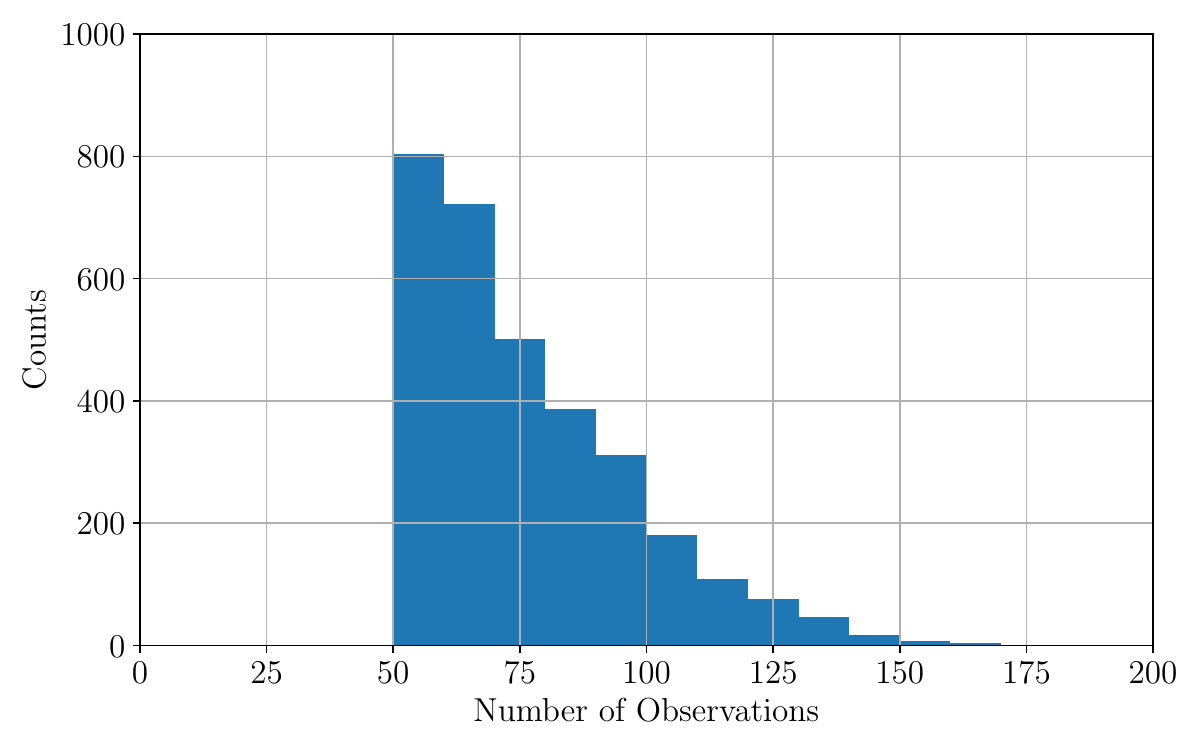}
}
\subfigure[\tessdata ($\mu=382.8$)]{
\includegraphics[width=0.31\textwidth]{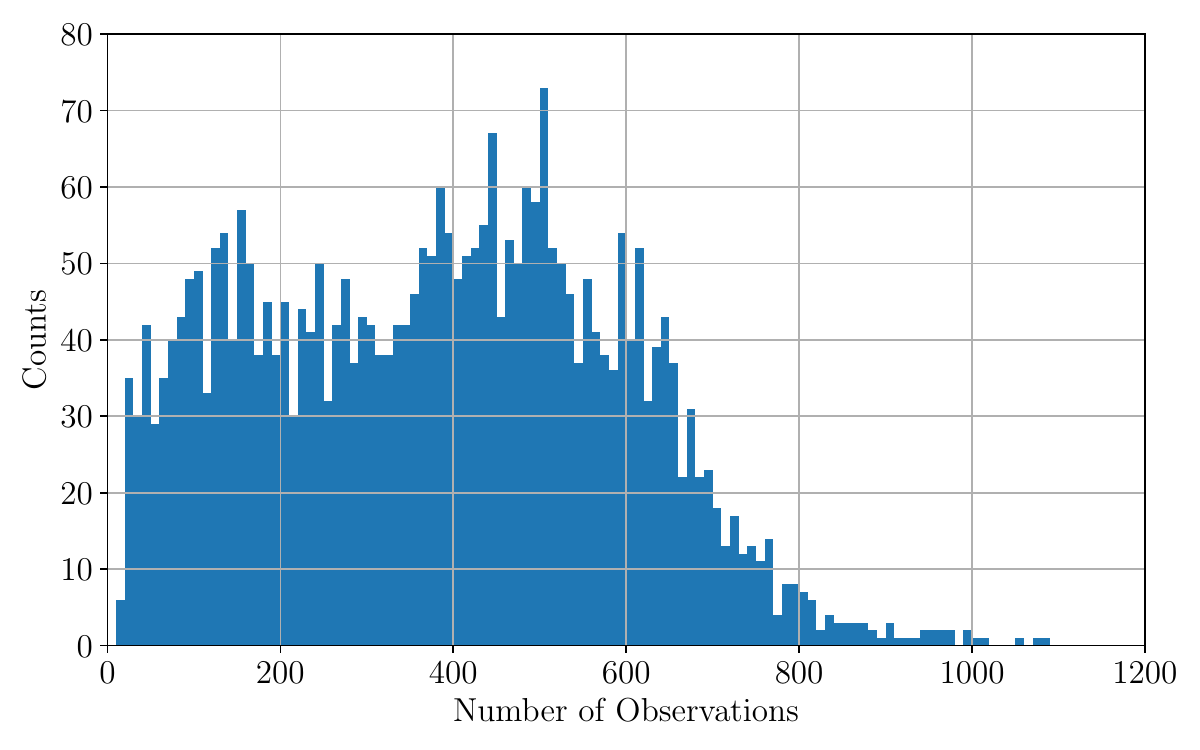}
}
\subfigure[\intersectsnapstess ($\mu=458.9$)]{
\includegraphics[width=0.31\textwidth]{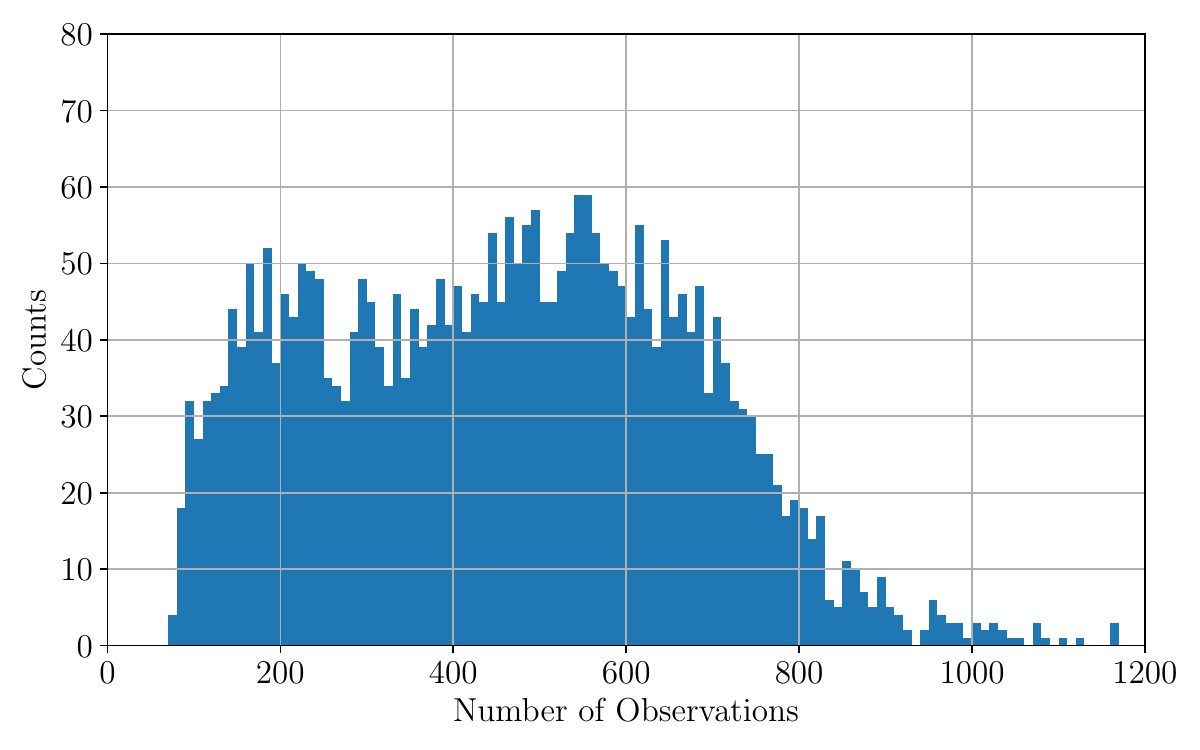}
}

    \caption{The distribution of the number of observations for the objects in the (a) \snapshotdata, (b) \tessdata, and (c) \intersectsnapstess datasets, limited to the $n=3168$ objects appearing in both \snapshotdata and \tessdata. The bin width is 10 observations in each histogram.}
   \label{fig:dataset_observation_histograms}
\end{figure*} 

The SNAPShot1 (\snapshotdata) dataset~\citep{trilling2023solar} contains ZTF~\citep{2019PASP..131a8002B} observations of small bodies, with the vast majority being main belt asteroids. In the \snapshotdata data release, only those objects with $\geq$51 observations have an assigned derived rotation period. This observation threshold was selected because a sufficient number of observations are needed to reliably derive a rotation period. Of those objects with $\geq$51 observations, there are a total of 2,145,478 observations across 28,638 objects. A real/bogus score~\citep{Duev2019} was used to discard observations with high uncertainties (e.g., non-point sources and observations with poor subtractions were removed using this method). For more information on data processing see~\citet{trilling2023solar}.

\tessdata contains 28,878 objects with a total of 5,637,892 observations, where the data was $\sigma$ clipped to remove outliers in the observational record for each object~\citep[for more detail on data processing see][]{mcneill2023tess}. Unlike the ZTF data in \snapshotdata, lightcurves were derived for all objects regardless of the number of observations~\citep{mcneill2023tess}. Instead of using a threshold of 51 observations for \snapshotdata, \citet{mcneill2023tess} assigned a confidence score to each object that described the probability that the derived rotation period is correct.  


Figures~\ref{fig:dataset_observation_histograms}(a)~and~(b) report the distribution of the number of observations for each object in the \snapshotdata and \tessdata datasets, respectively, limited to those objects that are common to both datasets.   Because the  goal of this paper is to combine the two datasets together, we do not enforce a lower limit number of observations required of each object in the \tessdata dataset, as even few observations in \tessdata for an object may help supplement the observational record for \snapshotdata. However, as shown in Figure~\ref{fig:dataset_observation_histograms}(b), the mean number of observations per object in \tessdata is $\mu=382.8$ and so there are a small fraction of objects that have few observations.

\subsection{Combining \snapshotdata and \tessdata}
We combine the two data records for \snapshotdata and \tessdata by taking the intersection of the objects in both datasets. This yields a total of $n=3168$ objects and as described above, we denote this dataset as \intersectsnapstess. We calculate $H$
magnitude for each observation in SS1 by correcting for phase and
distance. The same procedure is conducted for \tessdata. To combine the two datasets for each object, we compute the mean of the magnitudes in \tessdata and then use this to offset the corresponding observational record in the \snapshotdata dataset. We elected to normalize to \tessdata because there are more observations on average in the \tessdata dataset than the \snapshotdata dataset. The photometric errors are unchanged when combining the magnitudes from both datasets. The observational records for each object often contain several apparitions. The apparitions vary between objects, where \snapshotdata and \tessdata apparitions can be temporally disjoint, or there can be overlap where \snapshotdata observations occur during the same apparition as the \tessdata observations. Several examples of unphased photometry will illustrate this in Section~\ref{sec:example_lightcurves_and_implications}. Figures~\ref{fig:dataset_observation_histograms}(c) plots the distribution of objects in the combined \intersectsnapstess dataset where the mean number of observations per object is $\mu=458.9$.








As we will show later in Section~\ref{sec:sigma_clipping}, we also examine $\sigma$ clipping the \intersectsnapstess dataset at the 2$\sigma$ and 3$\sigma$ levels. The $\sigma$ clipped variants of this dataset contain a total of 1,384,659 and 1,449,961 observations, respectively.

\section{Rotation Period Derivation}\label{sec:lomb_scargle}
To derive the rotation periods of asteroids we use the Lomb-Scargle Periodogram (\lsp)~\citep{lomb1976least, 1982ApJ...263..835S}. We employ the GPU implementation outlined in~\citet{gowanlock2021fast} that uses the generalized variant of \lsp that uses the floating mean method and considers photometric error when fitting the lightcurve. 

While the periods derived for \tessdata that are outlined by~\citet{mcneill2023tess} also used the generalized \lsp algorithm with the photometric error, the rotation periods reported in \citet{trilling2023solar} ignored the photometric error. Consequently, to ensure consistency between the methods used to derive the rotation periods across all datasets, we re-derived the rotation periods for each object in \snapshotdata using the same approach.

Across all datasets, for each object we execute \lsp and search a uniformly spaced frequency grid of $n_f=10^6$ frequencies. For the \snapshotdata and \intersectsnapstess datasets, we search in the frequency range $[f_{min}, f_{max})$ where $f_{min}=0.0048$~day$^{-1}$ and $f_{max}=24.0$~day$^{-1}$, respectively. This corresponds to rotation periods of $\approx$2--10,000~h, where the rotation period of an asteroid is twice the lightcurve period derived by \lsp (one lightcurve period would only capture half of an asteroid's rotation).   

For each of the $n_f=10^6$ frequencies searched for each object, \lsp outputs a power value. In this paper, the frequency with the highest power in the periodogram corresponds to the lightcurve period, which is doubled to compute the rotation period. In all that follows, when we refer to the period of an asteroid, we are referring to the \emph{rotation period} and not the lightcurve period. This quantity will be reported in hours throughout the paper.

\subsection{Period Distribution of \intersectsnapstess}\label{sec:period_dist_intersection_SNAPSTESS}

\begin{figure*}[!t]
\centering
\includegraphics[width=1\textwidth]{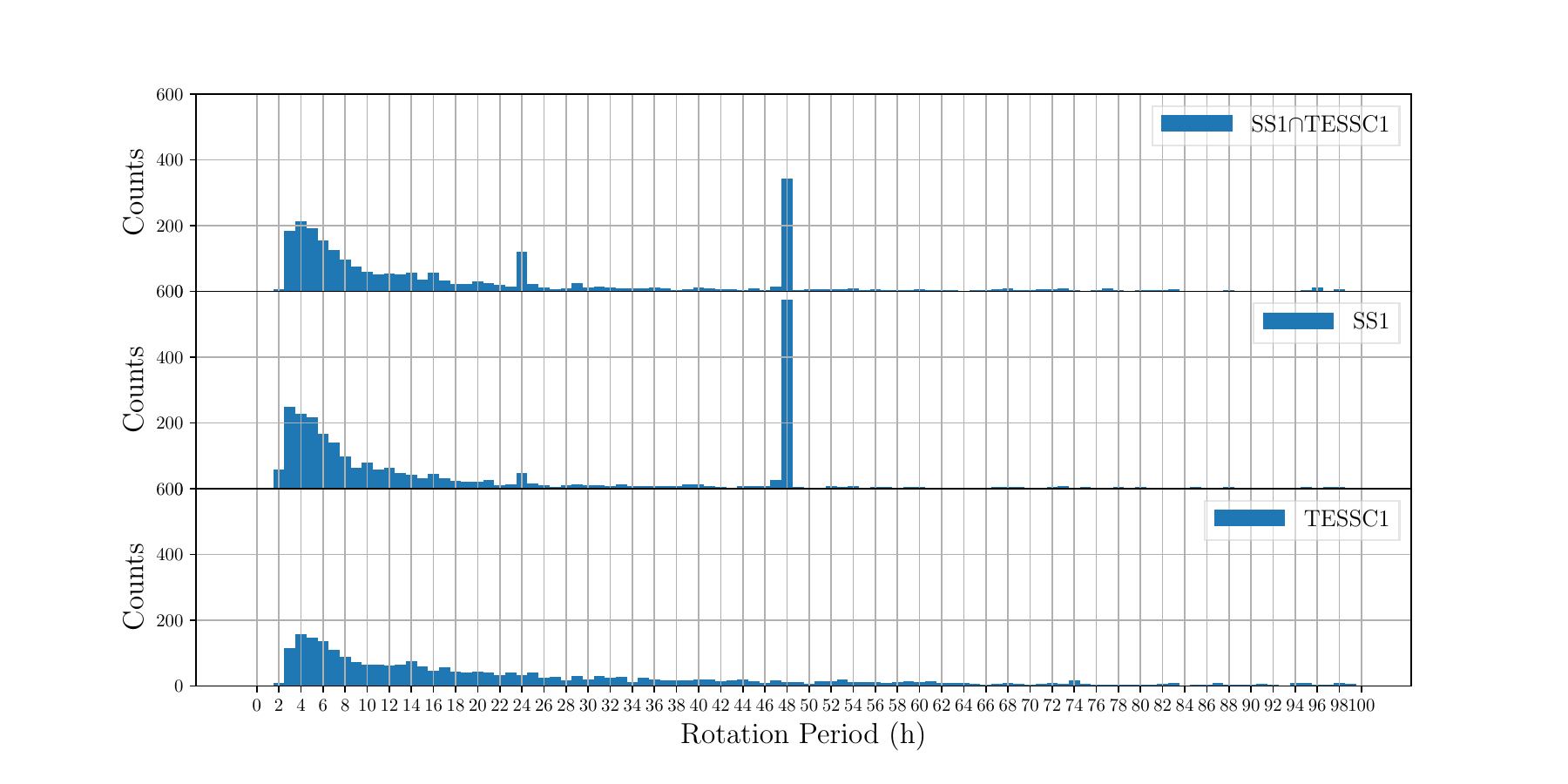}
    \caption{The period distribution for the $n=3168$ objects in \intersectsnapstess. The bottom to top panels show the period distribution using only the observations in \tessdata, using only the observations in \snapshotdata, and the period distribution when combining the observational records of both datasets, respectively. Although we search for solutions up to 10,000~h, we only show solutions up to 100~h because there are so few solutions $>$100~h.}
   \label{fig:ZTF_TESS_period_distribution}
\end{figure*}

Using the \lsp we derive the rotation periods of all $n=3168$ asteroids in \intersectsnapstess by selecting the period corresponding to the frequency with the greatest power in the periodogram. 

Figure~\ref{fig:ZTF_TESS_period_distribution} shows the distribution of rotation periods, and we observe that the period distribution for \snapshotdata (middle panel) has a large spike at 48 hours, which is clearly an alias generated by the diurnal signature described in Section~\ref{sec:motivation}. Contrasting this with \tessdata (bottom panel) we observe that there are no spikes at typical aliases, which is to be expected as the space-based observatory is not impacted by a diurnal observing schedule. When we combine the observational records together to obtain \intersectsnapstess, we find that there is an overabundance of objects having a 48~h period and a smaller overabundance at 24~h. This is surprising as we would expect that \tessdata may eliminate a large fraction of period solutions that are derived at the abovementioned aliases. While the number of aliases has decreased when comparing the middle panel to the top panel of Figure~\ref{fig:ZTF_TESS_period_distribution}, the overabundance of period solutions at aliases in \intersectsnapstess clearly indicates that the ground-based ZTF data (\snapshotdata) negatively impacts the derived rotation period solutions.

\section{Comparison to LCDB and Improving Derived Period Fidelity}\label{sec:comparison_to_LCDB}
In this section, we compare the fidelity of our period solutions to those in the literature. We begin by comparing the period solutions in the literature to those derived from the \intersectsnapstess dataset which is our baseline approach (Section~\ref{sec:period_dist_intersection_SNAPSTESS}). Then we apply several optimizations to improve the confidence in our period solutions and determine to what extent we can improve derived period fidelity over the baseline. 

\subsection{Baseline Approach}\label{sec:baseline}

\begin{deluxetable*}{l|r|r|r|r|r}[!t]
\tablecaption{The fraction of objects in three datasets, \tessdata, \snapshotdata, \intersectsnapstess , that match the period solutions in \lcdbdata. There are $n=222$ objects in \intersectlcdbdata.  The upper bound refers to whether one of the \tessdata, \snapshotdata, or \intersectsnapstess datasets contains the correct period solution. We report exact matches to \lcdbdata and aliased matches that are within a factor $0.5\times$$\pm3\%$ or $2\times$$\pm3\%$ of the \lcdbdata period solution.}\label{tab:matchfraclcdb}
\tablewidth{\columnwidth}
\tabletypesize{\footnotesize}
\tablehead{
\colhead{Dataset Name}&\colhead{$n$}&\colhead{Match Fraction (exact)}&\colhead{Number (exact)}&\colhead{Match Fraction (incl. aliased)}&\colhead{Number (incl. aliased)}
}
\startdata
\snapshotdata&222&0.599&133&0.608&135\\
\tessdata&222&0.739&164&0.806&179\\
\intersectsnapstess&222&0.784&174&0.829&184\\\hline
Upper Bound&222&0.820&182&0.878&195\\\hline
\enddata
\end{deluxetable*}





The Light Curve Data Base (\lcdbdata) contains asteroid rotation periods reported in the literature and a confidence level ($U$). In this paper we only consider objects with a quality score of $U=3$ which are unambiguous rotation period solutions where there is full sampling of the lightcurve~\citep{warner2009}.  Taking the intersection of the $n=3168$ objects in \intersectsnapstess with the objects in \lcdbdata we are left with a total of 222 objects. This sample of 222 objects in \lcdbdata is used to determine whether we obtain correct rotation periods, and we refer to this sample of objects as \intersectlcdbdata. 

We define an \emph{exact match} to \lcdbdata when an object has a rotation period within 3\% of the \lcdbdata period. Often there will be several power spikes in the periodogram where two or more spikes will yield nearly identical power values. Often these occur at periods that are a factor of half or double the true rotation period defined by \lcdbdata; therefore, we also report when our periods are a factor of $0.5$$\times\pm3\%$ or $2$$\times\pm3\%$ of the \lcdbdata period solution. We refer to these as \emph{aliased matches}, although they should not be confused with aliases that are a function of the diurnal observing schedule of a ground-based telescope (Section~\ref{sec:motivation}). 

Table~\ref{tab:matchfraclcdb} reports the fraction of matches to \lcdbdata for \snapshotdata, \tessdata, and \intersectsnapstess. We find that the ZTF dataset (\snapshotdata), has an exact match percentage of $\approx$60\%, whereas \tessdata has an exact match percentage of $\approx$74\%. This is to be expected, as there are more observations per object in \tessdata compared to \snapshotdata (Figures~\ref{fig:dataset_observation_histograms}(a)~and~(b)), which helps improve the overall match fraction. Furthermore, as described in Section~\ref{sec:motivation}, the \tessdata dataset is not susceptible to aliases that are common in \snapshotdata. We find that combining \snapshotdata with \tessdata (\intersectsnapstess) yields the greatest exact match percentage of $\approx$78\%. 

As described in Section~\ref{sec:lomb_scargle}, we re-derived the rotation periods for the \snapshotdata dataset outlined in~\citet{trilling2023solar} to include the photometric error and use the generalized variant of \lsp. We found a $\approx$60\% match percentage to LCDB. This match percentage is consistent with \citet{trilling2023solar} which found that for an \lcdbdata quality code $U=3-$ or better, they achieve a match percentage with \lcdbdata of 67\%, where they define an exact match to \lcdbdata as a derived rotation period being within 10\% of the \lcdbdata rotation period. Given that we use a 3\% threshold in this paper, and that the samples are different (we only examine those objects in \intersectsnapstess), our 60\% match percentage is roughly consistent with the 67\% match percentage reported by~\citet{trilling2023solar}.

Table~\ref{tab:matchfraclcdb} also reports the upper bound match fraction. We define the upper bound as whether one of the three datasets (\snapshotdata, \tessdata or \intersectsnapstess) obtains the correct period solution for an object. We find that the upper bound exact match fraction is 82\%, implying that for 18\% of the objects, none of the three datasets are able to derive the correct period solution. As we will show in future sections, the 18\% of objects that we do not derive the correct period solution for are largely low amplitude objects.

Regarding exact matches, our combined \intersectsnapstess dataset is within $\approx$4.4\% of the upper bound and when we include the aliased period solutions that are a factor $0.5\times$$\pm3\%$ or $2\times$$\pm3\%$ of the \lcdbdata period solution, our results are within $\approx$5.6\% of the upper bound. 



\begin{figure*}[!t]
\centering

\includegraphics[width=1\textwidth]{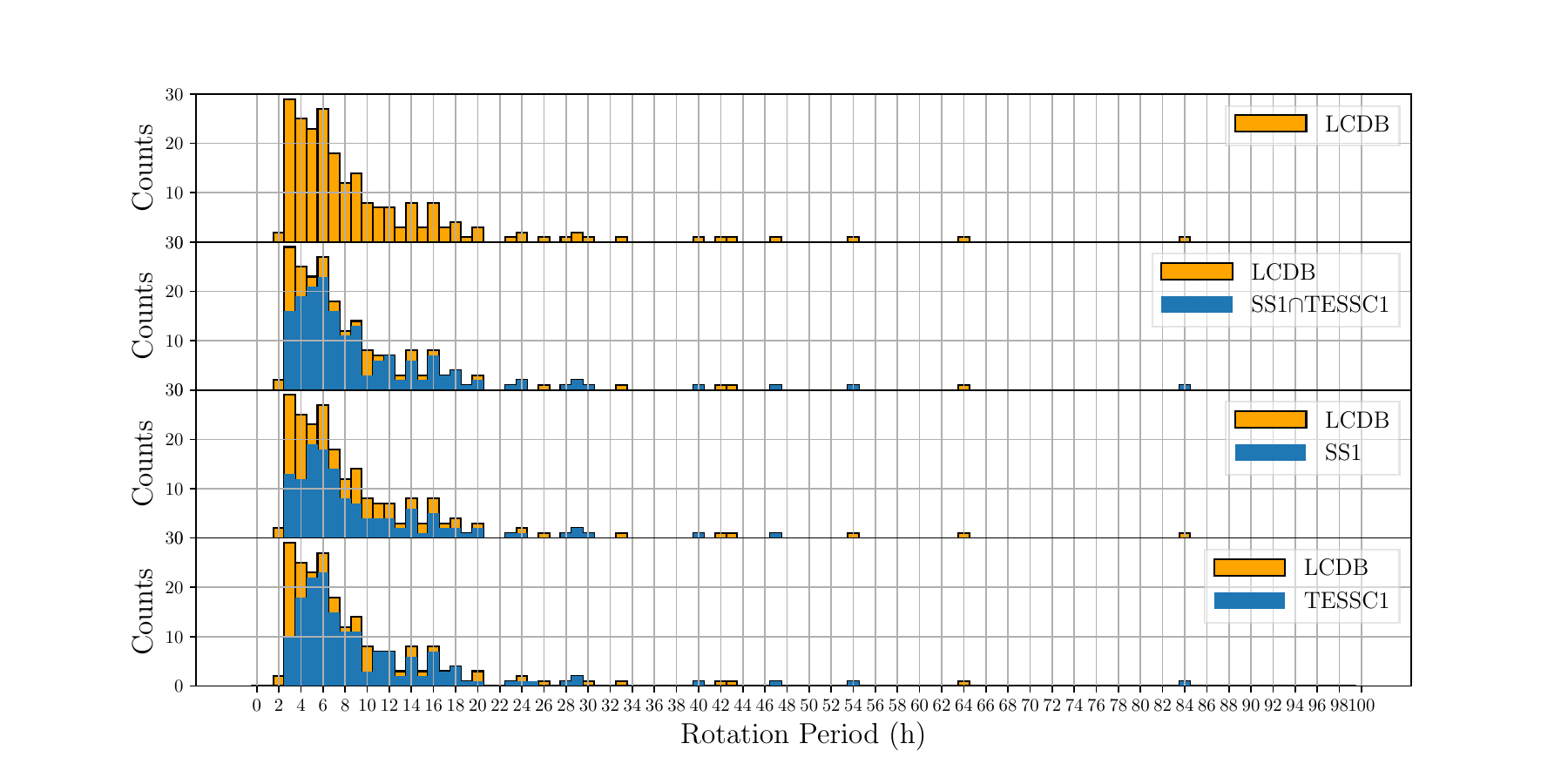}

    \caption{The distribution of correct rotation periods as a function of period for the 222 objects in \intersectlcdbdata across the four datasets is shown. Each rotation period bin is one hour, and the bins are centered on integer period values. The top panel shows the \lcdbdata period distribution, where each object has been assigned an unambiguously correct rotation period. The three lower panels from top to bottom are as follows: \intersectsnapstess, \snapshotdata, and \tessdata, where the (correct) \lcdbdata distribution is plotted behind the period distribution shown for each of these three datasets. Only those objects that have an exact match the \lcdbdata period are plotted.}
   \label{fig:LCDB_period_distribution_correct}
\end{figure*}

Figure~\ref{fig:LCDB_period_distribution_correct} shows histograms limited to objects with the exact correct rotation period as summarized in Table~\ref{tab:matchfraclcdb} as a function of rotation period. By plotting the histogram as a function of rotation period, we can observe whether any of the datasets yield period solutions that are deficient at a particular period range.  Plotted in the top panel is the true period distribution from \lcdbdata, and this histogram is plotted behind the three other histograms in the bottom three panels. This makes it straightforward to observe at what rotation periods the three datasets are deficient at finding the correct rotation period, which is denoted by an excess orange outlined bar compared to the blue bar at a given period.

Overall, we find that none of the datasets are able to detect the periods in the smallest bin at 2~h and the datasets are generally unable to recover many of the periods at $2<p<4$~h. This is consistent with~\citet{mcneill2023tess} that found that TESS rotation periods were unreliable at $\leq3$~h and \citet{trilling2023solar} found a similar result. Deriving the periods for fast rotating asteroids is difficult with TESS because of the 30 minute cadence of TESS full-frame images. In the case of ZTF, sparse photometry makes it challenging to constrain asteroid rotation periods when they are rapidly rotating, and it is for this reason that \citet{trilling2023solar} did not search for rotation periods $<2$~h. At rotation periods $\gtrsim 4$~h, we find no discernible trends; however, because there are only 222 objects in the sample, it is possible that there are trends beyond $\gtrsim 4$~h, but these cannot be recovered due to small number statistics.

\subsection{Improving Confidence in Period Solutions}\label{sec:improving_confidence}

As shown in Section~\ref{sec:baseline}, the baseline approach using \intersectsnapstess achieves a match percentage of 78.4-82.9\%. We highlight that obtaining a period match rate of $\sim$80\% is outstanding; therefore, improving the match rate much beyond this level will only recover periods at the margins in few cases. Thus, in what follows, we attempt to recover periods in these marginal cases, while simultaneously improving the match fraction such that we have greater confidence in the derived period solutions. The results are summarized in Table~\ref{tab:methods}.

\begin{deluxetable*}{llrrrrrrrl}
\tablecaption{Comparison of the fraction and number of matches of the objects in \intersectlcdbdata as a function of optimizations (1)--(7). $n_{LCDB}$ refers to the number of objects in \intersectlcdbdata after discarding objects with periods below the period cutoff ($p_{min}$) and $n_{amp}$ refers to the number of objects in the sample after discarding those with lightcurve amplitudes below the threshold using method (7). When $n_{LCDB}$ and $n_{amp}$ are equivalent, this implies that no amplitude threshold was utilized. The match fraction is the total number of objects with a correct rotation period as a ratio of $n_{amp}$. The utility column refers to whether the optimization should be used (\checkmark) or not (\xmark).}\label{tab:methods}
\tablewidth{\columnwidth}
\tabletypesize{\scriptsize}
\tablehead{
&&&&&\colhead{Exact}&\colhead{Exact}&\colhead{Incl. Aliased}&\colhead{Incl. Aliased}&\\
\colhead{Method}&Sec.&\colhead{$n_{LCDB}$}&\colhead{$p_{min}$}&\colhead{$n_{amp}$}&\colhead{Match Frac.}&\colhead{Number}&\colhead{Match Frac.}&\colhead{Number}&\colhead{Utility}
}

\startdata
Baseline: \intersectsnapstess (Table~\ref{tab:matchfraclcdb})&&222&$\geq 2$~h&222&0.784&174&0.829&184\\\hline
(1) $2\sigma$ clipping&\ref{sec:sigma_clipping}&222&$\geq 2$~h&222&0.797&177&0.842&187&\checkmark\\
(2) $3\sigma$ clipping&\ref{sec:sigma_clipping}&222&$\geq 2$~h&222&0.788&175&0.833&185&\xmark\\
(3) Periodogram masking: Aliases (0.1 hour mask width)&\ref{sec:period_masking_aliases}&222&$\geq 2$~h&222&0.788&175&0.833&185&\checkmark\\
(4) Periodogram masking: $p < 0.9(t_{window})$~h&\ref{sec:period_masking_window}&222&$\geq 2$~h&222&0.784&174&0.829&184&\xmark\\
(5) Replace with \tessdata period at high confidence level&\ref{sec:replace_TESSC1}&222&$\geq 2$~h&222&0.793&176&0.838&186&\checkmark\\
(6) Replace with \snapshotdata period at high confidence level&\ref{sec:replace_SS1}&222&$\geq 2$~h&222&0.779&173&0.824&183&\xmark\\
(7) Excluding amplitudes $\leq 0.075$&\ref{sec:low_amplitude_object_exclusion}&222&$\geq 2$~h&203&0.793&161&0.823&167&\checkmark\\\hline
\{1, 3, 5, 7\} Combined &\ref{sec:combined_optimizations}&222&$\geq 2$~h&196&0.827&162&0.852&167&\xmark\\
\{1, 3, 5, 7\} Combined &\ref{sec:combined_optimizations}&206&$\geq 3$~h&182&0.868&158&0.890&162&\checkmark\\
\{1, 3, 5, 7\} Combined &\ref{sec:combined_optimizations}&173&$\geq 4$~h&151&0.881&133&0.907&137&\xmark\\\hline
\enddata
\end{deluxetable*}

\subsubsection{Sigma Clipping}\label{sec:sigma_clipping}
The \snapshotdata dataset reported in~\citet{trilling2023solar} has been filtered upstream from the SNAPS broker by the ANTARES broker~\citep{ANTARES,2021AJ....161..107M}. In this process, observations with low real-bogus scores are removed from consideration. Therefore, the observational records for each object are largely free from contamination by poor observations. The \tessdata dataset reported in~\citet{mcneill2023tess} has been $\sigma$ clipped to remove outliers in the observational records for each object. 

While each of the \snapshotdata and \tessdata datasets have individually had outliers removed from the observational records for each object, when combining the datasets together to create \intersectsnapstess, new outliers in the combined observational record may be introduced. Consequently, we investigate removing outliers by removing observations with magnitudes that exceed 2$\sigma$ or 3$\sigma$ from the median magnitude. We do this using the sigma clip function in Astropy~\citep{robitaille2013astropy}, {\tt astropy.stats.sigma\_clip}, with a single iteration. 

We compare the match fraction to \lcdbdata for both 2$\sigma$ and 3$\sigma$ levels. Table~\ref{tab:methods} shows that 3$\sigma$ clipping (the standard level) is able to recover the correct period for an additional object over the baseline for both exact and aliased period solutions. Furthermore, 2$\sigma$ is able to recover the correct period for 3 additional objects over the baseline for both exact and aliased period solutions. This demonstrates that new outliers were introduced when combining the datasets, and that $\sigma$ clipping can improve the total match fraction. We caution that $\sigma$-clipping below the 2$\sigma$ level will remove observations from objects that have a high lightcurve amplitude, so we do not investigate removing outliers below the 2$\sigma$ level.

\subsubsection{Periodogram Masking: Aliases}\label{sec:period_masking_aliases}
The diurnal observing schedule of ground-based telescopes impart a periodic signal into the periodogram. Excluding periods at aliases such as $p\in$\{16, 24, 48\}~h is straightforward --- if the peak power in the periodogram is at one of these aliases then a secondary peak is selected. The drawback of this method is that a true period at these aliases are rejected despite the fact that asteroids may truly have these rotation periods (e.g., see Figure~\ref{fig:ZTF_TESS_period_distribution}, lower panel or Figure~\ref{fig:LCDB_period_distribution_correct}, top panel). 

Recently, \citet{Erasmus2021} used masking to detect slowly rotating asteroids.  \citet{Coughlin2021} used the masking method when examining variable stars in ZTF DR2\footnote{https://www.ztf.caltech.edu/ztf-public-releases.html}. \citet{KRAMER2023} compared several methods that remove aliases and found that masking
is both the simplest and most effective method on the Legacy Survey of Space and Time (LSST) Solar System Products Data Base (SSPDB)~\citep{SSPDB2021}, which is used to simulate the LSST observational records, and was competitive with the Monte Carlo method on the \snapshotdata dataset. Consequently we examine the use of the masking method here.

\begin{figure}[!t]
\centering

\includegraphics[width=1\columnwidth]{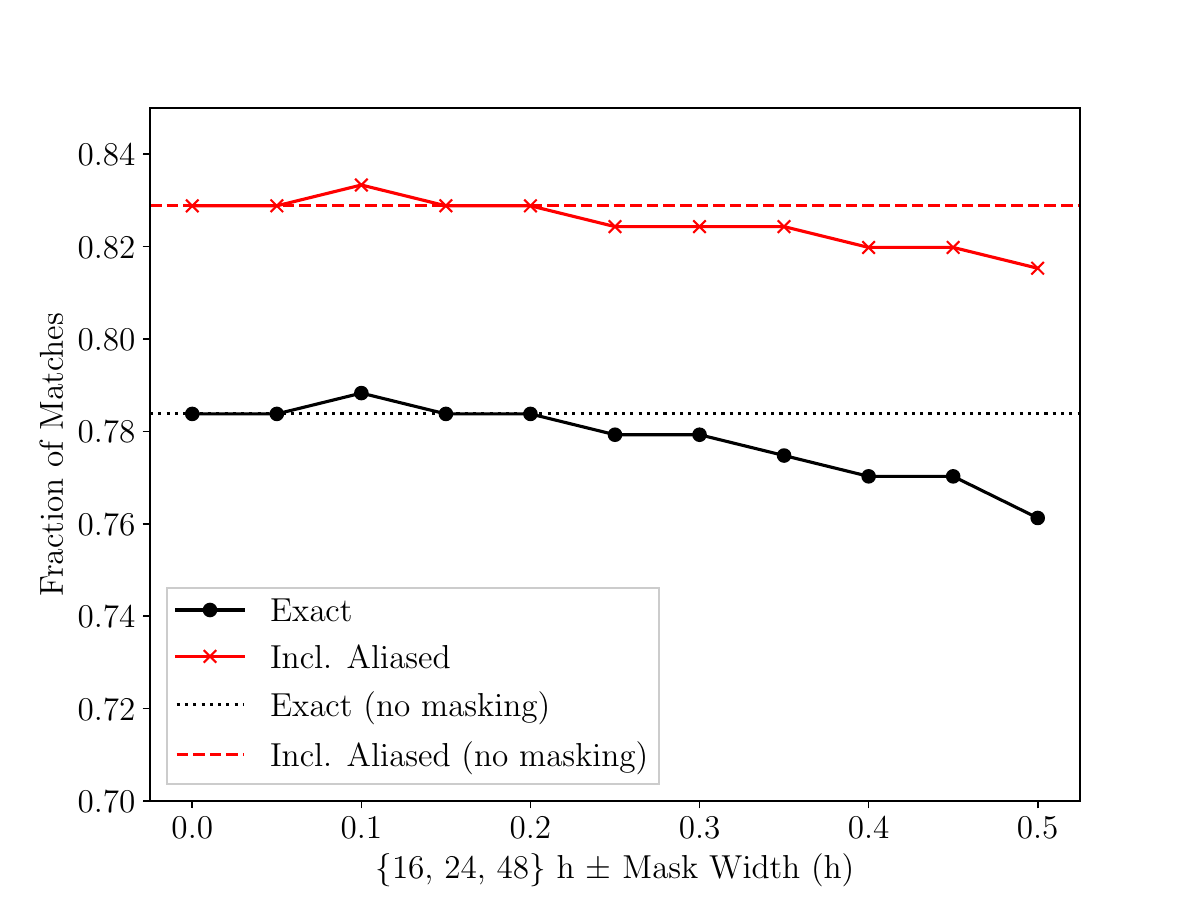}
    \caption{The fraction of exact and aliased matches is shown as a function of mask width. Masks are applied at $p\in$\{16, 24, 48\}~h. The mask ranges are reported as $\pm$ the mask width on the horizontal axis. For example, a mask width of 0.1 around the 16~h alias refers to masking the range 15.9--16.1~h. The mask with of 0 refers to the match fraction without any masking. For clarity, the exact and aliased match fractions without masking are shown as the black dotted and red dashed horizontal lines, respectively.}
   \label{fig:masking}
\end{figure}

Figure~\ref{fig:masking} shows the match fraction to \lcdbdata as a function of the mask width around  the \{16, 24, 48\}~h aliases. Here, the mask widths are centered on the abovementioned aliases and are reported as $\pm$ the mask width as shown on the horizontal axis. We find that masking does not significantly improve the overall match fraction. Also, as the mask width increases the total match fraction decreases. This is expected as we increasingly eliminate larger fractions of the frequency range that may contain the true period of an object.

This result is consistent with that of~\citet{KRAMER2023} which showed that masking only improved the match percentage from 64.8\% to 65.8\% on the \snapshotdata dataset. The reason this occurs is that the secondary peak selected outside of the masked region of the periodogram is often incorrect. Despite this result on the \snapshotdata dataset, \citet{KRAMER2023} showed that on the SSPDB dataset, masking improved the match percentage from 57.9\% to 74.5\% demonstrating that the method is expected to work well for the forthcoming LSST catalog. 

We reiterate that masking is irrelevant for the \tessdata dataset as it does not suffer from a diurnal observing schedule, so when we combine the \snapshotdata and \tessdata datasets, \tessdata may eliminate aliases which diminishes the utility of the masking method. Despite this, we expect that recovering the periods for a few additional (marginal) objects is worthwhile even if the number of objects is not statistically significant here. Thus, when we combine optimizations, we use this method and apply masks at \{16, 24, 48\}$\pm 0.1$~h.

\subsubsection{Periodogram Masking: $p \geq 0.9(t_{window})$}\label{sec:period_masking_window}
Ground-based observatories, such as ZTF and the forthcoming Rubin Observatory, will have a long observational baseline for each object. This will allow deriving long rotation periods for a small fraction of asteroids. In contrast, \tessdata has a much shorter observational baseline for each object. The telescope stares at a sector and then transitions to the next sector, where it is unlikely that an asteroid is captured in more than one sector. As a result, the maximum rotation period that can be detected with \tessdata is roughly 816 hours (see~\citet{mcneill2023tess} for more information). Let $t_{window}=t_{end}-t_{start}$, where $t_{start}$ and $t_{end}$ refer to the first and last times that the object has been observed. Then, if $p>t_{window}$ for an object, this implies that the lightcurve will be poorly sampled as only a partial lightcurve will be produced. Consequently, we mask all periods for an object where $p\geq0.9(t_{window})$ such that we reject periods that are within 10\% of the observing window, as periods derived within the last 10\% of the observing window are well-known to be unreliable~\citep{mcneill2023tess}.

From Table~\ref{tab:methods}, we find that this method has no impact on the total match fraction. In this set of 222 objects, there were no instances where $p\geq0.9(t_{window})$. While this filter was important for deriving asteroid periods for \tessdata, when the data record is combined with \snapshotdata, each object is assigned a much larger $t_{window}$. Thus, this method is unnecessary when applied to the \intersectsnapstess dataset.

\subsubsection{Period Replacement with High Confidence \tessdata Solutions}\label{sec:replace_TESSC1}
\citet{mcneill2023tess} showed that objects having a sufficient number of observations and \lsp power are more likely to produce the correct period solution than objects having few observations and low \lsp power. This is intuitive, as the greater the number of observations, the more likely that a good model fit is obtained, and furthermore, greater \lsp power typically indicates that the periodogram has a high signal-to-noise ratio and that the lightcurve has a reasonably high amplitude. \citet{mcneill2023tess} showed that without assigning a threshold for the \lsp power, they achieved a match fraction to \lcdbdata of 0.65. In contrast, for those objects observed $\geq$200 times and having a \lsp power $\geq$0.2, the match fraction increases to 0.85, and this further increases to 0.91 if aliased periods are also considered.

Given the above, if a period in the \tessdata dataset has a high confidence of being correct ($\geq$200 observations and an \lsp power $\geq$0.2), then we simply select this period instead of using the period derived from \intersectsnapstess. This approach also has the benefit of potentially validating solutions at the \{16, 24, 48\}~h aliases. From Table~\ref{tab:methods} we find that this method is able to recover two additional objects relative to the baseline.

\subsubsection{Period Replacement with High Confidence \snapshotdata Solutions}\label{sec:replace_SS1}
\begin{figure}[!t]
\centering
\subfigure[]{

      \includegraphics[width=1\columnwidth]{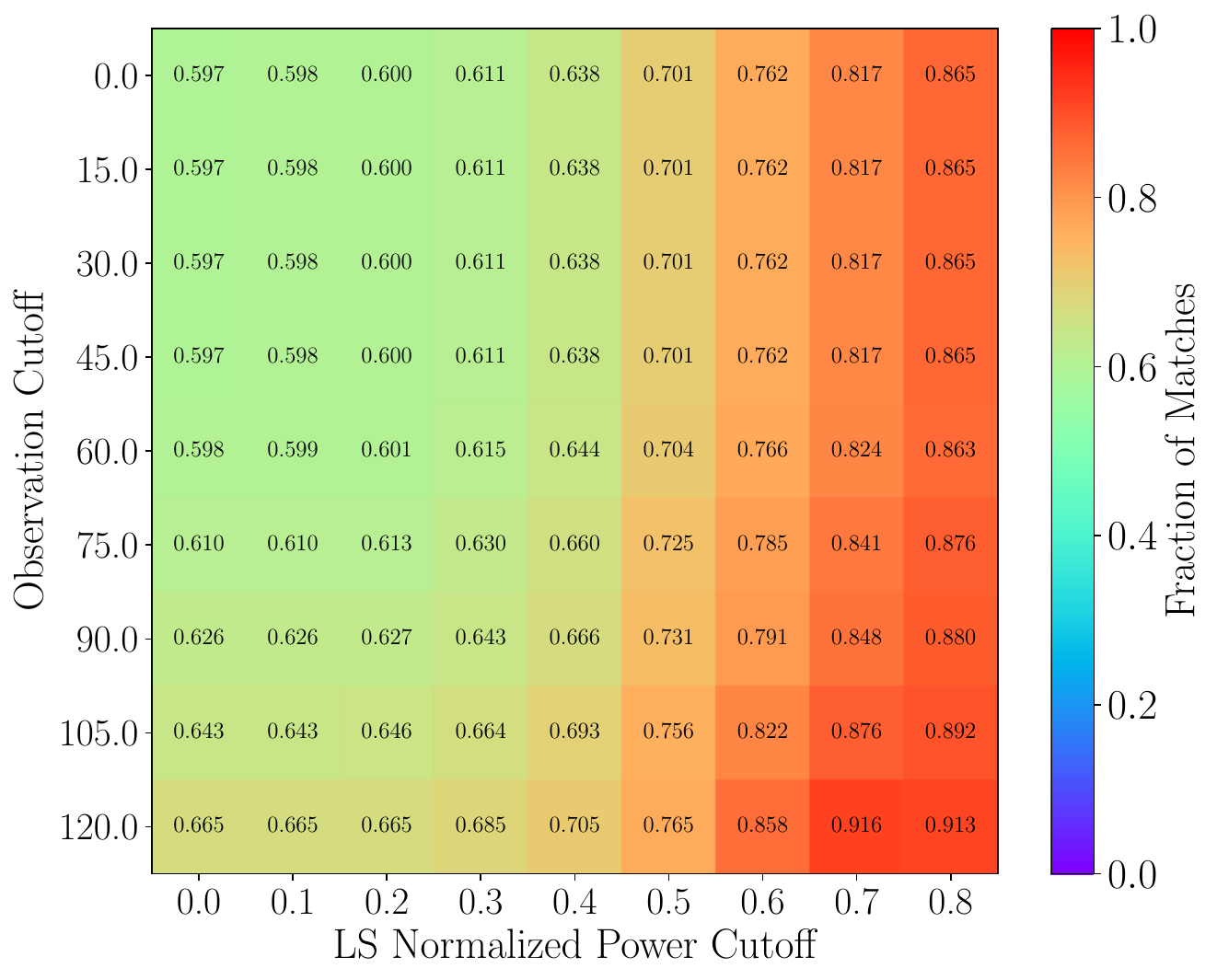}
       
}
\subfigure[]{

       \includegraphics[width=1\columnwidth]{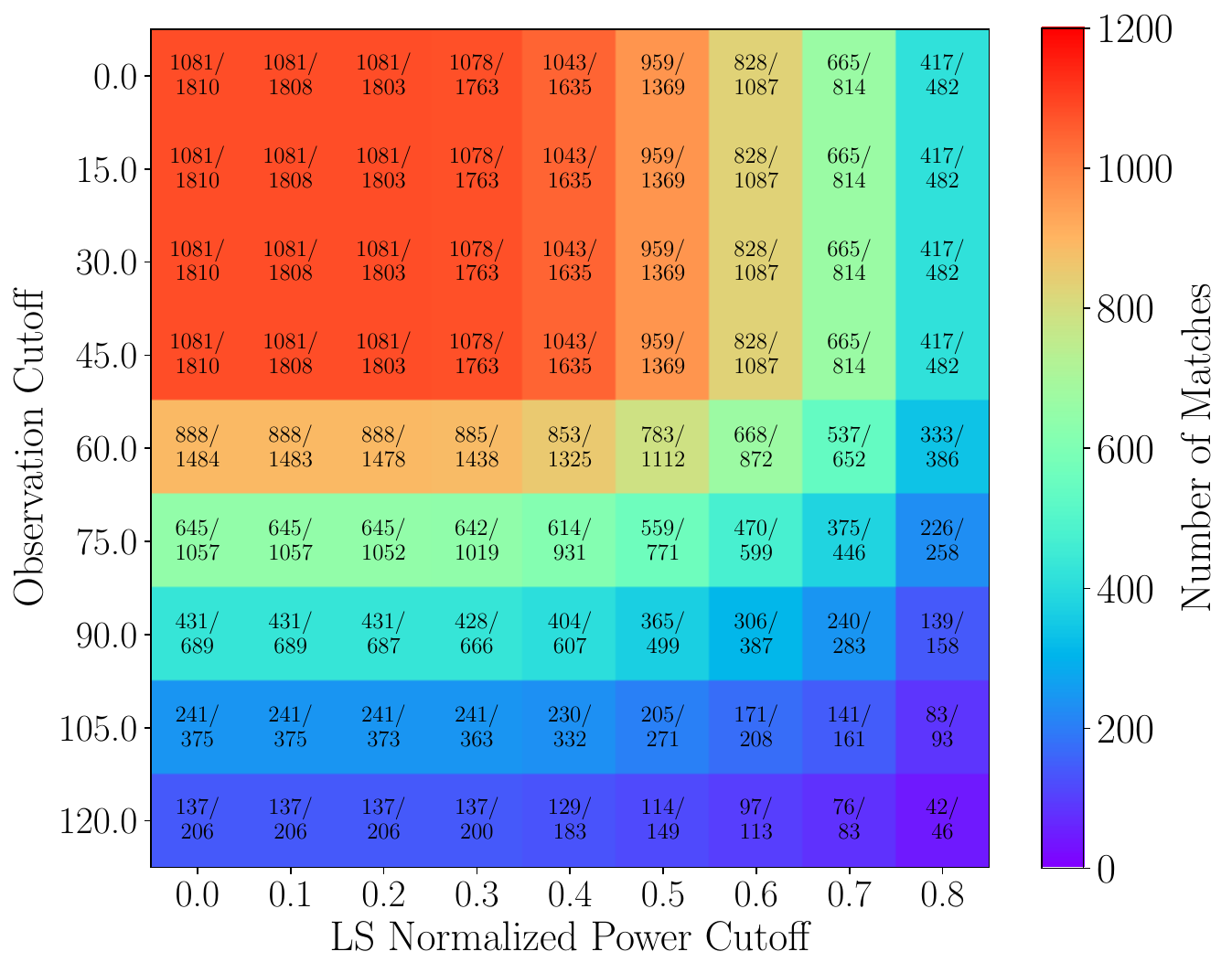}
       
}  		
    \caption{Cumulative heatmaps showing the rotation period match fraction (and number) of objects in \snapshotdata as a function of both the number of observations and the \lsp power. In the sample, there are a total of 1,810 objects which are those that are found in \lcdbdata ($U=3$) and \snapshotdata. (a) The exact rotation period match; and, (b) the number of objects in the bins shown in (a).}
   \label{fig:snapshot1_heatmap}
\end{figure}

Similarly to period replacement with high confidence \tessdata solutions, we also examine replacement using \snapshotdata. Figure~\ref{fig:snapshot1_heatmap} shows a similar heatmap to that outlined in \citet{mcneill2023tess} which plots a cumulative heatmap of period matches to \lcdbdata as a function of the observation threshold and \lsp power threshold. With unconstrained observational and \lsp power thresholds we obtain an exact match fraction of 0.597, but this increases to 0.85 when there are $\gtrsim$100 observations and an \lsp power of 0.7. 

If a period in the \snapshotdata dataset has a high confidence of being correct ($\geq$105 observations and an \lsp power $\geq$0.7), then we select this period instead of using the period derived from \intersectsnapstess. From Table~\ref{tab:methods} we find that this method does not recover any additional correct rotation periods relative to the baseline. Furthermore, based on Figure~\ref{fig:snapshot1_heatmap}(b), this only captures a small fraction of the objects in the sample because the number of observations and \lsp power thresholds need to be very high which limits its utility for period replacement. Consequently, when we combine optimizations later, we do not use this method. 

Note that future SNAPS data releases will contain significantly more ZTF observations per object which will likely increase the utility of this approach.

\begin{figure}[!t]
\centering
\subfigure[]{

       \includegraphics[width=0.9\columnwidth]{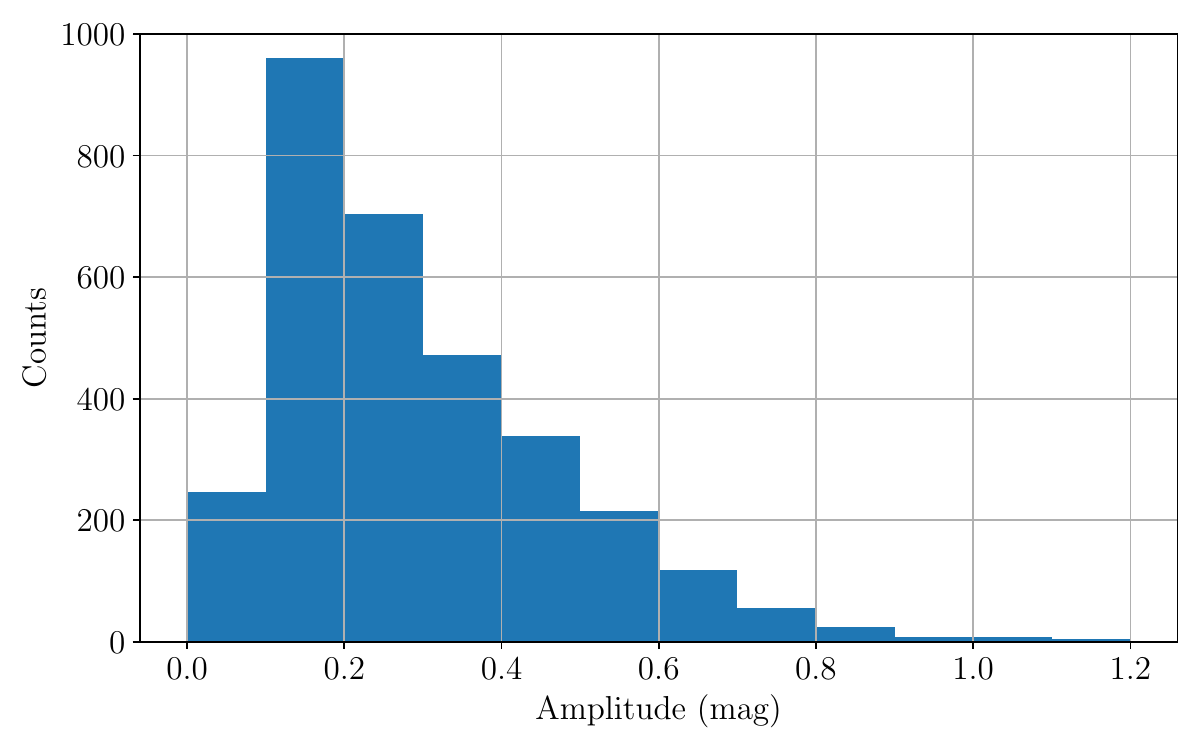}

}
\subfigure[]{

       \includegraphics[width=0.9\columnwidth]{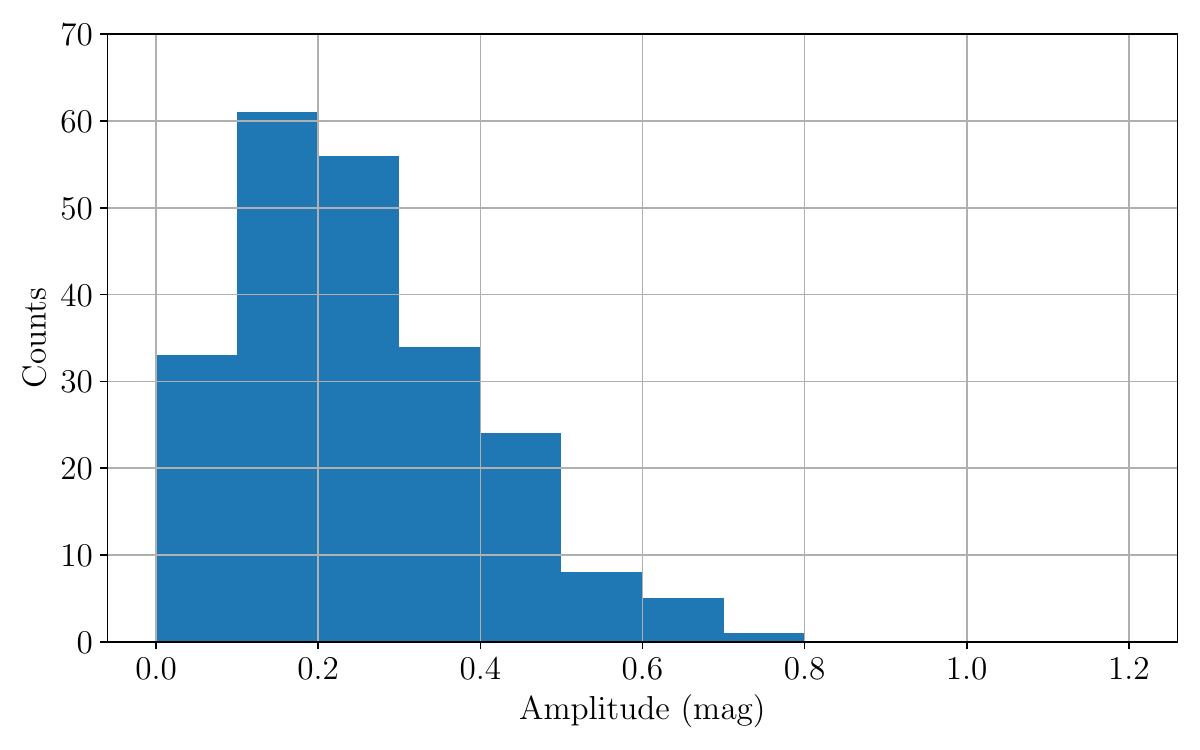}

}       
\subfigure[]{

       \includegraphics[width=0.9\columnwidth]{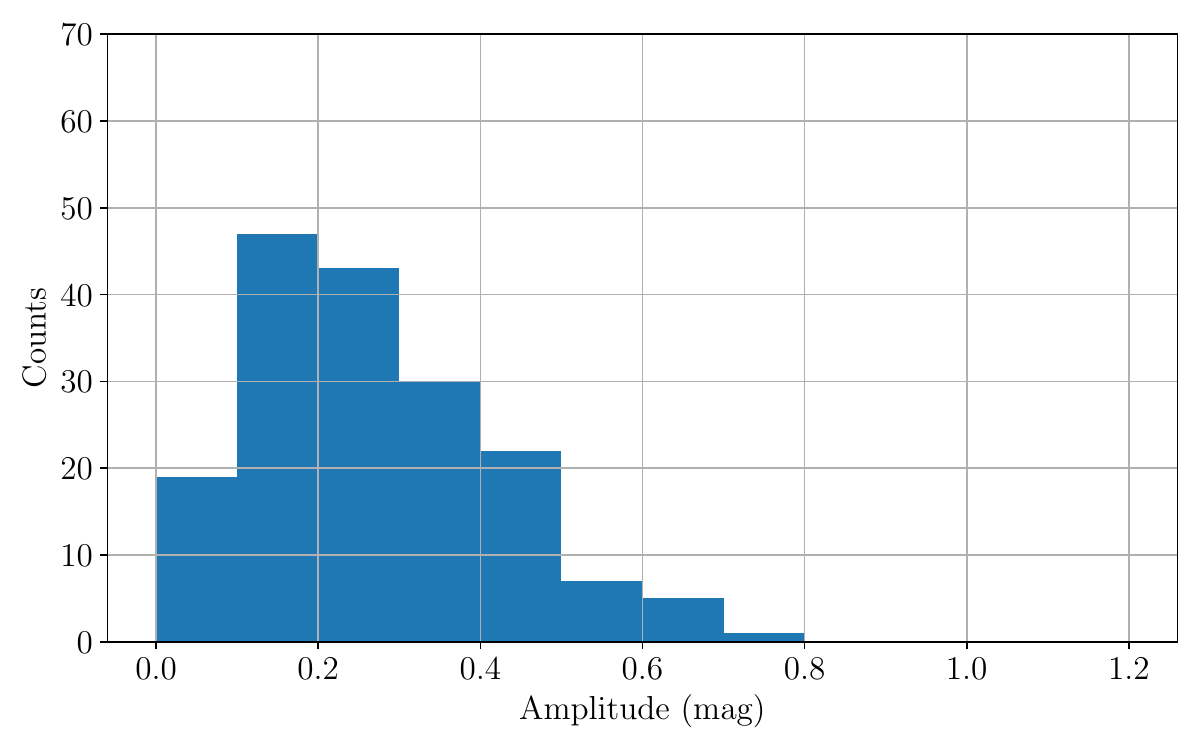}

}       
    \caption{Lightcurve amplitude distributions in 0.1 magnitude bins. (a) Amplitudes of the 3168 objects in the \intersectsnapstess dataset. (b) Amplitudes of the 222 objects in the \intersectlcdbdata dataset. (c) is the same as (b) except limited to the 174 exact matches to the \lcdbdata and so (b) and (c) are plotted on the same scale for comparison purposes.}
   \label{fig:amplitude_histograms}
\end{figure}

\subsubsection{Excluding Low Amplitude Objects}\label{sec:low_amplitude_object_exclusion}
We now examine the lightcurve amplitude distribution of the objects in the datasets. Amplitudes are derived using the sinusoidal fit of the lightcurve  for each object using the peak and trough of the curve. On average, this will underestimate the amplitudes of asteroids, as a sinusoid will often not capture the peak and trough within a lightcurve. However, this is preferable to alternative methods, such as taking the difference between the minimum and maximum magnitudes in a lightcurve which may dramatically overestimate the amplitude due to outlying measurements with low and/or high magnitudes.

Figure~\ref{fig:amplitude_histograms}(a) plots the amplitude distribution of the 3168 objects in the \intersectsnapstess dataset, which shows that there is a deficit of amplitudes $\leq 0.1$ mag. This is due to bias in period finding, where asteroids with a spherical morphology will have a lightcurve amplitude of $\sim$0 mag, and thus it is not possible to derive rotation periods for these objects. Furthermore, lightcurves that are fit with a poor period may produce a low signal-to-noise periodogram which often generates a relatively flat (low amplitude) fit to the time series. Also, asteroids with a pole-on orientation of any shape can produce a flat lightcurve. Consequently, the deficit of amplitudes $\leq 0.1$ mag is largely unphysical and it is difficult to know the true distribution of objects with amplitudes $\leq 0.1$ mag.

Comparing Figure~\ref{fig:amplitude_histograms}(b) on the \intersectlcdbdata dataset to Figure~\ref{fig:amplitude_histograms}(a), we find that there are a greater fraction of amplitudes $\leq 0.1$ mag. This is expected as the \intersectlcdbdata sample contains objects vetted from \lcdbdata and hence the sample should be a closer match to the real amplitude distribution which cannot be determined due to the abovementioned presence of asteroids that are roughly spherical in shape. Figure~\ref{fig:amplitude_histograms}(c) shows the amplitude distribution of the \intersectlcdbdata dataset that is limited only to the 174 objects that are an exact match to \lcdbdata. Because all objects have been assigned the correct period, we plot this as a sanity check to demonstrate that a similar deficit of amplitudes $\leq 0.1$ mag is present as shown in Figure~\ref{fig:amplitude_histograms}(a)-(b). If this deficit was not shown in this plot, then this would indicate that spherical objects and poor lightcurve fits are not the cause of the deficit described above.

An additional challenge when fitting lightcurves is photometric error. We find that many of the observations in the \tessdata dataset have a photometric error $>$0.1~mag  (the photometric error with \snapshotdata is lower) and so if we assume that the average error is roughly 0.1~mag, then we would not expect to be sensitive to objects with amplitudes $\leq 0.1$~mag.

\begin{figure}[!t]
\centering 
       \includegraphics[width=0.9\columnwidth]{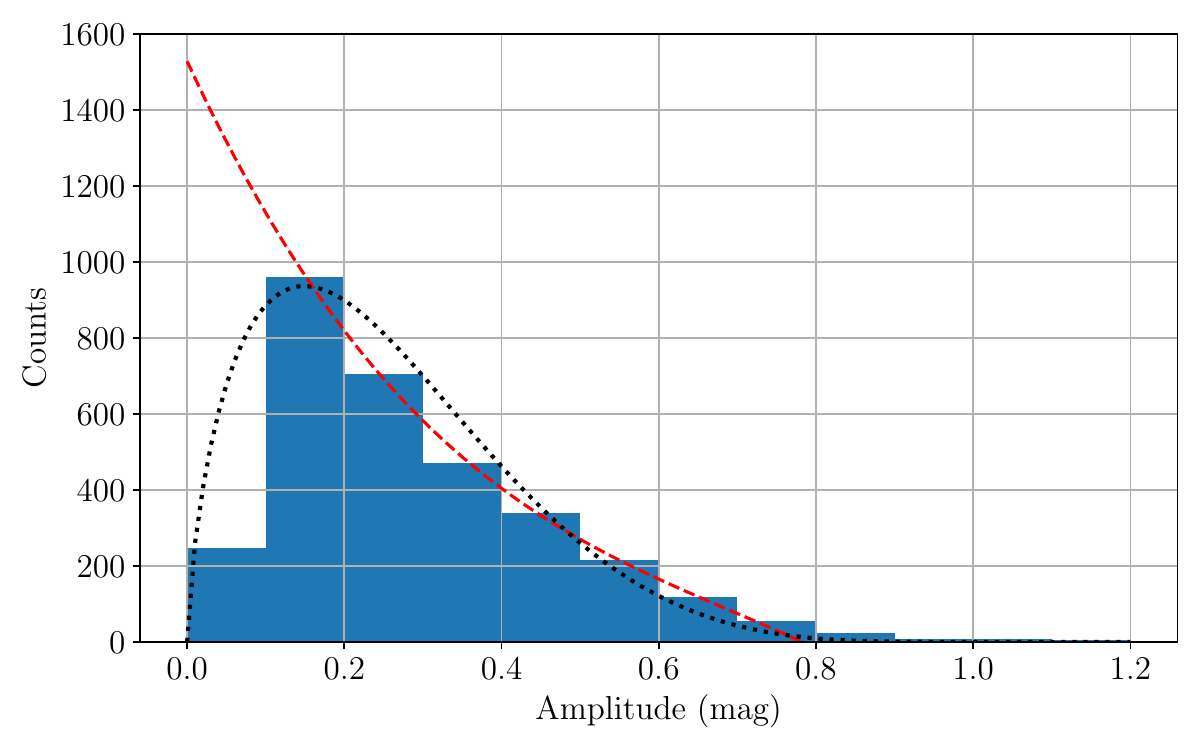}   
    \caption{The same histogram as in Figure~\ref{fig:amplitude_histograms}(a) except where two example models are shown that hypothesize the distribution of asteroids with amplitudes $<$0.1 mag. The red dashed curve shows the scenario where spherical-like objects are the most abundant. Using a $\beta$-distribution, the black dotted curve shows the scenario where there are fewer objects at $<$0.1~mag than at 0.1--0.2~mag.}
   \label{fig:amplitude_histograms_model}
\end{figure}

Figure~\ref{fig:amplitude_histograms_model} shows two plausible scenarios for the amplitude distribution at $<0.1$~mag. We describe two hypotheses as follows:
\begin{enumerate}
\item \emph{``Sphere-Abundant Hypothesis'':} The first scenario (red dashed curve) shows the case where the amplitude histogram should increase monotonically with decreasing amplitude. If this hypothesis is correct, this implies that there are numerous asteroids that are spherical in morphology and that the amplitude distribution will be difficult to accurately constrain.
\item \emph{``Sphere-Limited Hypothesis'':} The second scenario (black dotted curve) is where there are fewer objects with amplitudes $<0.1$~mag than at 0.1--0.2~mag, implying that the vast majority of asteroids are likely to be at least slightly elongated\footnote{While not directly applicable to main belt asteroids, this hypothesis was proposed for Trans-Neptunian Objects (TNOs)~\citep{BernardinelliTNO2023}.}.  If this hypothesis is correct, then the amplitude distributions with a deficit at $<$0.1~mag shown in Figure~\ref{fig:amplitude_histograms} is indicative of the true distribution.
\end{enumerate}

In the context of the above, we describe implications for rotation period derivation. If we assume that the ``Sphere-Abundant Hypothesis'' is correct, then this implies that we are unable to correctly assign $\sim$25\% of the objects their correct amplitude, and thus, it is likely that the derived rotation period  for a significant fraction of these objects is also incorrect. One important caveat to note is that it is possible to derive the correct rotation period for an asteroid but still derive a poor amplitude (i.e., asteroid lightcurves may not always be best fit by \lsp sinusoids); therefore, we would not expect that 25\% of the rotation periods are incorrect based on an analysis of lightcurve amplitudes alone. If the ``Sphere-Limited Hypothesis'' is correct, then this implies that lightcurve amplitudes cannot be the dominant reason that we obtain incorrect period solutions.

Recall that the baseline finds that there is a 0.784 exact match fraction of rotation periods (Table~\ref{tab:methods}). Thus, up to $\approx 25\%$ of rotation periods may be incorrect which is consistent with the fraction of asteroids missing from the distribution having an amplitude $<$0.1 mag (assuming that the ``Sphere-Abundant Hypothesis'' is correct). We conclude that a substantial fraction of asteroids that have an unreliable derived rotation period may be due to spherical asteroids with rotation periods that are intractable to  constrain.



\begin{figure}[!t]
\centering
\subfigure[]{

       \includegraphics[width=0.9\columnwidth]{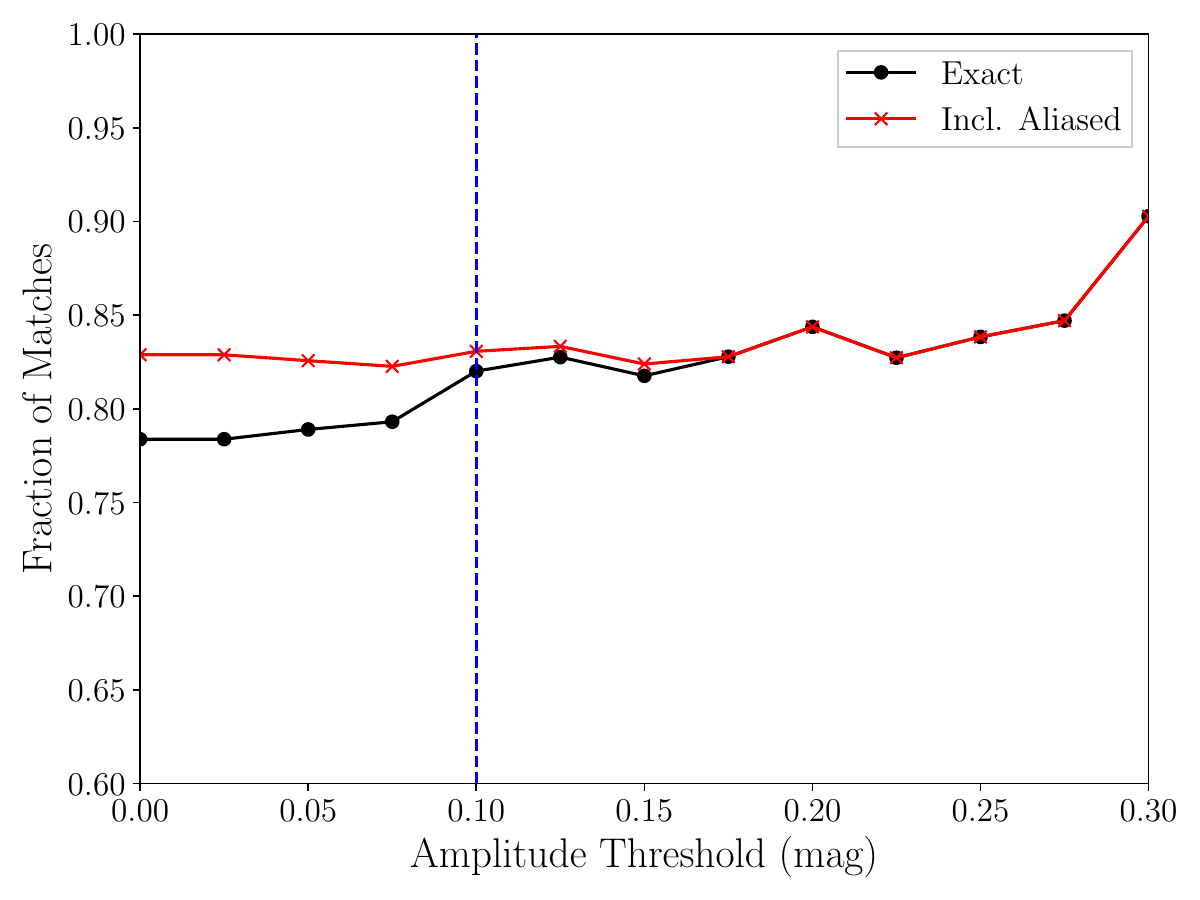}

}
\subfigure[]{

       \includegraphics[width=0.9\columnwidth]{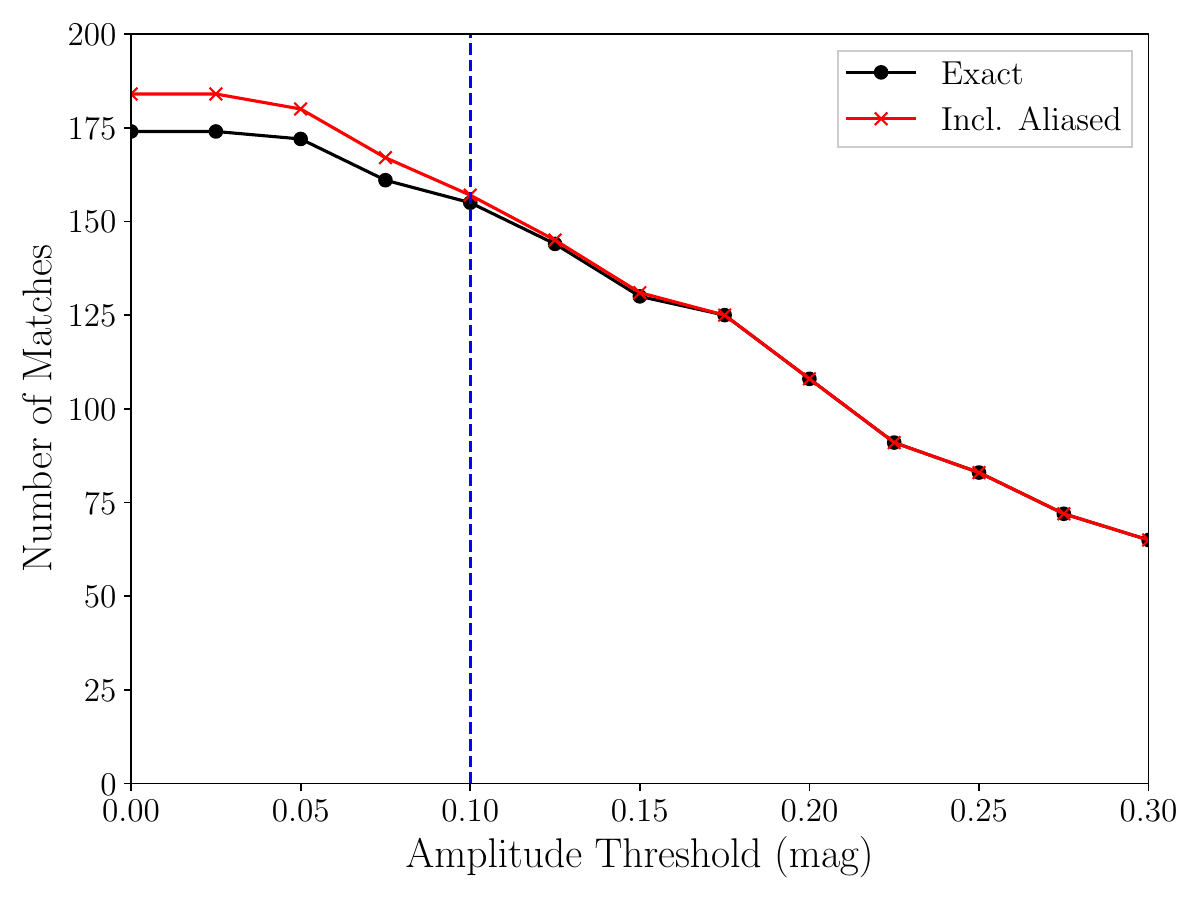}

}       
    \caption{(a) The fraction of rotation period matches (exact and including aliased) with \lcdbdata as a function of the amplitude threshold. (b) The number of objects corresponding to (a). The dashed blue horizontal line demarcates amplitudes at 0.1 mag as described in the text accompanying Figure~\ref{fig:amplitude_histograms}.}
   \label{fig:amplitude_match_fraction}
\end{figure}

Figure~\ref{fig:amplitude_match_fraction}(a) plots the fraction of objects that match \lcdbdata as a function of the amplitude cutoff. Unlike the other optimizations presented in this section, by making a cut on the amplitude, we lose some number of objects that have been assigned a correct rotation period, and so we also plot the number of objects recovered in Figure~\ref{fig:amplitude_match_fraction}(b). Consequently, we aim to select an amplitude cutoff that reaches a trade-off between yielding a high match fraction without eliminating too many objects from the sample with a correct rotation period. For instance, in an extreme case, we could achieve a $\sim$90\% match fraction by removing objects having an amplitude $<$0.3 mag, but we would only recover 65 asteroids.  

Figure~\ref{fig:amplitude_match_fraction}(a) reveals an interesting trend regarding exact vs. aliased matches where we find that the match fraction for both exact and aliased matches converge when we exclude objects with amplitudes $<$0.175 mag. This implies that the \lsp period finding algorithm produces an ambiguous solution (within a factor $0.5\times$$\pm3\%$ or $2\times$$\pm3\%$ of the \lcdbdata period solution) when there is a low amplitude, and this is to be expected as low amplitude period solutions also produce low power periodograms which implies that the peak power in the periodogram is not significantly differentiated from powers at other candidate frequencies. But when the amplitude is high, the ambiguity seems to be reduced or eliminated.

Based on the above, we select an amplitude threshold that removes all objects from the sample at $<0.075$~mag which allows for some of the objects in the 0-0.1 mag bin to be included in the sample. This reaches a trade-off between maximizing the match fraction and the number of objects that have been assigned a correct rotation period. With this amplitude threshold, we find that there are a total 203 objects (thus 19 objects are excluded), and this yields an exact (aliased) match fraction of 0.793 (0.823) corresponding to 161 (167) objects, respectively. 


\subsubsection{Summary: Combining Optimizations and Quantifying Period Confidence}\label{sec:combined_optimizations}

\begin{figure*}[!t]
\centering
\subfigure[]{

       \includegraphics[width=0.48\textwidth]{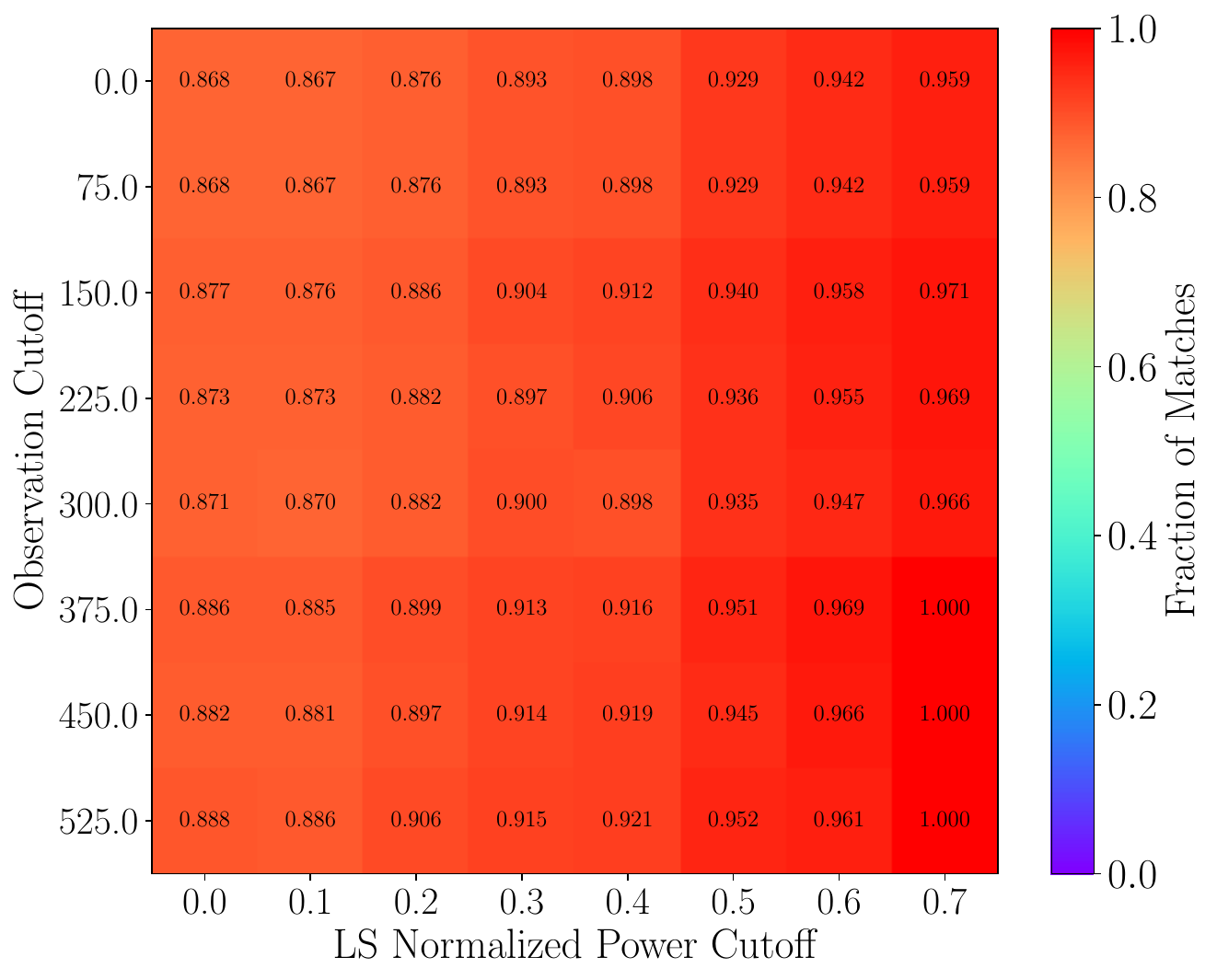}

}
\subfigure[]{

       \includegraphics[width=0.48\textwidth]{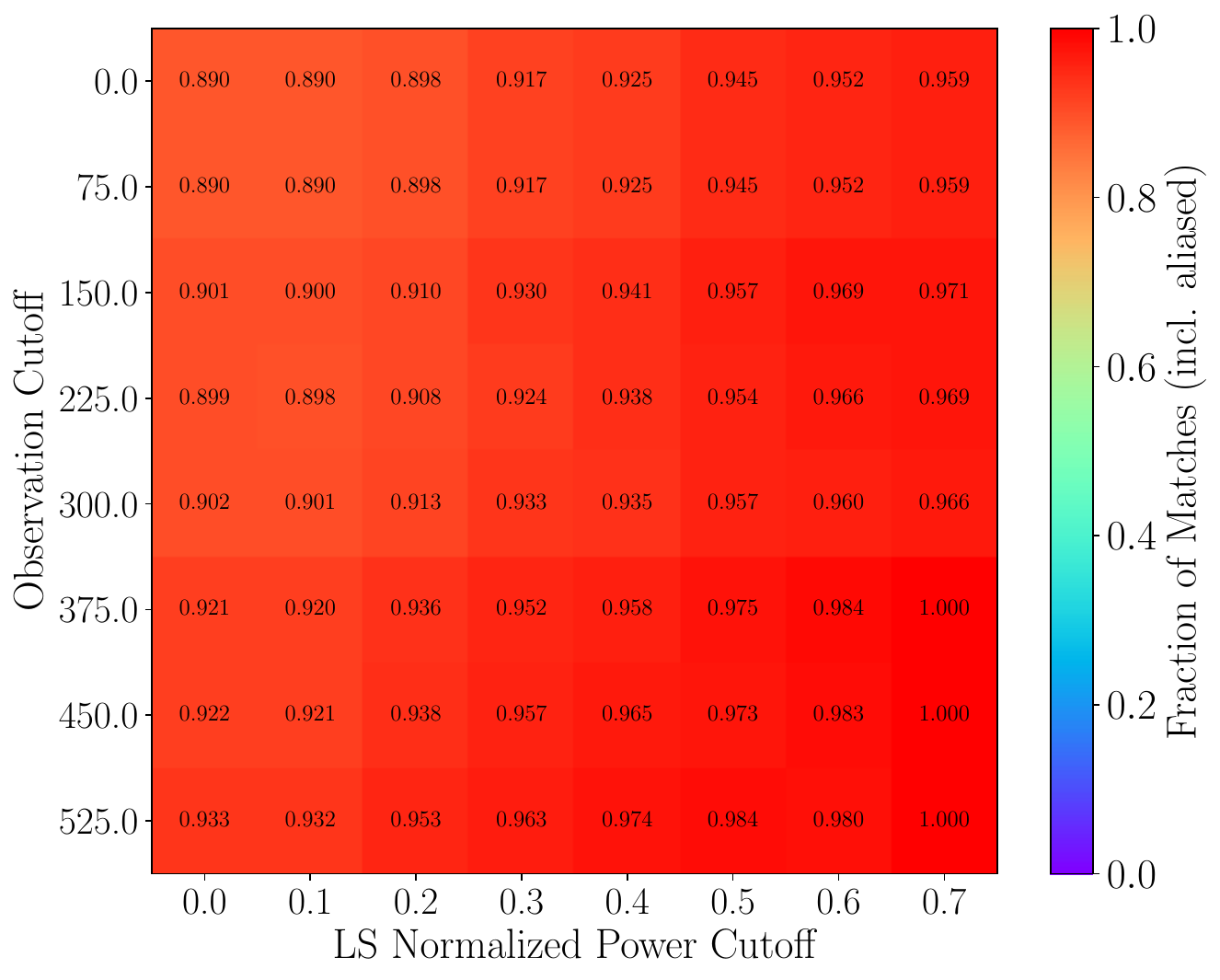}

}
\subfigure[]{

       \includegraphics[width=0.48\textwidth]{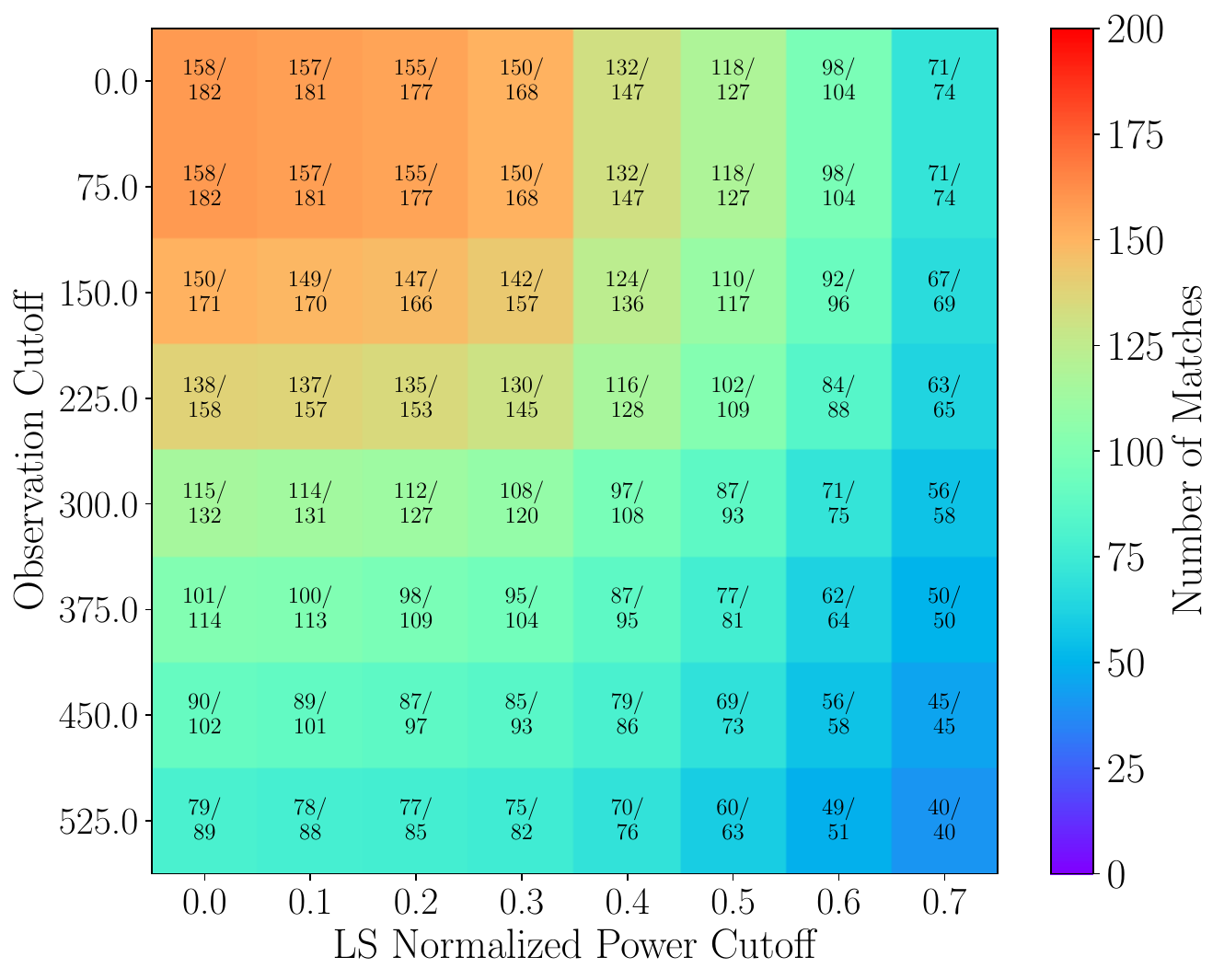}

}
\subfigure[]{

       \includegraphics[width=0.48\textwidth]{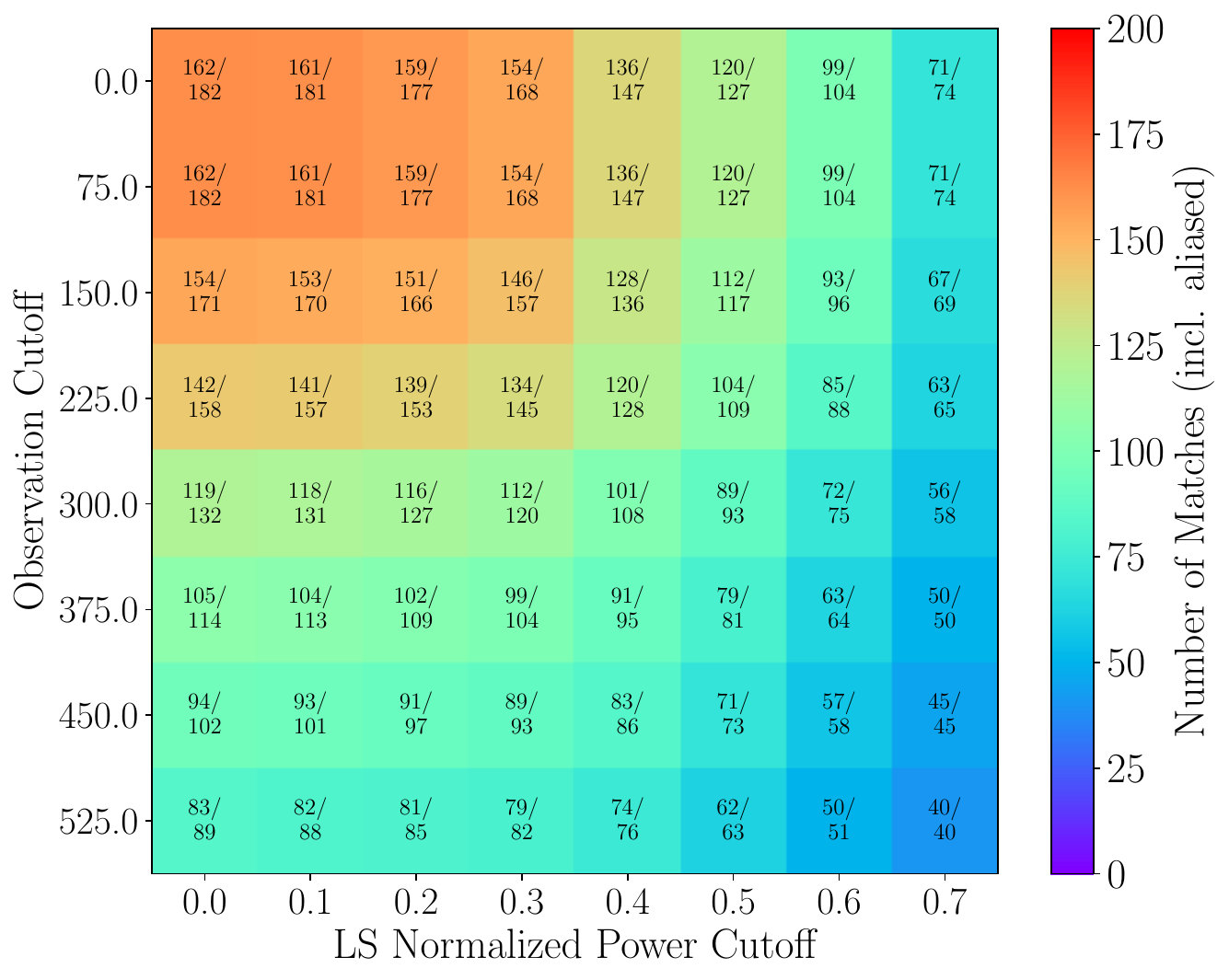}

}
    \caption{Cumulative heatmaps showing the rotation period match fraction (and number) of objects in \intersectlcdbdata as a function of both the number of observations and the \lsp power. The heatmap uses the data from the combined optimizations outlined in Table~\ref{tab:methods} with $p_{min}=3$~h. Panels (a)~and~(c) show the exact rotation period match and number, respectively. Panels (b)~and~(d) are the same as (a)~and~(c) but include the aliased rotation periods as well. }
   \label{fig:heatmap_confidence_period_gtr3h}
\end{figure*}

\begin{figure*}[!t]
\centering
\subfigure[]{

       \includegraphics[width=0.48\textwidth]{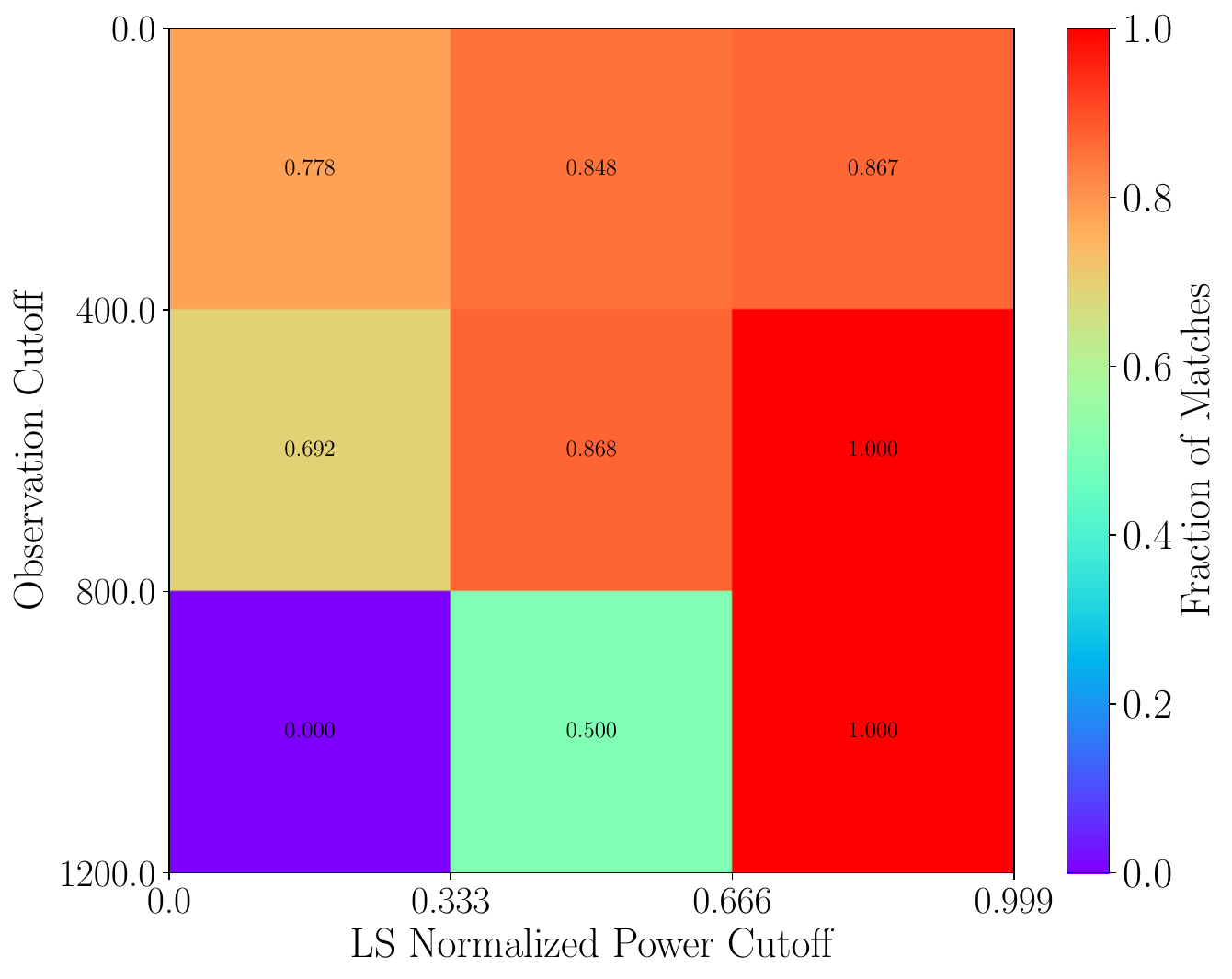}

}
\subfigure[]{

       \includegraphics[width=0.48\textwidth]{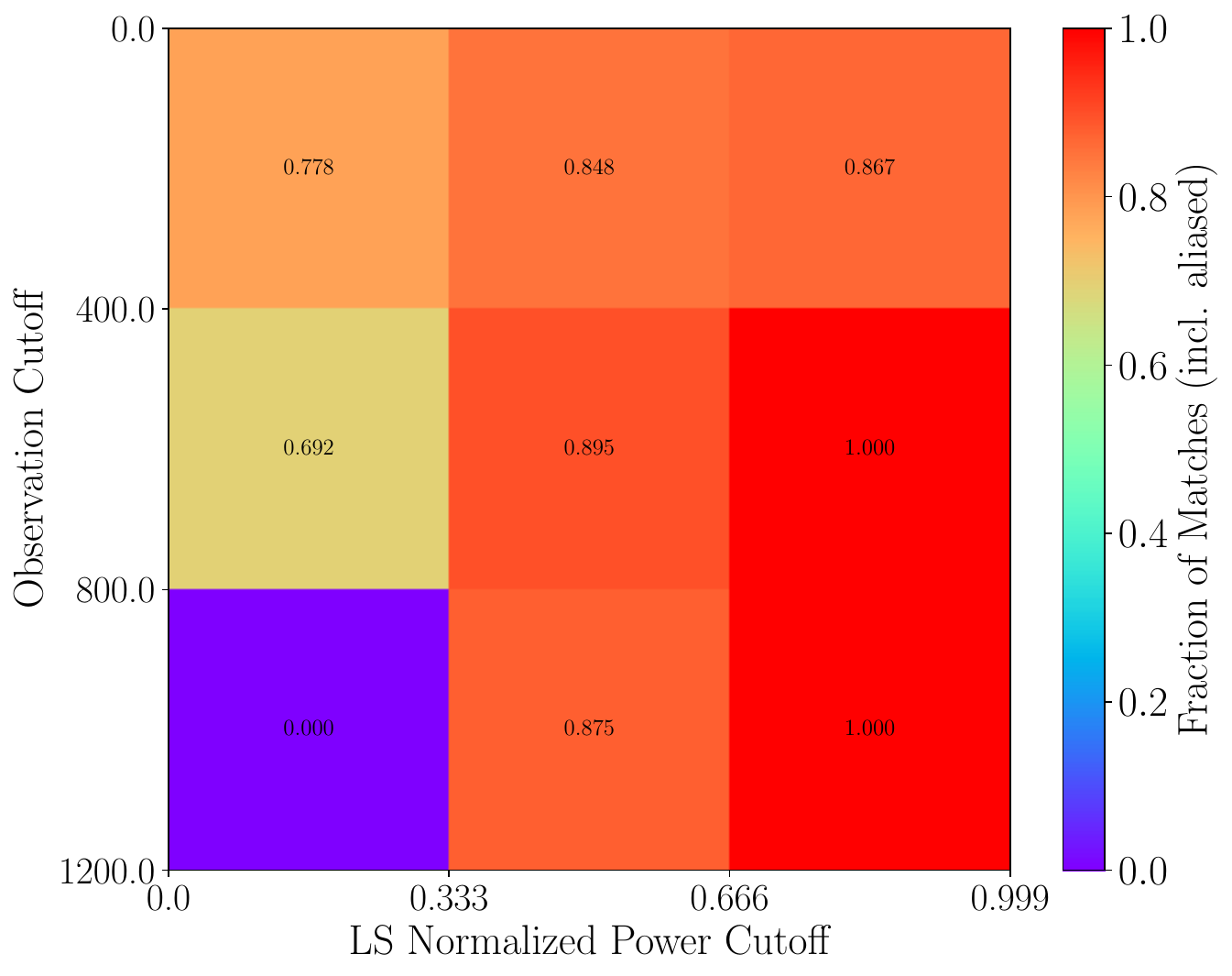}

}
\subfigure[]{

       \includegraphics[width=0.48\textwidth]{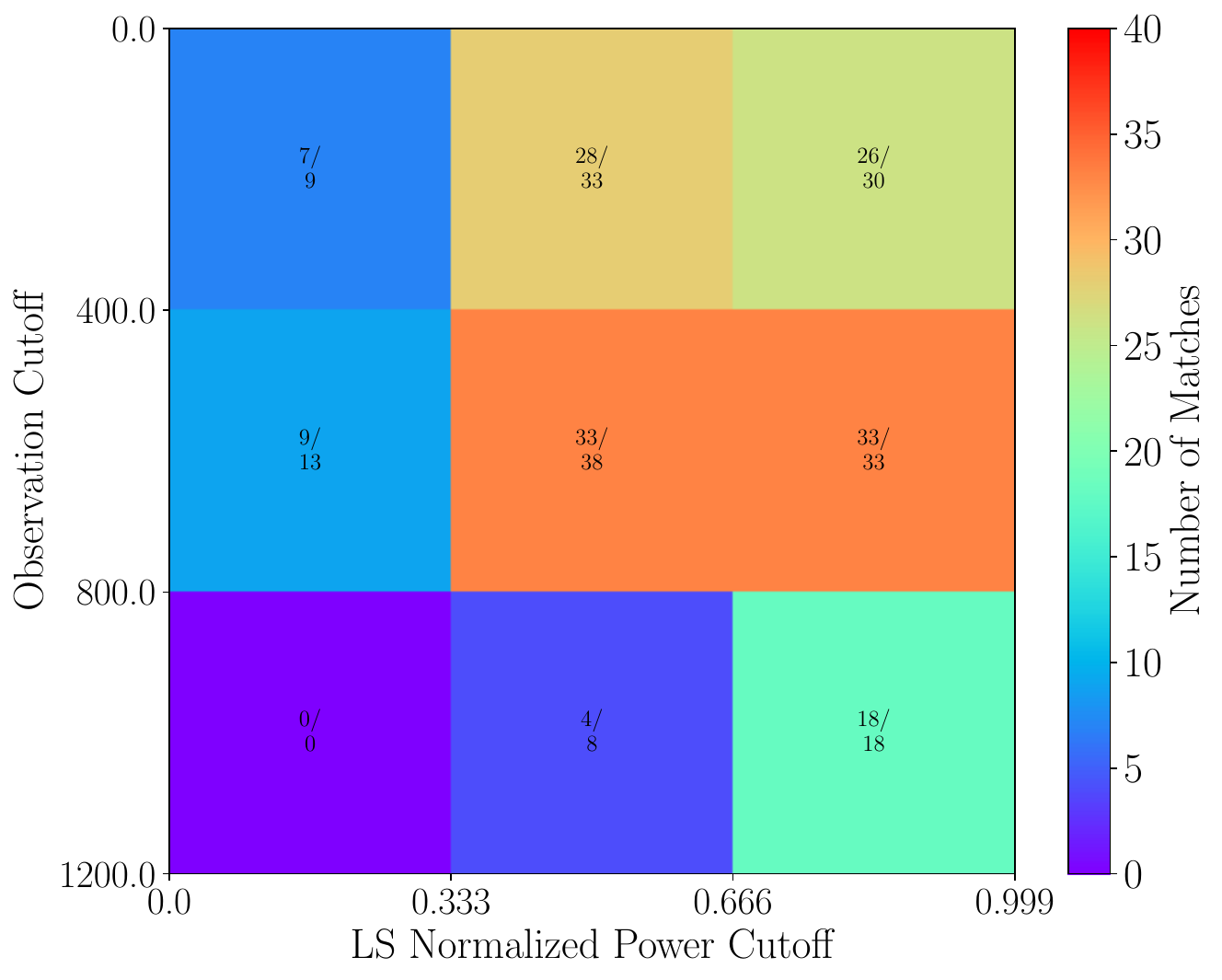}

}
\subfigure[]{

       \includegraphics[width=0.48\textwidth]{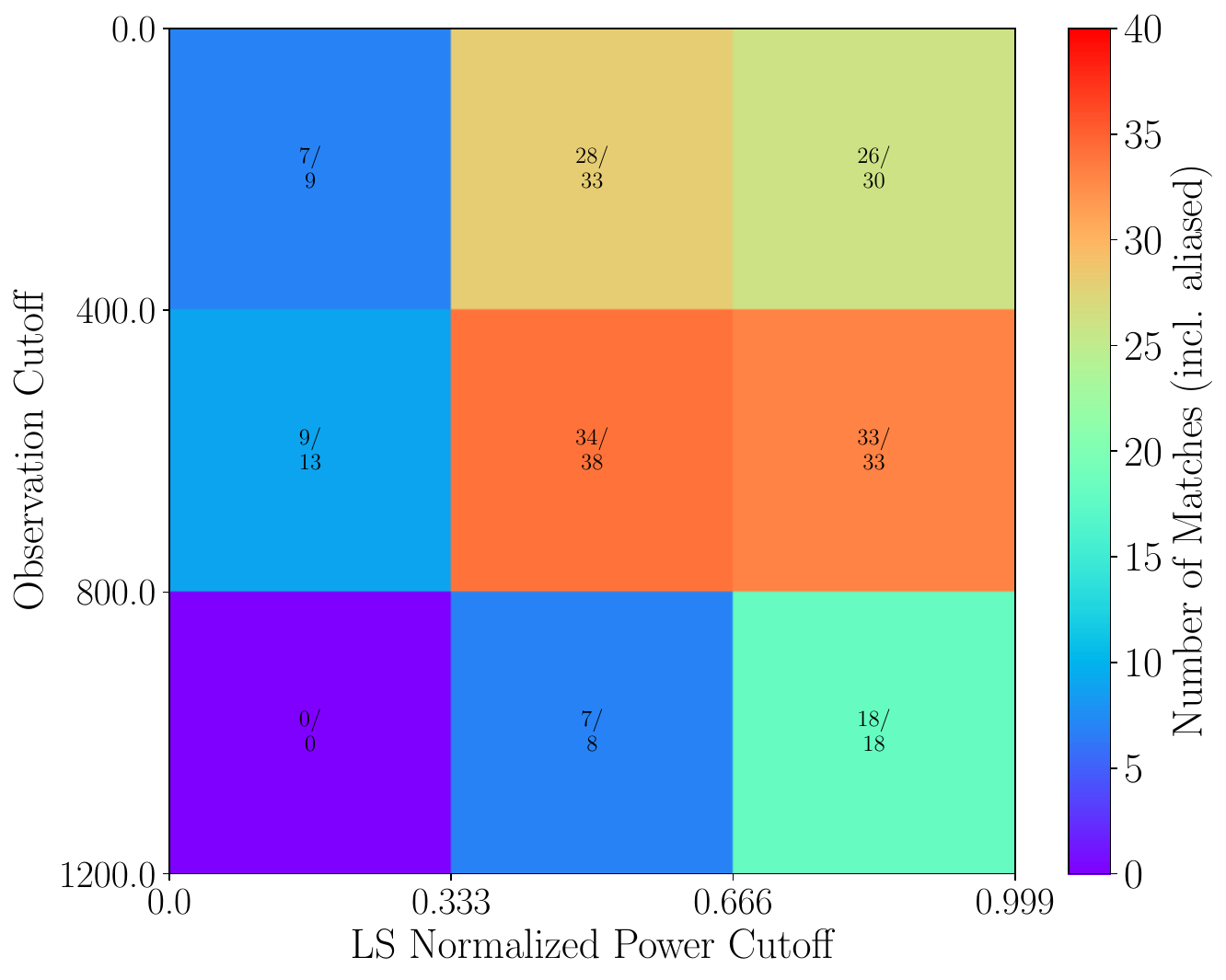}

}
    \caption{The same as Figure~\ref{fig:heatmap_confidence_period_gtr3h} but showing the differential heatmap. Because there are only 182 objects in the sample, the differential heatmap is poorly sampled and is thus very sparse. Consequently, we plot the heatmap using large bins such that we may obtain a significant number of objects in each bin. Caution is warranted as the number of objects in the bins remains low (e.g., there are no objects with $800<p<1200$~h and an \lsp power 0-0.333).}
   \label{fig:heatmap_confidence_period_gtr3h_differential}
\end{figure*}

The methods described above are used to recover the rotation periods of asteroids at the margins where simply combining the \snapshotdata and \tessdata datasets are unable to produce the correct period. If we select and combine the methods that were deemed to have utility above, which are (1), (3), (5) and (7) in Table~\ref{tab:methods}, then we obtain an exact match fraction of 0.827, which increases to 0.852 when aliased solutions are considered correct. 

Recall that the minimum rotation period search range is $p_{min}=2$~h. However, \citet{mcneill2023tess} demonstrated that periods derived with \tessdata are unreliable at  $p<3$~h. Consequently, we examine the fraction of correctly derived rotation periods when we exclude objects in \lcdbdata having $p<3$~h, which yields a total of 206 objects in the sample. Of those 206 objects, 182 had a lightcurve amplitude $>0.075$~mag.  We find that the exact and aliased match fraction is 0.868 and 0.890, respectively, which corresponds to recovering the rotation periods of 158 and 162 asteroids. Thus, we are able to further increase the match fraction using $p_{min}=3$~h compared to $p_{min}=2$~h. We also examined setting $p_{min}=4$~h but this yields a dramatic loss in the number of correct asteroid rotation periods recovered (see Table~\ref{tab:methods}). 

Based on this analysis, we use the combined optimizations with $p_{min}=3$~h because it yields a trade-off between maximizing the match fraction vs. the number of correct periods recovered.

To reiterate, we are able to recover the rotation periods with a 0.890 confidence when aliased solutions are included. However, the optimizations do not completely explain which properties of a lightcurve may indicate that it has a high probability of being correct. To this end, Figure~\ref{fig:heatmap_confidence_period_gtr3h}(a) plots the exact match fraction using the set of optimizations and $p_{min}=3$~h, as a function of the number of times an object has been observed and the \lsp power, where a higher \lsp power typically implies that there is a high signal-to-noise ratio when selecting the maximum power in the periodogram for each object. The heatmap is cumulative so the top left corner contains all of the objects that are correct (158) without any cut on the number of observations or \lsp power, and the number of objects in each bin decreases as the threshold on the number of observations or \lsp power increases.

We find that we can obtain a 100\% match to \lcdbdata when we have at least 375 observations and a \lsp power of 0.7. Figure~\ref{fig:heatmap_confidence_period_gtr3h}(b) shows the number of objects in the bins indicating that there are 50 objects with correct rotation periods that have been observed at least 375 times and have a \lsp power $\geq$0.7. While this is a very high observational threshold to obtain a perfect match fraction, a 0.90 match fraction can be obtained using more reasonable constraints: $\geq$ 150 observations with a \lsp power $\geq$0.3. Figure~\ref{fig:heatmap_confidence_period_gtr3h}(b)~and~(d) are the same as Figure~\ref{fig:heatmap_confidence_period_gtr3h}(a)~and~(c) except that it includes aliased matches as well.

The heatmaps above showed the cumulative match fraction and number of matches for each grid cell where there are fewer objects in the bins as the observation threshold and \lsp power increase. To investigate which is more critical --- the number of observations or \lsp power --- and to better understand the impact of these parameters on rotation period discovery, we now examine differential heatmaps, where each grid cell only includes objects with a range of observations and \lsp powers which are minimum and maximum values that define each grid cell. Thus, compared to the cumulative heatmaps, the number of objects in each bin will not monotonically decrease as the number of observations and \lsp power increases.

Figure~\ref{fig:heatmap_confidence_period_gtr3h_differential} shows a differential version of the heatmap shown in Figure~\ref{fig:heatmap_confidence_period_gtr3h}. This allows us to observe which bins obtain a match fraction below the average across all of the bins (i.e., 0.868 without any observational or \lsp power cutoff). There are only 182 objects in the sample and so the differential heatmap is poorly sampled --- thus, to draw any meaningful conclusions we increased the bin size compared to Figure~\ref{fig:heatmap_confidence_period_gtr3h}. Considering exact matches, we find that when there are at least 400 observations and a \lsp power $\geq 0.666$ that we achieve a perfect match to \lcdbdata. The three bins shown in Figure~\ref{fig:heatmap_confidence_period_gtr3h_differential}(a) at a \lsp power $<$0.333 are poorly sampled as shown in Figure~\ref{fig:heatmap_confidence_period_gtr3h_differential}(c) and so while we would expect that the match fraction would increase as we increase the number of observations, we observe the opposite. In this \lsp power range, there are only 7, 9, and 0 objects in the bins, respectively, and so small number statistics limits our ability to interpret this result. Lastly, comparing Figure~\ref{fig:heatmap_confidence_period_gtr3h_differential}(a)~to~(b) we find that the bin with 800-1200 observations and an \lsp power of 0.333-0.666 has a 0.500 match fraction whereas it has a 0.875 match fraction when aliased matches are included. This shows that the 3 objects in this bin were assigned a period that was a factor $0.5\times$$\pm3\%$ or $2\times$$\pm3\%$ of the \lcdbdata period solution, and so the 0.500 match fraction is misleading in Figure~\ref{fig:heatmap_confidence_period_gtr3h_differential}(a) as objects with $\geq$800 observations and an \lsp power $\geq$0.333 should be considered reliable.

\subsection{A Conservative Approach to Biasing Against Incorrect Rotation Periods}\label{sec:conservative_classifier}

Sections~\ref{sec:baseline}--\ref{sec:improving_confidence} reported the rotation period match fraction as compared to \lcdbdata. The combination of methods summarized in Table~\ref{tab:methods} aims to maximize the number of objects with correct periods. However, it does not consider biasing against incorrect rotation periods. We address this drawback by proposing a metric that reports a null result if there is a sufficiently high probability that a period is incorrect.

To include biasing against incorrect rotation periods, we reframe the analysis as a binary classification problem. Each object has three properties that determine whether the period is likely to be correct (lightcurve amplitude, \lsp power, and number of observations) and based on these properties we assign each object a predicted label which is either ``incorrect period'' or ``correct period''. Then, we compare the period solutions to \lcdbdata to determine whether the labels are correct, where we aim to minimize the number of objects assigned a correct label that are incorrect (a false positive). The metrics described below are those used in supervised machine learning for assessing the quality of a model's ability to predict binary class labels~\citep{cabrera2017deep}. For clarity, we note that we are not training a machine learning model, rather we are incorporating metrics from the field of machine learning that are useful for our purposes.

We outline four possibilities regarding the outcome of binary classification as follows:
\begin{enumerate}
\item \emph{True Positive (TP):} An object was assigned a correct label, and the rotation period is correct.
\item \emph{False Positive (FP):} An object was assigned a correct label, but the rotation period is incorrect.
\item \emph{True Negative (TN):} An object was assigned an incorrect label, and the rotation period is incorrect.
\item \emph{False Negative (FN):} An object was assigned an incorrect label, but the rotation period is correct.
\end{enumerate}	

Using the TP, FP, TN, and FN rates, the following metrics can be computed, where 1.0 is the best (maximum) value for each.

\begin{itemize}
\item Accuracy: (TP+TN)/(TP+FP+TN+FN)
\item Precision: TP/(TP+FP) 
\item Recall: TP/(TP+FN)
\end{itemize}

Observe that the accuracy metric is very similar to the results reported in	Sections~\ref{sec:baseline}--\ref{sec:improving_confidence} that examine the match fraction to \lcdbdata, and so we are largely uninterested in this metric here. The recall metric does not consider false positives and so it is not very useful in this context. The precision metric is the most appropriate, as it biases against reporting false positives. In other words, if we have a precision value close to 1.0, this suggests that the number of false positives (objects assigned a correct label but have an incorrect rotation period) have been minimized. 

Observing Figures~\ref{fig:amplitude_match_fraction}--\ref{fig:heatmap_confidence_period_gtr3h_differential}, which show the relationship between match fraction and lightcurve amplitude, \lsp power, and number of observations, it is clear that perfect precision can be obtained with sufficiently high thresholds for these three properties, as there are several instances that yield a 1.0 match fraction and therefore do not have any false positives. 

For an object to be assigned a correct label, it must have a sufficiently high amplitude, number of observations, and \lsp power, which are the properties examined in Section~\ref{sec:improving_confidence}. We use the periods derived by optimizations ``\{1, 3, 5, 7\} Combined'' from Table~\ref{tab:methods} with $p_{min}=3$~h (a total of 182 objects). In the analysis, we only consider exact matches and exclude aliased matches. 

We perform a grid search over the parameter space to select thresholds for the following parameters: lightcurve amplitude, \lsp power, and number of observations. We then use these thresholds to determine whether the object is assigned a predicted correct or incorrect class label. The range of amplitudes searched is [0.075, 0.5] with a step size of 0.025~mag, the range of number of observations searched is [50, 400] with a step size of 10 observations, and the range of \lsp power values searched is [0, 0.5] with a step size of 0.05. Each object is assigned a correct label only if it meets all three thresholds. 

\begin{figure}[!t]
\centering
\includegraphics[width=1\columnwidth]{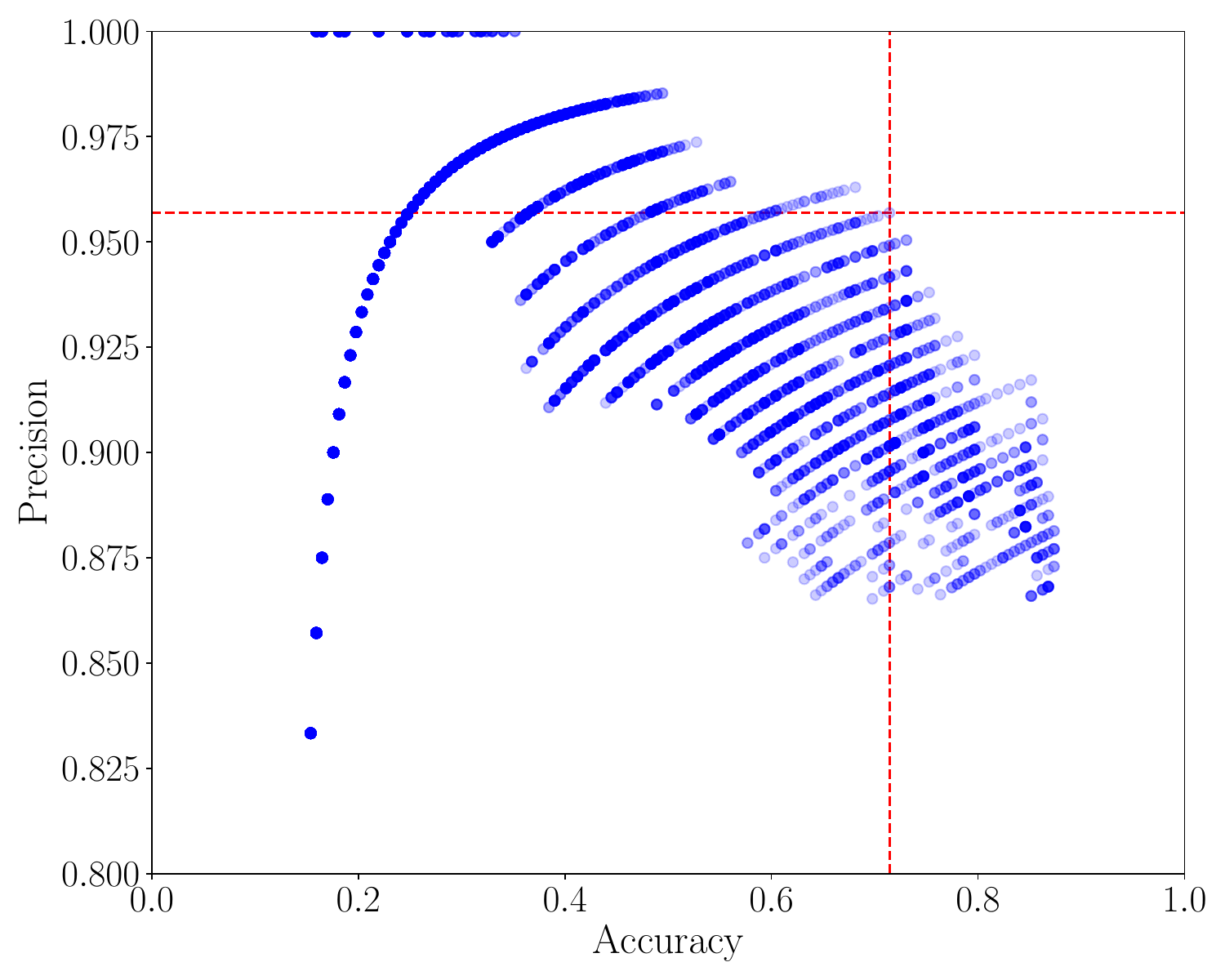}
    \caption{The precision as a function of accuracy for  the grid search described in the text. There are a total of 7128 configurations that were searched to find those that optimize the precision metric. Many of the configurations yield the same pair of precision and accuracy values and so the markers on the plot are translucent such that those with greater intensity denote duplicate configurations. The red dashed lines show an example configuration with a precision of 0.957 and an accuracy of 0.714.}
   \label{fig:scatter_precision_vs_accuracy}
\end{figure}

To summarize, we searched 18 amplitude, 36 observation, and 11 \lsp power values, yielding a total of 7128 configurations. Each of these configurations assigns all objects a class label which are used to derive accuracy, precision, and recall statistics. Figure~\ref{fig:scatter_precision_vs_accuracy} plots the relationship between accuracy and precision for each of the configurations searched, where we are interested in maximizing the precision as it minimizes false negatives. We make the following observations regarding Figure~\ref{fig:scatter_precision_vs_accuracy}:

\begin{enumerate}
\item Higher accuracy is obtained at the expense of lower precision. There are several configurations where perfect (1.0) precision is obtained; however, low accuracy implies that there are many false negatives, which would have the effect of recovering far fewer objects than at higher accuracy thresholds. So perfect precision is not ideal unless one is only interested in correctly deriving the periods of few objects.
\item There are several configurations that achieve very high precision and reasonably high accuracy, and one of the configurations is denoted by the dotted red lines in the figure. This example configuration has a precision of 0.957 and an accuracy of 0.714, which is achieved with the following thresholds. An amplitude of 0.125~mag, 120 observations, and an \lsp power of 0.5. This configuration has a TP rate of 111 and a FP rate of 5.
\end{enumerate}

Using the above thresholds, we confidently assign incorrect and correct labels to objects that minimize false negatives while simultaneously recovering a large fraction of the periods of objects. In Section~\ref{sec:comparison_full_intersection}, we examine both the match fraction to \lcdbdata and this conservative approach that biases against incorrect periods when applied to all objects on the combined dataset (\intersectsnapstess).

\subsection{Summary: Comparison of Methods and the Utility of ZTF and TESS Observational Data}
In this section, we demonstrated several methods that can be applied to improve the fraction of matches to \lcdbdata, and these methods included $\sigma$ clipping, using periodogram masking, replacing the derived period in \intersectlcdbdata with the \tessdata period when there is a high probability that the \tessdata period is correct, and using a lightcurve amplitude threshold. The fraction of matches to \lcdbdata using the heatmap analysis (Figures~\ref{fig:heatmap_confidence_period_gtr3h}--\ref{fig:heatmap_confidence_period_gtr3h_differential}) describes when a derived rotation period is likely to be correct; however, it does not bias against assigning an incorrect rotation period to an object. In contrast, the conservative binary classification method biases against assigning an object an incorrect rotation period (a false positive), and so when the classifier reports that an object's classification is incorrect, then this should be considered a null result and the period cannot be trusted. The benefit of using the binary classifier over simply using the fraction of matches to \lcdbdata is that there is a higher probability that the classifier will identify correct periods. The drawback of this method is that it favors precision over accuracy and so several correct rotation periods will be labeled as incorrect (false negatives). In summary, to employ these methods on large catalogs, rather than only reporting a single confidence metric, it is preferable to report the total match fraction, the match fraction as a function of \lsp power and number of observations (Figures~\ref{fig:heatmap_confidence_period_gtr3h}-\ref{fig:heatmap_confidence_period_gtr3h_differential}), and the conservative binary classification approach that reports correct rotation periods with very high probability.

Recall from Section~\ref{sec:baseline} that the match fraction by simply combining \intersectsnapstess is 78.4\% and 82.9\% for exact and aliased matches, respectively. Therefore, the optimizations summarized in Section~\ref{sec:combined_optimizations} are only able to improve the overall match fraction in a small number of remaining cases, as the $\sim$80\% match fraction is remarkable given the drawbacks of the datasets. The baseline approach shows that the match fraction increases substantially when comparing \snapshotdata and \intersectsnapstess whereas the match fraction has a smaller increase when comparing \tessdata to \intersectsnapstess  (Table~\ref{tab:matchfraclcdb}). Therefore, the benefits of using \tessdata are clear, but it is not abundantly clear what the benefits are to adding \snapshotdata to \tessdata.

We discuss why incorporating \snapshotdata is important.  Recall from Section~\ref{sec:baseline} that \tessdata is not sensitive to $p<4$~h periods because of the 30 minute cadence of the full frame TESS images. Using the \intersectlcdbdata dataset with optimizations ``\{1, 3, 5, 7\} Combined'' with $p_{min}=3$~h (182 total objects), there are a total of 31 objects in \lcdbdata with 3--4~h rotation periods. Of these 31 objects, \tessdata correctly derives 19 of the rotation periods (considering only exact matches), but there are 6 additional objects that \intersectsnapstess derives correctly for which \tessdata derives incorrectly. This shows that  incorporating \snapshotdata significantly improves period derivation in this period range. 

This trend can be observed in Figure~\ref{fig:LCDB_period_distribution_correct}, where \snapshotdata recovers more rotation periods within the second bin than \tessdata. It is noteworthy that the true asteroid rotation period distribution has a high fraction of objects with periods in this range, and therefore, sensitivity to $p\lesssim 4$~h is of great importance. This is a concrete example of where \snapshotdata is better than and can complement \tessdata, and in Section~\ref{sec:example_lightcurves_and_implications} we show an example lightcurve for an object that is derived correctly with \snapshotdata but not \tessdata with a period of 10.215~h. To summarize, based on our analysis, \snapshotdata has great utility at $p\sim2-4$~h whereas \tessdata does not, \snapshotdata is sensitive to long period asteroids whereas \tessdata is not because of its small observing window, and \snapshotdata has moderate utility for intermediate rotation periods compared to \tessdata.

\section{Period Derivation On The Combined ZTF and TESS Dataset}\label{sec:comparison_full_intersection}

Now that we have examined maximizing the rotation period match fraction by comparing our solutions to those reported in \lcdbdata, we apply the combined methods in Section~\ref{sec:combined_optimizations} to the 3168 objects that are included in the \intersectsnapstess dataset. In particular, we use configuration ``\{1, 3, 5, 7\} Combined'' from Table~\ref{tab:methods} with $p_{min}=3$~h.

\begin{figure*}[!t]
\centering
\includegraphics[width=0.8\textwidth]{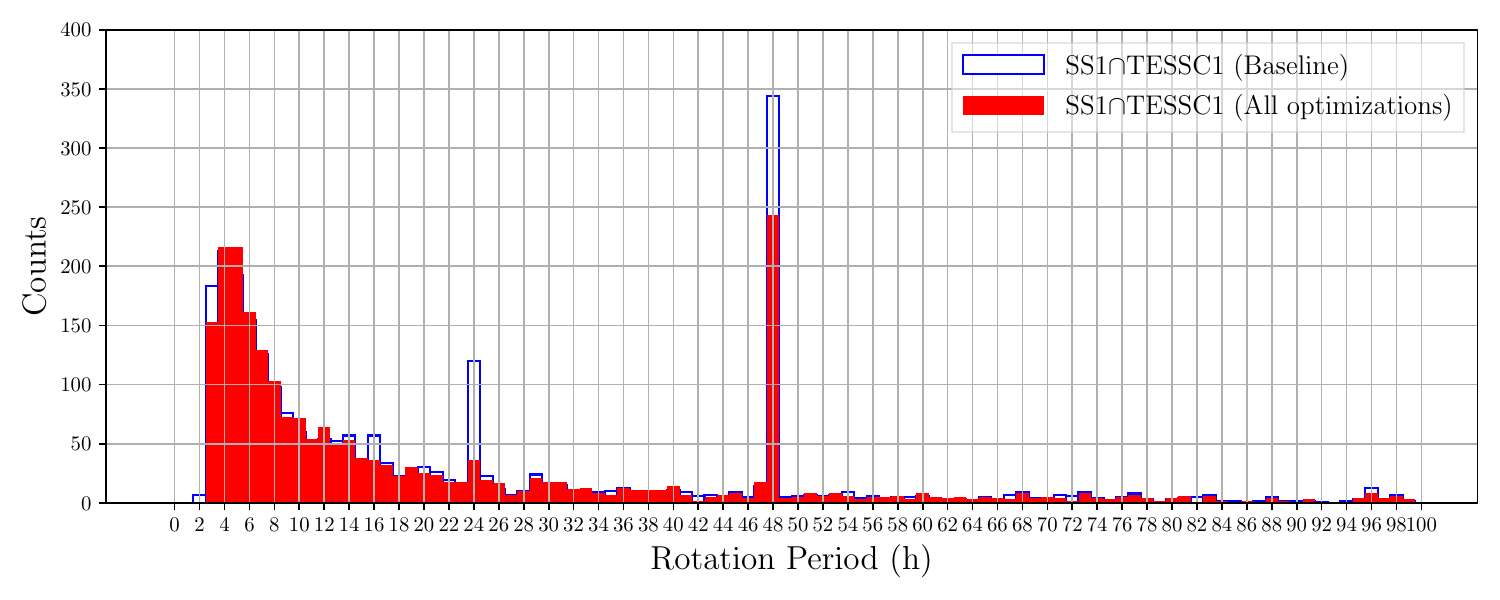}
    \caption{Rotation period distribution of \intersectsnapstess comparing the baseline approach to the set of optimizations ``\{1, 3, 5, 7\} Combined'' with $p\geq$3~h from Table~\ref{tab:methods}.}
    
   \label{fig:histogram_after_filtering}
\end{figure*}

\begin{figure*}[!t]
\centering
\includegraphics[width=0.8\textwidth]{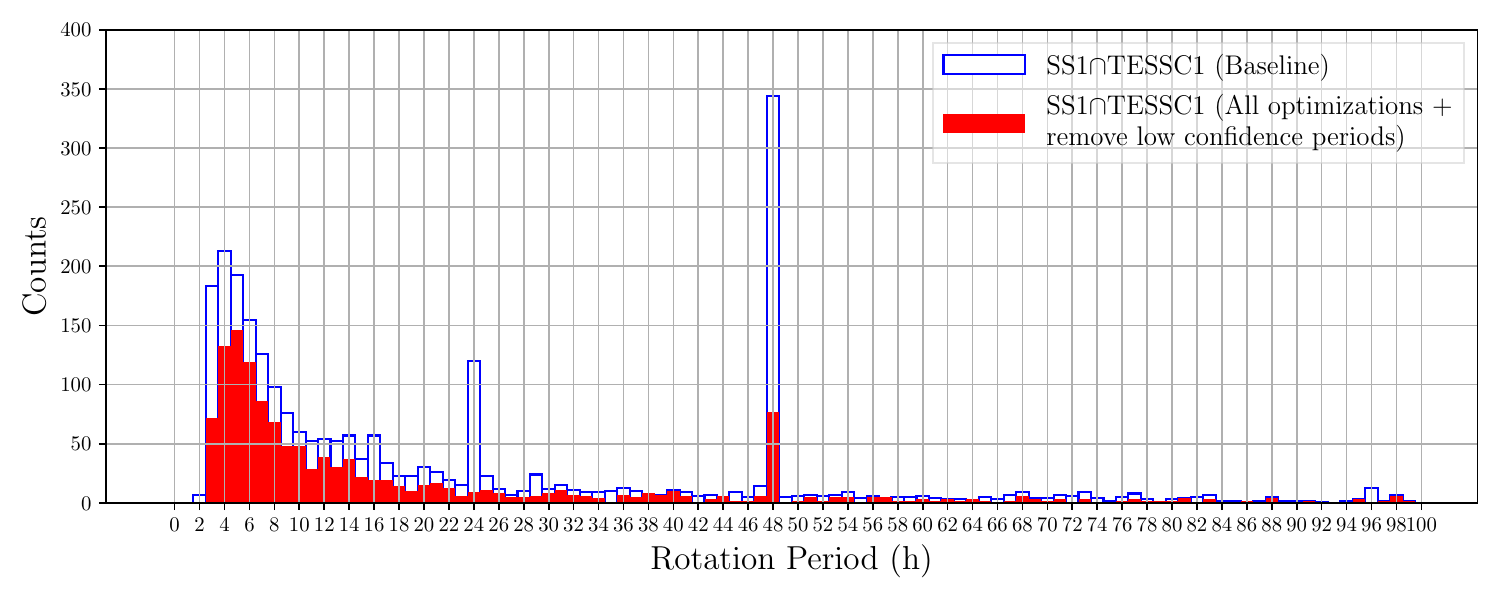}

    \caption{The same as Figure~\ref{fig:histogram_after_filtering} but where low confidence solutions have been removed.}
   \label{fig:histogram_after_filtering_and_rm_low_confidence_periods}
\end{figure*}

\begin{figure*}[!t]
\centering
\includegraphics[width=0.8\textwidth]{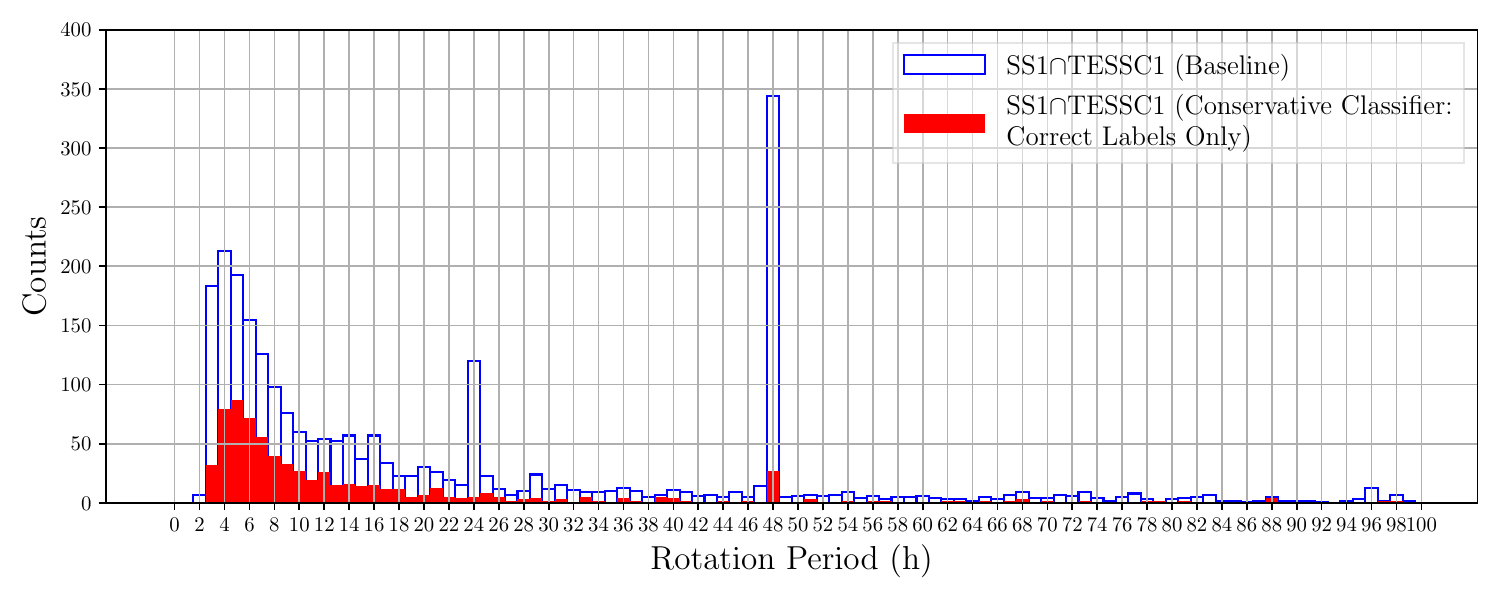}
    \caption{ The same as Figure~\ref{fig:histogram_after_filtering} but where we only show those classified as correct with the conservative binary classifier described in Section~\ref{sec:conservative_classifier}. The thresholds are as follows: an amplitude of 0.125~mag, 120 observations, and an \lsp power of 0.5.}
   \label{fig:histogram_after_filtering_and_conservative_classifier}
\end{figure*}

\subsection{Rotation Period Distribution}
In this section, we examine three different perspectives on period recovery as follows. $(i)$ Using the combined optimizations summarized in Table~\ref{tab:methods}. $(ii)$ Using the \lsp power in the differential heatmap (Figure~\ref{fig:heatmap_confidence_period_gtr3h_differential}) to determine which periods are considered reliable. $(iii)$ Using the conservative binary classification approach outlined in Section~\ref{sec:conservative_classifier}.

Figure~\ref{fig:histogram_after_filtering} plots a histogram of the rotation periods comparing the baseline approach of combining \tessdata and \snapshotdata (\intersectsnapstess) to the distribution after applying the optimizations ``\{1, 3, 5, 7\} Combined'' with $p\geq$3~h from Table~\ref{tab:methods}.  We observe that we have reduced solutions at the aliases: the 16~h alias is no longer visible, the 24~h alias is visible but much reduced, and the 48~h alias is still pronounced. Table~\ref{tab:aliases} summarizes the number of solutions at aliases in Figure~\ref{fig:histogram_after_filtering}.

\begin{deluxetable*}{r|r|r|r}
\tablecaption{The number of aliases for the baseline approach (no filtering) and using the combined methods ``\{1, 3, 5, 7\} Combined'' with $p_{min}\geq$3~h in Table~\ref{tab:methods}, and the combined method with low confidence period removal. The fractions in parentheses refer to the fraction of the sample at the reported alias.}\label{tab:aliases}
\tablewidth{\textwidth}
\tabletypesize{\footnotesize}
\tablehead{
\colhead{Alias (h)}&
\colhead{Baseline ($n=3168$)}&
\colhead{Combined ($n=3008$)}& 
\colhead{Combined and Low Confidence Removal ($n=1623$)}
}
\startdata
16$\pm$0.1& 57 (0.018)& 36 (0.012)& 19 (0.012)\\
24$\pm$0.1&120 (0.038)& 36 (0.012)&  9 (0.006)\\
48$\pm$0.1&344 (0.109)&243 (0.081)& 77 (0.047)\\
\enddata
\end{deluxetable*}

Recall that there are only $n_{LCDB}=206$ objects in \lcdbdata that overlap with our sample (Table~\ref{tab:methods}). Thus, if the $n=3168$ objects in \intersectsnapstess were drawn from the same distribution as the $n_{LCDB}=206$ objects in \intersectlcdbdata then the match fraction values in the cumulative heatmap (Figure~\ref{fig:heatmap_confidence_period_gtr3h}) would be directly applicable to the objects in \intersectsnapstess. However, we know from the histogram shown in Figure~\ref{fig:histogram_after_filtering} that our period distribution is not the same, which is clearly indicated by the overabundance of aliased solutions. Thus, the differential heatmap (Figure~\ref{fig:heatmap_confidence_period_gtr3h_differential}) is a better statistical tool for determining whether we should be confident in a given solution despite the abovementioned challenges regarding small number statistics.

Consider from Figure~\ref{fig:heatmap_confidence_period_gtr3h_differential}(b) that objects with an \lsp power $<0.333$ are unreliable and objects with a \lsp power $\geq 0.333$ have a minimum confidence level of 0.848. Figure~\ref{fig:histogram_after_filtering_and_rm_low_confidence_periods} shows the same histogram as Figure~\ref{fig:histogram_after_filtering} where all periods with a \lsp power $<0.333$ are removed from the sample. This yields a total of $n=1623$ objects, indicating that 1545 period solutions were removed because they have a low \lsp power. For these $n=1623$ objects, the minimum confidence that we assign is 0.848. Comparing Figure~\ref{fig:histogram_after_filtering} to Figure~\ref{fig:histogram_after_filtering_and_rm_low_confidence_periods} shows that many  of the 1545 (removed) objects with an \lsp power $<0.333$ are those were located in the 48~h bin (11\% of the 1545 removed objects). We find that combining the optimizations and removing low confidence solutions is unable to completely eliminate aliased period solutions at 48~h, but it is much improved over both the baseline and the optimization set ``\{1, 3, 5, 7\} Combined'' that did not remove low confidence period solutions from the sample.

Section~\ref{sec:conservative_classifier} used a binary classification approach to label the rotation periods as either incorrect or correct based on amplitude, number of observations, and \lsp power. We find that there are a total of 897 asteroids assigned a correct label and 2271 assigned an incorrect label. The asteroids assigned a correct label are plotted in Figure~\ref{fig:histogram_after_filtering_and_conservative_classifier}. We find that compared to Figures~\ref{fig:histogram_after_filtering}~and~\ref{fig:histogram_after_filtering_and_rm_low_confidence_periods}, there is only a slight overabundance of periods at 48~h, indicating that the binary classifier filters out many of the aliased solutions. Recall that this conservative approach trades precision for accuracy by penalizing false positives, and so the total number of objects recovered is much lower than in Figure~\ref{fig:histogram_after_filtering_and_rm_low_confidence_periods}, as these objects have a higher probability of being assigned correct rotation periods.

The Appendix (Section~\ref{sec:appendix_table}) reports the rotation periods of the $n=1623$ asteroids defined where all periods with a \lsp power $<0.333$ were removed (Figure~\ref{fig:histogram_after_filtering_and_rm_low_confidence_periods}). For each of these objects we also report the binary classification prediction, where 897 were labeled as correct and 726 were labeled as incorrect.





\subsection{Rotation Period Comparison with \tessdata and \snapshotdata}

\begin{figure*}[!t]
\centering

\includegraphics[width=1\textwidth]{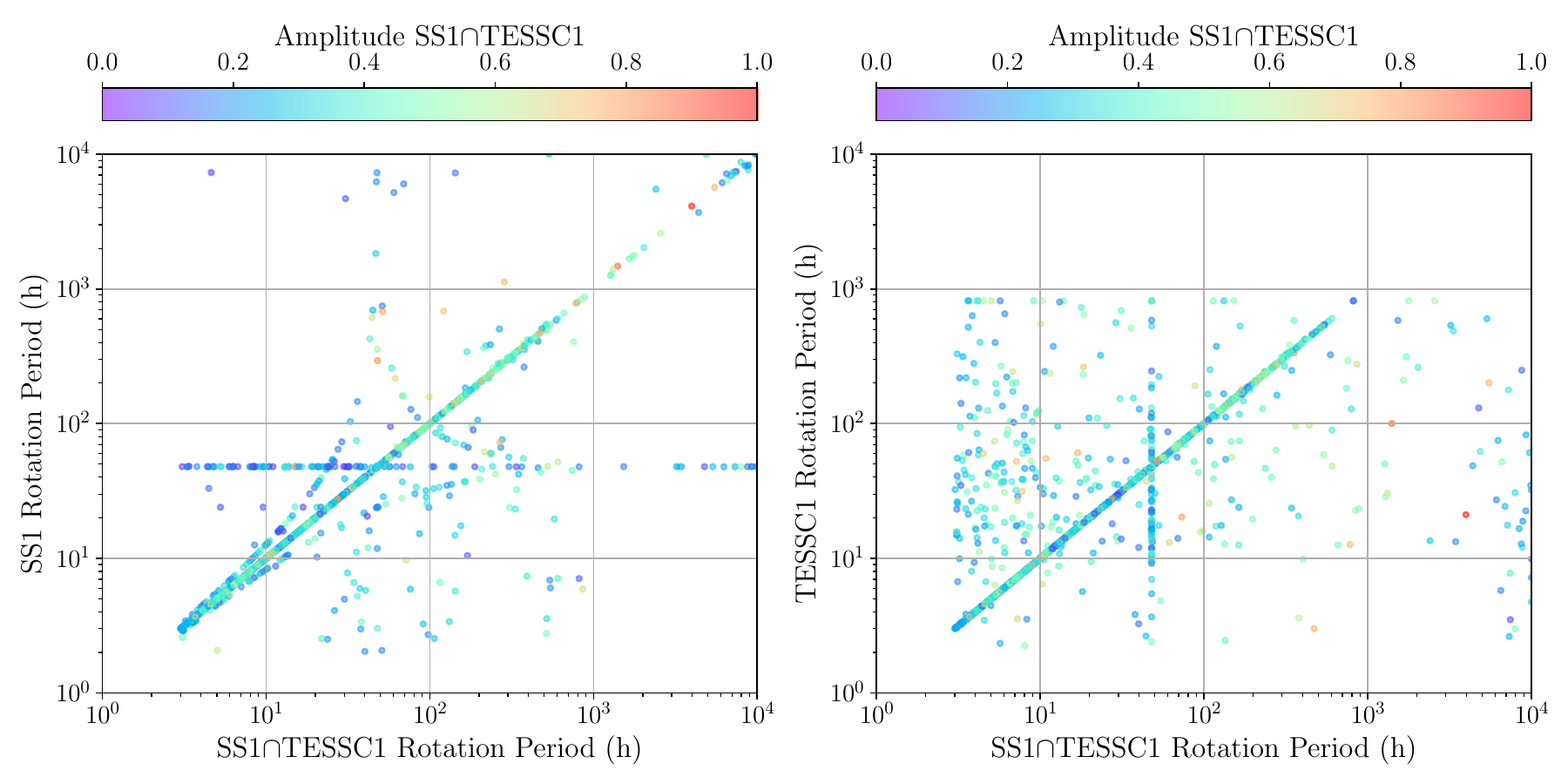}

    \caption{Comparison of rotation period distributions as scatterplots for \snapshotdata and \tessdata as a function of \intersectsnapstess, plotted in the left and right panels, respectively. The \intersectsnapstess  rotation periods and color coded amplitudes are from the $n=1623$ sample that has all optimizations applied and low confidence solutions removed. In contrast to prior figures, such as Figure~\ref{fig:histogram_after_filtering}, the plot is shown on a log scale such that the entire period range can be shown. Consistent periods between the two datasets appear on the diagonal line.}
   \label{fig:scatter_rotation_period_comparison}
\end{figure*}

We examine the rotation period distributions of \snapshotdata as a function of the \intersectsnapstess rotation period in the left panel of Figure~\ref{fig:scatter_rotation_period_comparison}. While a substantial fraction of periods are in agreement, we clearly observe the aliased solutions at 48~h in \snapshotdata that are not in agreement with \intersectsnapstess. Examining the \intersectsnapstess amplitudes of these objects, we find that they are generally fairly low at 48~h, as most are $<$0.25~mag. We also observe the effect of pseudo-aliases~\citep{VanderPlas2018} which appear as the curved triangle shape in the left panel. These are due to interactions between an object's real period and the strongest alias in the periodogram~\citep{2022FrASS...909771D,trilling2023solar}.

The right panel of Figure~\ref{fig:scatter_rotation_period_comparison} plots the rotation periods of \tessdata as a function of \intersectsnapstess. Here, the maximum rotation period for \tessdata is $\sim$816~h which is why none of the periods in \tessdata exceed  this value. We find that there are an overabundance of aliases clearly observed at 48~h in \intersectsnapstess, which is unsurprising as we showed this in Figure~\ref{fig:histogram_after_filtering_and_rm_low_confidence_periods} and in Table~\ref{tab:aliases}. However, of great interest is the observation that the pseudo-aliases are no longer visible when comparing \tessdata to \intersectsnapstess. This is because \tessdata does not have aliases and so the pseudoaliases are eliminated.

\subsection{Lightcurve Amplitude Distribution}

\begin{figure}[!t]
\centering

\includegraphics[width=1\columnwidth]{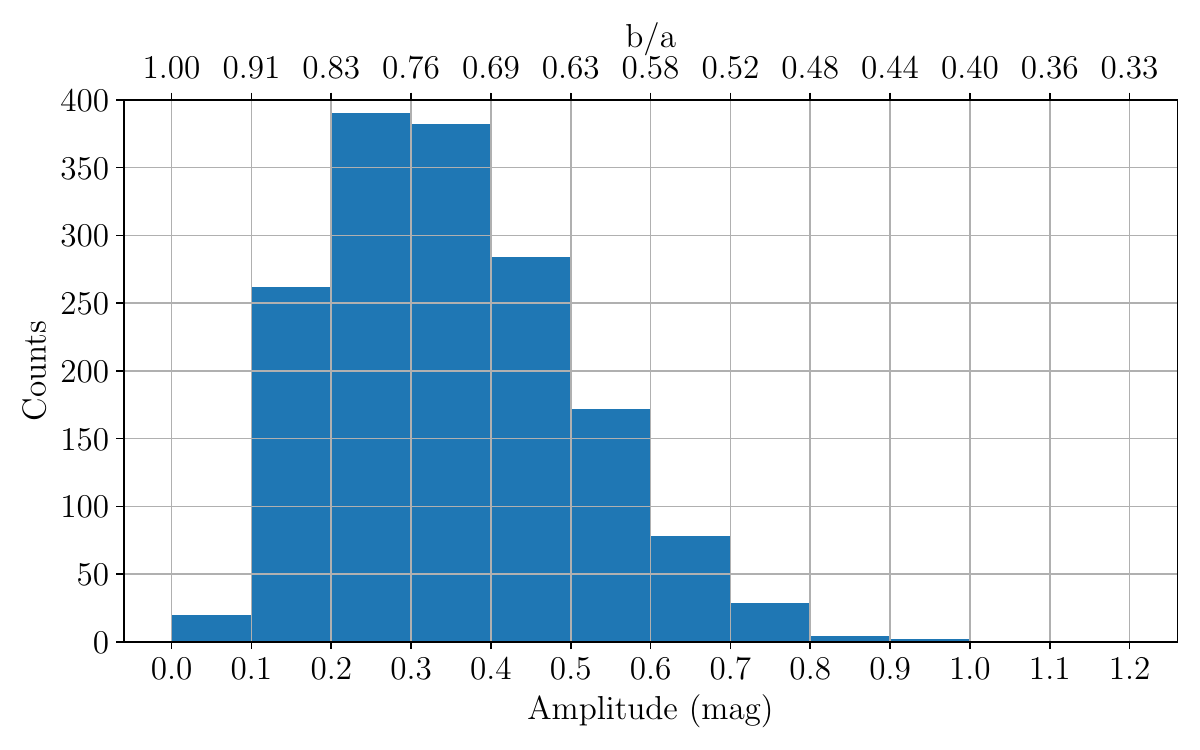}

    \caption{The amplitudes of the $n=1623$ sample of objects in \intersectsnapstess that have all optimizations applied and low confidence solutions removed. The top horizontal axis shows the $\frac{b}{a}$ ratio which is calculated directly from the amplitude ($\Delta$mag) where $\Delta$mag$ = -2.5 \log \frac{b}{a}$.}
   \label{fig:histogram_amplitude_ztftess_after_filter_rm_low_confidence}
\end{figure}

Figure~\ref{fig:histogram_amplitude_ztftess_after_filter_rm_low_confidence} plots the amplitudes of the $n=1623$ sample of objects in \intersectsnapstess. The figure is consistent with Figure~\ref{fig:amplitude_histograms}(a) except that there are a smaller fraction of objects with low amplitudes as we discarded many of them to maximize the fraction of matches in the sample. The shape of an asteroid can be constrained through $\Delta {\rm mag} = -2.5 \log \frac{b}{a}$, where $b$ and $a$ are the minor and major axis lengths of the ellipsoidal asteroid shape, respectively. There are few objects with $\frac{b}{a}\geq 0.9$, and a single asteroid cannot be this elongated; therefore, these asteroids are likely to be binary systems.

\subsection{Example Lightcurves and Implications}\label{sec:example_lightcurves_and_implications}
In all prior sections we described facets of lightcurve generation and fidelity without showing any example lightcurves. In this section we show a selection of interesting lightcurves from the $n=1623$ sample of objects in \intersectsnapstess. In the plots, we show the derived lightcurve of \intersectsnapstess without \tessdata replacement (if applicable), such that we can observe the lightcurves generated by the combined dataset in each instance.

\begin{figure}[!t]
\centering

\includegraphics[width=1\columnwidth]{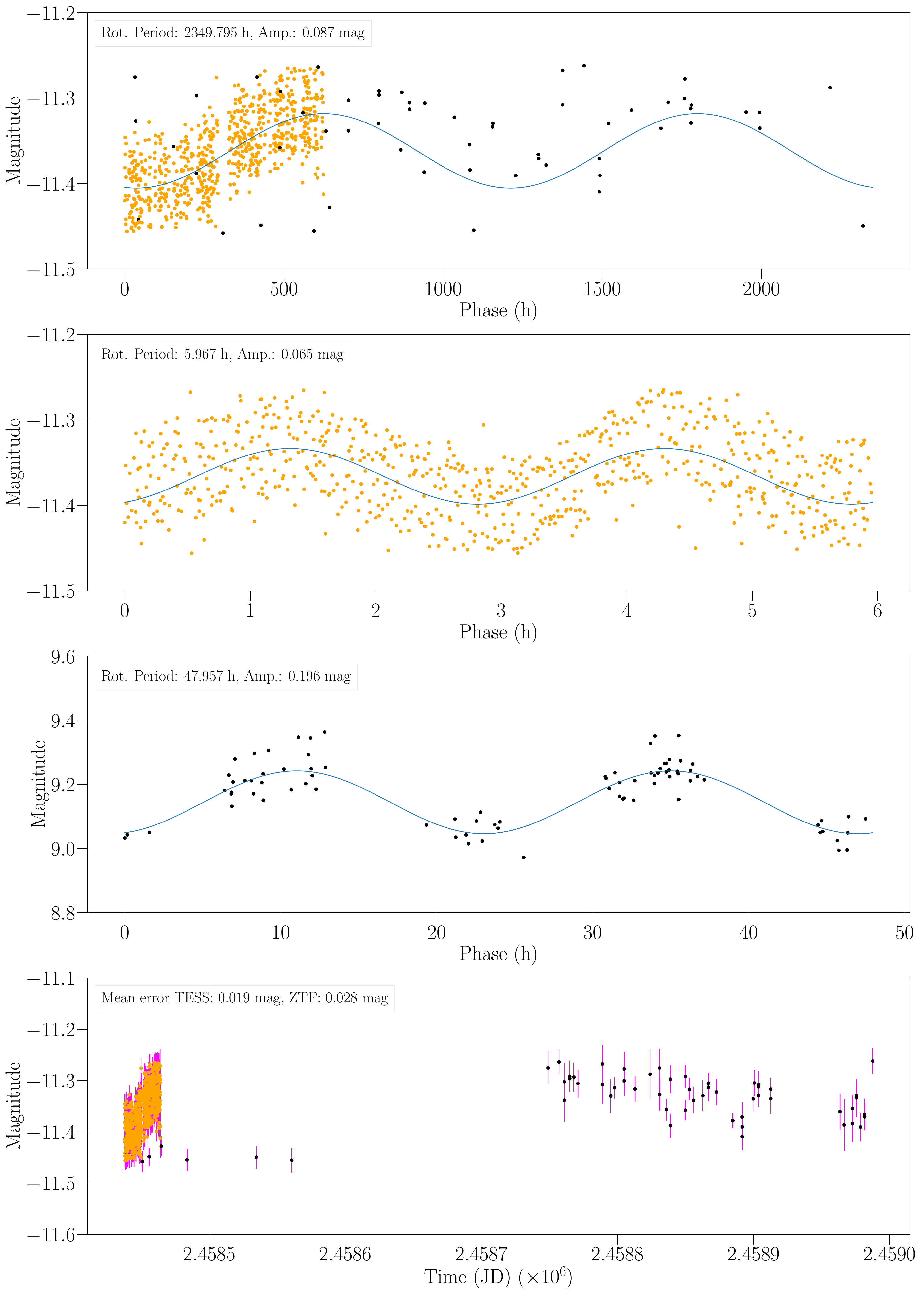}

    \caption{Lightcurves of object 339. Panels from top to bottom are as follows: \intersectsnapstess ($p=2349.795$~h), \tessdata ($p=5.967$~h), \snapshotdata ($p=47.957$~h), and the unphased data for \intersectsnapstess. Orange and black markers refer to observations from \tessdata and \snapshotdata, respectively. The bottom panel shows photometric errors that are plotted using magenta bars, and the markers are plotted on top of the error bars to prevent them from being obfuscated.}
   \label{fig:339_TESSC1_replacement}
\end{figure}

\textbf{Object 339:} Figure~\ref{fig:339_TESSC1_replacement} shows the lightcurves for object 339 for \intersectsnapstess ($p=2349.795$~h), \tessdata ($p=5.967$~h), and \snapshotdata ($p=47.957$~h), and the unphased \intersectsnapstess data with error bars ordered from top to bottom, respectively.  The lightcurve for \intersectsnapstess (top panel) has a rotation period of $p=2349.795$~h, but \tessdata is not sensitive to rotation periods $\gtrsim816$~h and so the lightcurve is poorly sampled after $\sim600$~h (i.e., $t_{window}$ for this object). The derived rotation period is a poor solution for this object. The second panel shows the data for \tessdata and we observe an excellent fit to the data. Here, the rotation period is $p=5.967$~h and \lcdbdata reports $p=5.974$~h and so the rotation periods are in agreement. The third panel shows the lightcurve generated with \snapshotdata and we observe a very poor fit to the data. This lightcurve has a rotation period of 47.957~h and is an example of a quintessential alias where the only signal that can be detected by \lsp is a function of the diurnal observing schedule. Using the optimizations described in the prior sections, the \intersectsnapstess period of $p=2349.795$~h was replaced by the \tessdata $p=5.967$~h, which is the correct period. The bottom panel shows the unphased \intersectsnapstess data where we observe that some of the \snapshotdata observations temporally overlap the \tessdata data, but most of the \snapshotdata data is in a separate apparition.

\begin{figure}[!t]
\centering

\includegraphics[width=1\columnwidth]{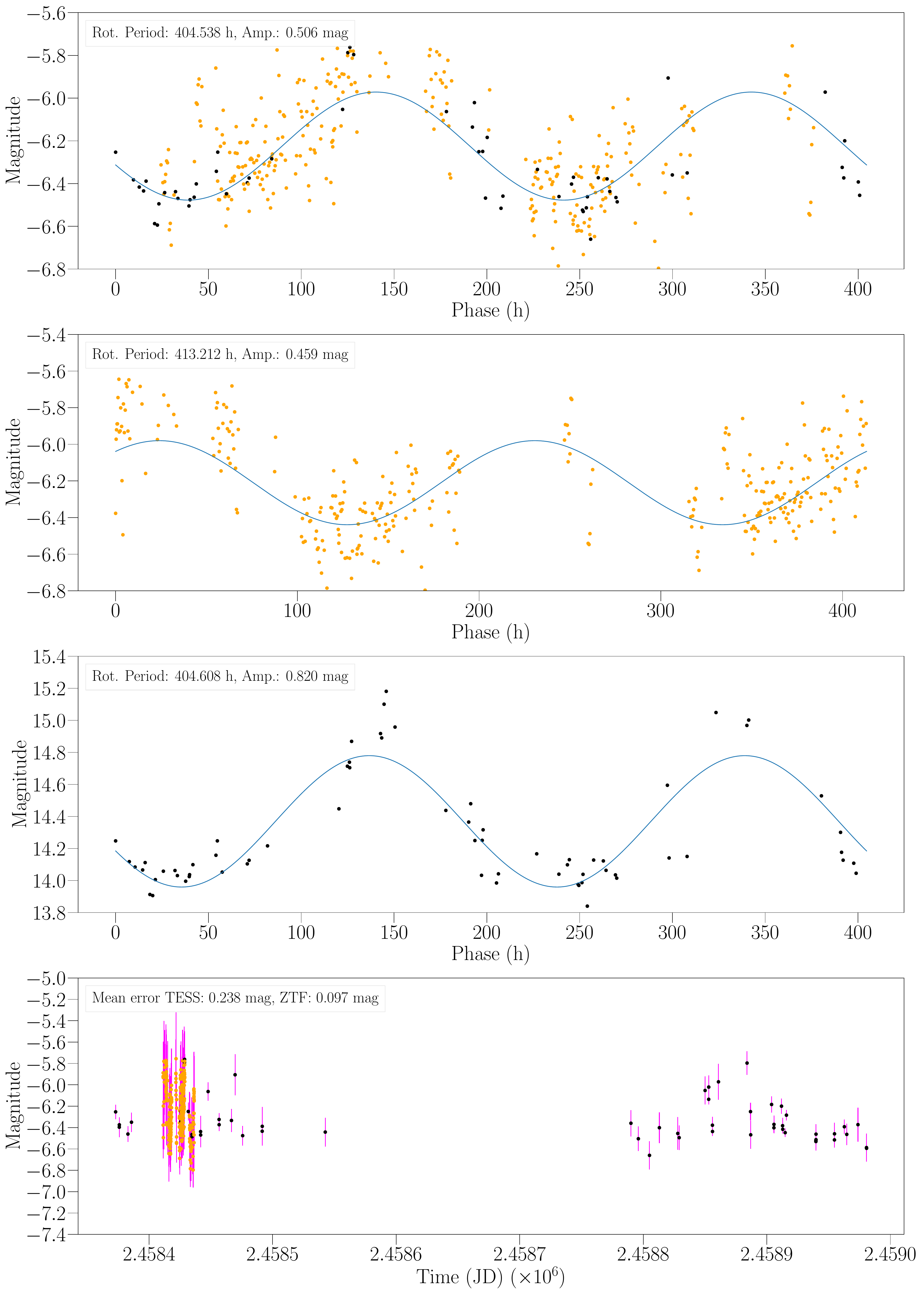}

    \caption{Lightcurves of object 18975. Panels from top to bottom are as follows: \intersectsnapstess ($p=404.538$~h), \tessdata ($p=413.212$~h), \snapshotdata ($p=404.608$~h), and the unphased data for \intersectsnapstess.}
   \label{fig:18975_fills_in_the_LC}
\end{figure}

\textbf{Object 18975:} Figure~\ref{fig:18975_fills_in_the_LC} reports the lightcurves for object 18975. This object is not reported in \lcdbdata. The rotation periods are $p=404.538$~h, $p=413.212$~h, and $p=404.608$~h for \intersectsnapstess, \tessdata, and \snapshotdata, respectively. All of the rotation periods are in agreement, and the object appears to be a slowly rotating asteroid. This set of lightcurves is interesting because if we consider the \tessdata (middle panel) and \snapshotdata (bottom panel) lightcurves on their own, one may be skeptical that the derived rotation periods are accurate because the lightcurves are poorly sampled (i.e., there are large gaps in the folded lightcurve where no observations are present). However, the lightcurve is more robustly sampled in \intersectsnapstess (top panel) and so this is an example where combining the two datasets allows us to have greater confidence in the solution. In our sample of $n=1623$ objects there are numerous examples of this including the case where the three period solutions disagree due to poor sampling of the folded \tessdata and \snapshotdata lightcurves, but the \intersectsnapstess lightcurve is more compelling due to better sampling.

\begin{figure}[!t]
\centering

\includegraphics[width=1\columnwidth]{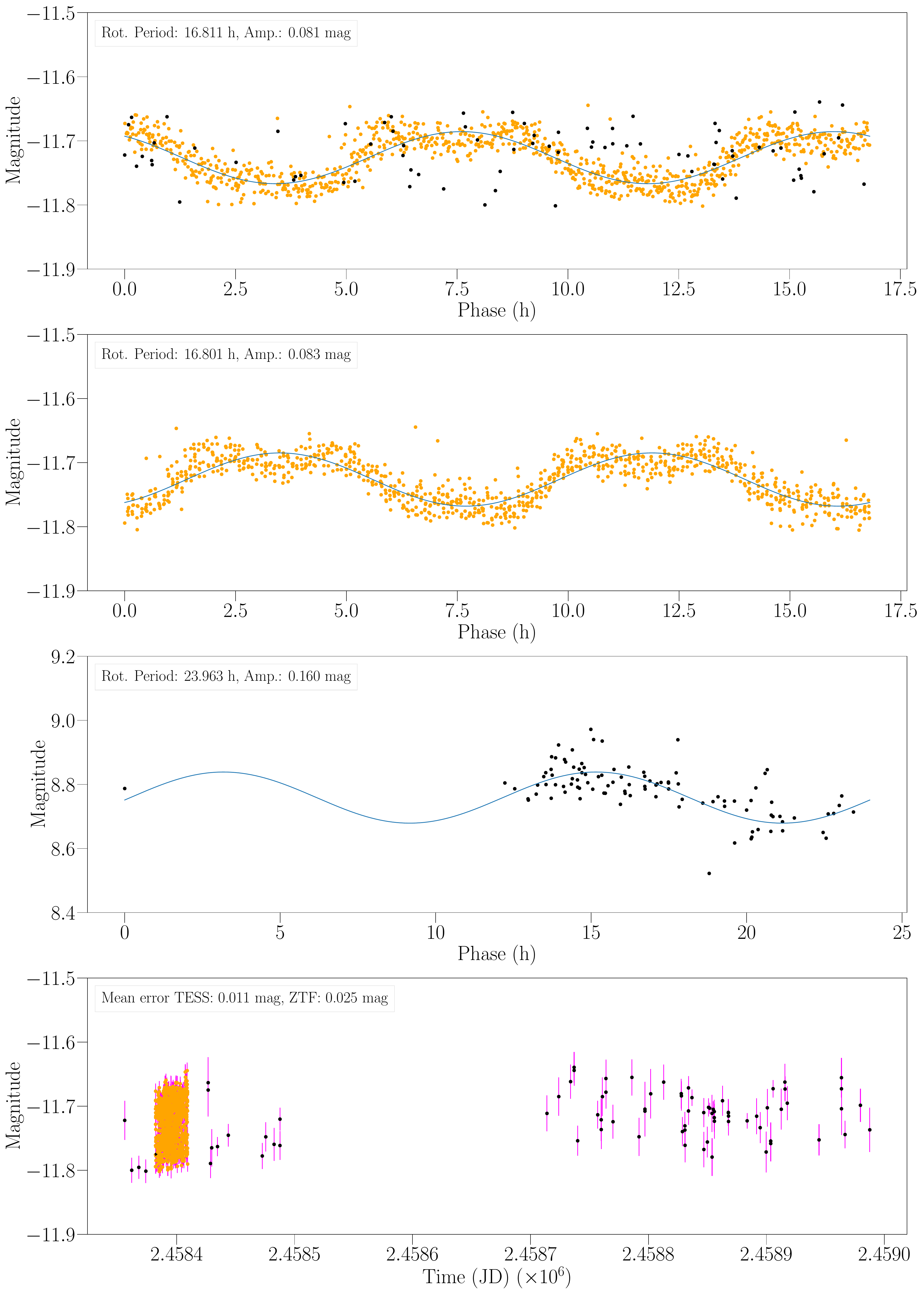}

    \caption{Lightcurves of object 363. Panels from top to bottom are as follows: \intersectsnapstess ($p=16.811$~h), \tessdata ($p=16.801$~h), \snapshotdata ($p=23.963$~h), and the unphased data for \intersectsnapstess. The \lcdbdata rotation period is reported as 8.401~h, which is a factor 0.5 of the \tessdata derived rotation period solution of 16.801~h (middle panel). This lightcurve is also remarkable because it has extra structure at the peaks instead of simply having a sinusoidal shape. }
   \label{fig:363_low_amplitude}
\end{figure}

\textbf{Object 363:} Figure~\ref{fig:363_low_amplitude} is interesting for two reasons. First, \snapshotdata (bottom panel) generates a lightcurve that is clearly very poorly sampled; however, we find that \tessdata (middle panel) has an excellent fit of $p=16.801$~h. When we examine \intersectsnapstess (top panel) we find that for this rotation period of $p=16.811$~h, the \snapshotdata observations clearly do not contribute to the goodness of the fit. In other words, using only the \tessdata observational record produces a better lightcurve than \intersectsnapstess, but the solution provided by \intersectsnapstess yields a good fit  despite the noise imparted by the \snapshotdata observations. The second interesting observation is that the rotation period for \tessdata (middle panel) is $p=16.801$~h and the reported rotation period in \lcdbdata is $p=8.401$~h, which is a factor 0.5 of our solution. This is considered an aliased match, where the derived rotation period is a factor of two greater than the rotation period reported in the literature. We attribute this discrepancy to the very low amplitude of the object, which is 0.083~mag using the \tessdata rotation period, and as shown in Figure~\ref{fig:amplitude_match_fraction}, low amplitudes may lead to rotation period matches that are incorrect by a factor 0.5 or 2 of the rotation period reported by \lcdbdata.

\begin{figure}[!t]
\centering

\includegraphics[width=1\columnwidth]{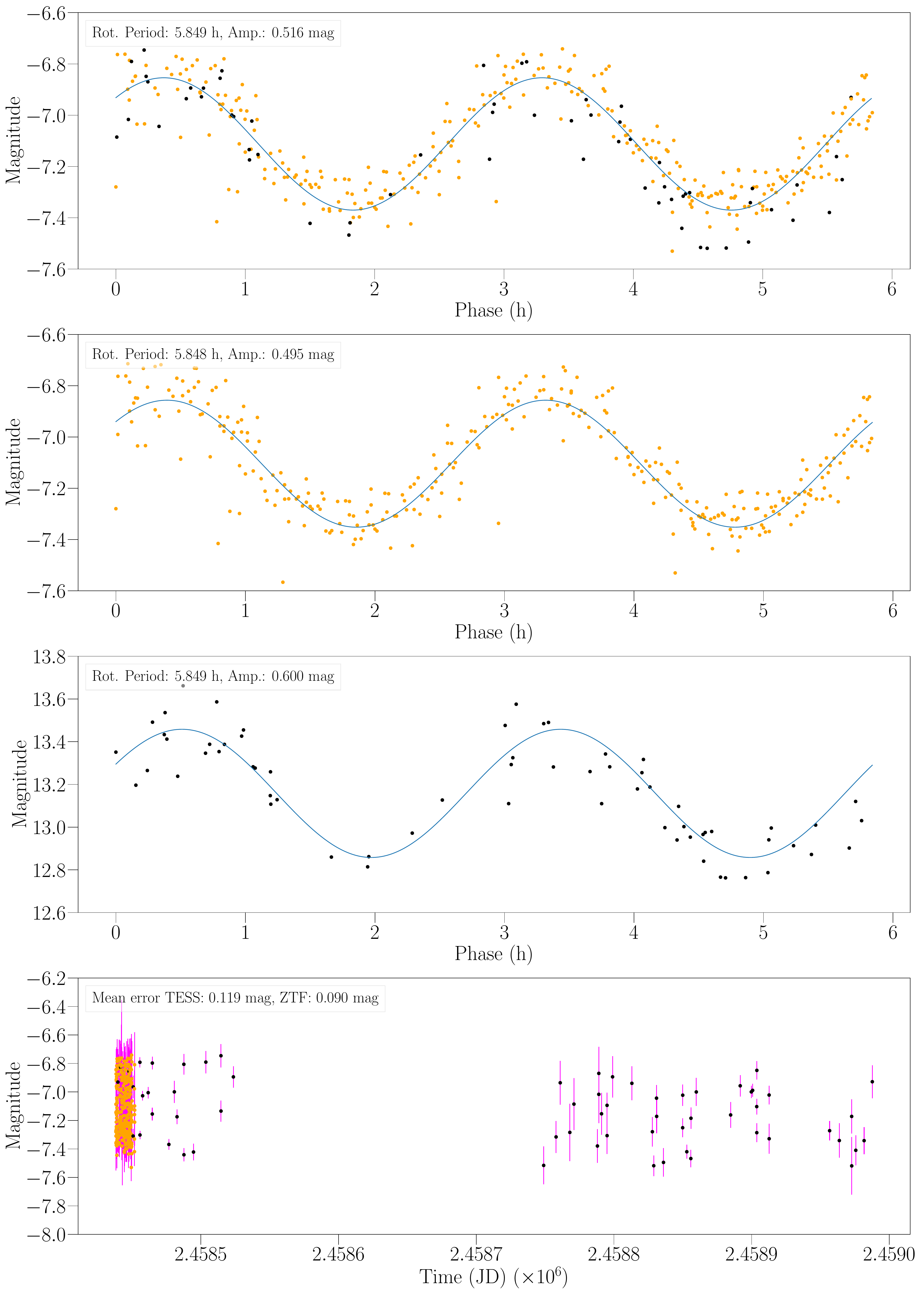}

    \caption{Lightcurves of object 14376. Panels from top to bottom are as follows: \intersectsnapstess ($p=5.849$~h), \tessdata ($p=5.848$~h), \snapshotdata ($p=5.849$~h), and the unphased data for \intersectsnapstess.}
   \label{fig:14376_seems_correct_but_wrong}
\end{figure}

\textbf{Object 14376:} Figure~\ref{fig:14376_seems_correct_but_wrong} shows that all three lightcurves are in agreement as they all report a rotation period of $p\approx 5.85$~h. Based on a manual inspection of the lightcurves, it appears that the rotation period of $p\approx 5.85$~h is plausible. However, \lcdbdata reports $p=5.614$~h which is in disagreement with our solution as we consider that rotation periods are matching if they are within 3\% of each other. Thus, while this fit is clearly reasonable, it is deemed incorrect in our analysis. This example shows that while many derived lightcurve solutions are incorrect for obvious reasons, such as poor sampling, even lightcurves with good fits may be incorrect. It is possible that the solution in \lcdbdata is incorrect, or another interesting possibility is that the rotation period has changed. This further demonstrates the challenges of reliably deriving rotation periods from sparse or dense photometry.

\begin{figure}[!t]
\centering

\includegraphics[width=1\columnwidth]{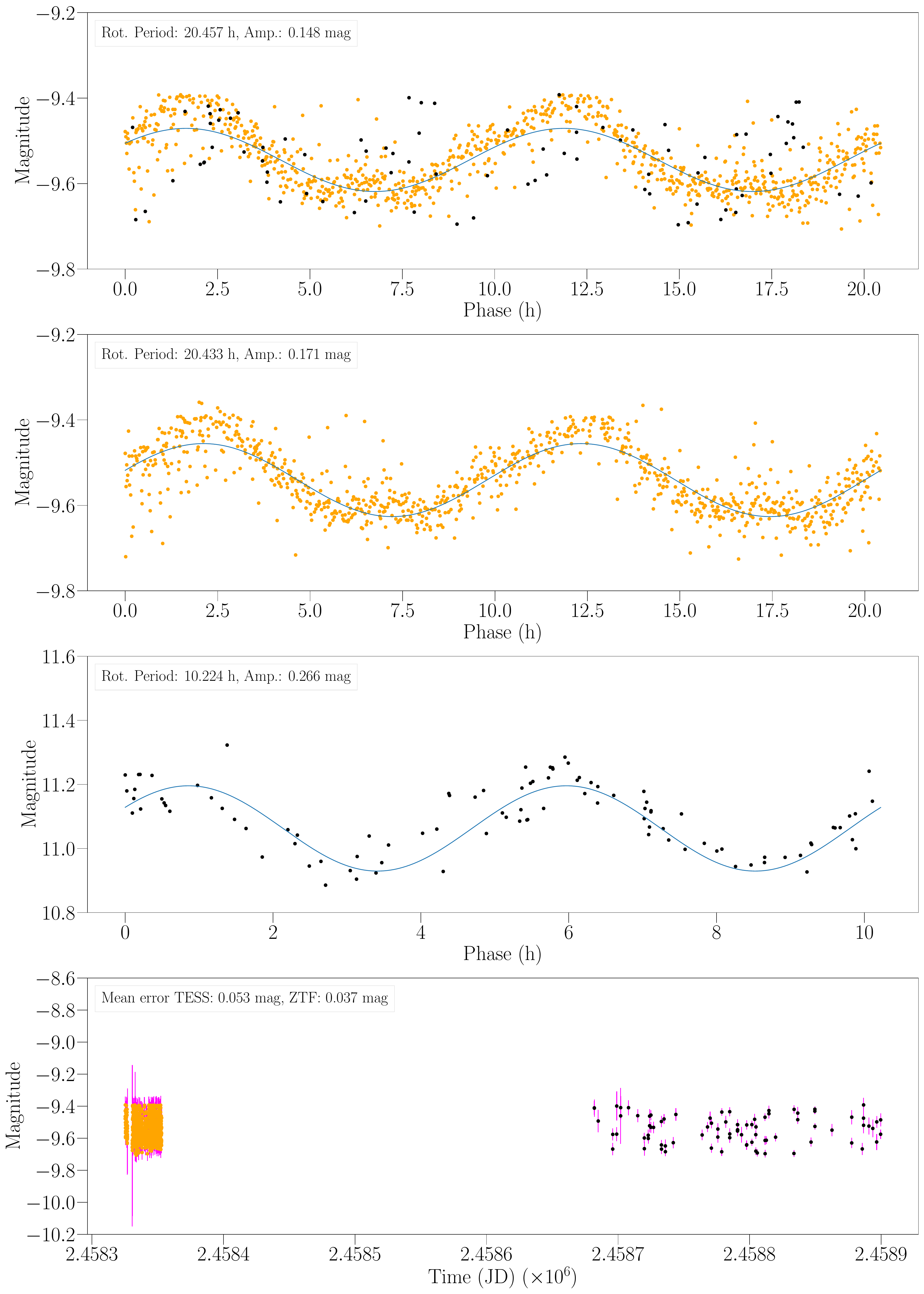}

    \caption{Lightcurves of object 3176. Panels from top to bottom are as follows: \intersectsnapstess ($p=20.457$~h), \tessdata ($p=20.433$~h), \snapshotdata ($p=10.224$~h), and the unphased data for \intersectsnapstess.}
   \label{fig:SS1_gets_correct}
\end{figure}

\textbf{Object 3176:} In the prior lightcurve figures we clearly observe that on average there are more data points in the observational records of \tessdata than in \snapshotdata. Additional observations for an object helps constrain the rotation period because it reduces sampling sparsity. Figure~\ref{fig:SS1_gets_correct} shows an example where \intersectsnapstess and \tessdata are in agreement and yield rotation periods of $p=20.457$~h and $p=20.433$~h, respectively. However, the \snapshotdata rotation period of $p=10.224$~h is an exact match to that reported in \lcdbdata which is $p=10.215$~h (the \intersectsnapstess and \tessdata rotation periods are aliased matches as they are a factor 2 greater than the \lcdbdata rotation period). This demonstrates that while fitting lightcurves with additional observations is generally beneficial, there are cases where fewer observations may yield a better derived rotation period. We also note that while the $p=10.224$~h rotation period for \snapshotdata (bottom panel) is clearly a good fit, the $p=20.457$~h rotation period (top panel) is not a good fit to the \snapshotdata data.

\subsection{Discussion}
We examined several methods for improving derived period fidelity when combining observational records from ground- and space-based facilities. We find that combining the data records from ZTF and TESS improves the overall match fraction with \lcdbdata and we applied the optimizations on the sample of objects in \intersectlcdbdata to the much larger \intersectsnapstess sample. In this section, we discuss implications for forthcoming Rubin Observatory operations, the use of the binary classification approach on other catalogs, and other optimizations beyond the scope of this paper.

\subsubsection{Rubin Observatory}
In the context of preparing for the Rubin Observatory, several of the optimizations and insights may need to be reexamined. We outline these as follows:

\begin{itemize}
\item The amplitude optimization should be reexamined as Rubin will be more sensitive to amplitudes $<$0.075~mag.
\item There are a number of lightcurves that we found to be non-sinusoidal (e.g., see Figure~\ref{fig:363_low_amplitude}). These lightcurves may be better fit by other period finding algorithms that are not constrained to sinusoids.  
\item We examined period replacement with \tessdata and \snapshotdata and found that the former was useful but the latter was not. We found that few objects in \snapshotdata had a sufficient number of observations and \lsp power (Figure~\ref{fig:snapshot1_heatmap}) to achieve an 85\% match fraction to \lcdbdata. The next \snaps data release will provide many more ZTF observations per object and so period replacement with this new dataset should be reexamined if augmenting Rubin data with ZTF data.
\item The binary classification approach will need to be reexamined to select a configuration that optimizes for precision and accuracy. For example, Rubin will have lower photometric error and so the observation threshold may be reduced compared to that used in this paper.
\end{itemize}

Furthermore, there are two implications for the Rubin Observatory that originate from the analysis of Figure~\ref{fig:SS1_gets_correct} that we outline as follows.
\begin{itemize}
 \item While having a larger number of observations for an object reduces the sparsity of the folded lightcurve which improves the fit, another factor that may be of great importance is having better quality (lower photometric uncertainty) observations. \citet{kramer2023submitted} showed that for the Rubin Observatory, an observing cadence that favors longer exposures and thus reduces the total number of observations for each object yields a higher rotation period match fraction to \lcdbdata compared to a cadence with shorter exposures and more observations per object. 
 \item Because we find that a moderate number of observations (Figure~\ref{fig:SS1_gets_correct} showing \snapshotdata) can yield accurate rotation periods, the Rubin Observatory will produce high fidelity lightcurves within the first few years of operation. Thus a major task will be to automate the characterization of lightcurve fidelity to curate lightcurves that have a high probability of having correct rotation periods and biasing against reporting incorrect rotation periods.
\end{itemize}

In the context of examining ZTF and SSPDB data, \citet{KRAMER2023} examined several methods for improving the overall rotation period match rate to \lcdbdata and they discuss the prospect of combining optimizations. We find that combining optimizations is only beneficial for capturing rotation periods in a small number of instances and instead combining the two datasets to produce \intersectsnapstess had a larger overall benefit on period fidelity. One caveat to this analysis is that the \lcdbdata is not a particularly large catalog and so we were only able to compare against 222 objects. Therefore, a larger reference sample would help us better understand the utility of the optimizations presented here.

It is clear that low amplitude asteroids are often those where LSP predicts a period at an alias, and for ZTF these appear at $p\in$\{16, 24, 48\}~h. Thus, these asteroids fundamentally cannot produce a lightcurve and are simply intractable for period derivation, which implies that the $\approx$85\% match fraction to \lcdbdata achieved in Table~\ref{tab:methods} may be close to the upper limit on derived period fidelity assuming reasonable constraints on the minimum amplitude and minimum rotation period.

\subsubsection{Binary Classification Approach: Application to Other Catalogs}
The binary classification approach employed here is used to bias against reporting incorrect asteroid rotation period solutions. The method can be applied to other catalogs where there is a baseline (or fiducial) set of correct rotation period solutions and another set of solutions of unknown quality. In this paper, our baseline dataset was \lcdbdata.

While we selected binary classification labels using thresholds for the \lsp power, lightcurve amplitude, and number of observations, these parameters will need to be reevaluated based on the input dataset of unknown quality. For example, observations obtained with lower photometric error may require lower observation and amplitude thresholds to obtain similar accuracy and precision levels reported in Section~\ref{sec:conservative_classifier}. Furthermore, it may be preferable to select accuracy and precision based on the receiver operator characteristic (ROC) curves which illustrates the relationship between the true positive and false positive rates.

Deriving a generalizable set of parameter thresholds for the binary classification approach is not possible, as we only considered two surveys in this paper and so the parameters were tailored to those surveys. However, a promising future research direction is to combine data from multiple surveys. If the surveys span a large range of parameter values, including mean photometric error and number of observations per object, it may be possible to extract generalizable parameters that yield high precision with minimal loss in accuracy. This research direction may benefit from a supervised machine learning approach, as the number of parameters needed will exceed the three parameters investigated here.

\subsubsection{Other Optimizations}
There may be several other avenues of investigation that may improve period fidelity. We focused on combining several well-known methods in Section~\ref{sec:comparison_to_LCDB} but also attempted two other less obvious methods that were not useful and so we did not report these results. 

We know that \lsp is good for fitting the lightcurves of asteroids because they are generally sinusoidal in shape. We examined using the \ss~\citep{friedman1984variable} algorithm which takes advantage of structure in the data which can be used to disambiguate periods detected near aliases. For example, \citet{gowanlock2022gpu} found that  \ss was better than \lsp for computing the periods of RR-Lyrae stars. However, we found that \lsp is better than \ss for computing the rotation periods of asteroids. Consequently this method was not presented here. 

Another approach to disambiguate aliases is to use Monte Carlo sampling of the observational record to produce a set of periodograms for each object. Combining the set of periodograms to select the frequency with the greatest combined power may reduce aliases~\citep{KRAMER2023}. We selected a sample of objects for which our methods were unable to derive the correct rotation period where most of these objects had aliased period solutions. However, we found that the Monte Carlo approach was unable to improve period fidelity in these instances. It is likely that this is because of the low amplitude problem that we outlined above --- many of these objects have very low amplitudes for which producing a lightcurve is intractable.

\section{Prospects for the Rubin Observatory}\label{sec:rubin_companion}



In the prior sections we examined several approaches to improve the fidelity of derived asteroid rotation periods using several optimizations in the context of combining data from ground- and space-based observatories. In this section, we examine a practical aspect of Rubin Observatory operations that is driven by the following question: \emph{To what extent can a companion (ground- or space-based) observatory significantly improve lightcurve fidelity and derived rotation periods of regularly varying sources observed with the Rubin Observatory, and how much telescope time is required?}

Despite the impressive capabilities of the Rubin Observatory, it will still  be limited by the diurnal observing schedule, which will produce aliases. By employing a companion telescope to augment the LSST catalog, aliases may be partially suppressed (as we observed examining the \intersectsnapstess dataset).  Because telescopic observations are expensive, it would be interesting to know how much observing time on a companion telescope is necessary to eliminate a substantial fraction of the rotation periods reported at the aliases generated by the Rubin Observatory's schedule and improve the overall yield of correct rotation period solutions. 

\subsection{A Future TESS-like Observatory}

Here we explore the benefit of including data from a future TESS-like observatory (which could indeed be TESS in an extended mission).
To assess the above question, we use the $n=222$ objects in \intersectlcdbdata. We use the entire observational record for each object in \snapshotdata (thus using ZTF data to simulate the LSST catalog) and then incrementally add \tessdata to \snapshotdata, where we are simulating using TESS (or a TESS-like facility) as a companion telescope to the Rubin Observatory. We do not apply any of the optimizations outlined in Section~\ref{sec:comparison_to_LCDB} as we are simulating incrementally adding observations to the ground-based observational records. As such, some of the optimizations would not be possible, such as knowing which observations to $\sigma$ clip without the full observational record for an object (e.g., with a small observational record for an object, we may inadvertently remove observations that are useful for fitting the lightcurve).

 The number of observations for each object in \tessdata varies. Therefore, when we combine the data between \snapshotdata and \tessdata, we use either the desired number of observations that we wish to add to \snapshotdata, or the total number of observations for an object, whichever is smaller. For example, if we want to use 100 \tessdata observations for an object, but the observational record only contains 90 entries for that object, then we are limited by the 90 observations and use the full observational record containing these observations.

\begin{figure}[!t]
\centering

\includegraphics[width=1\columnwidth]{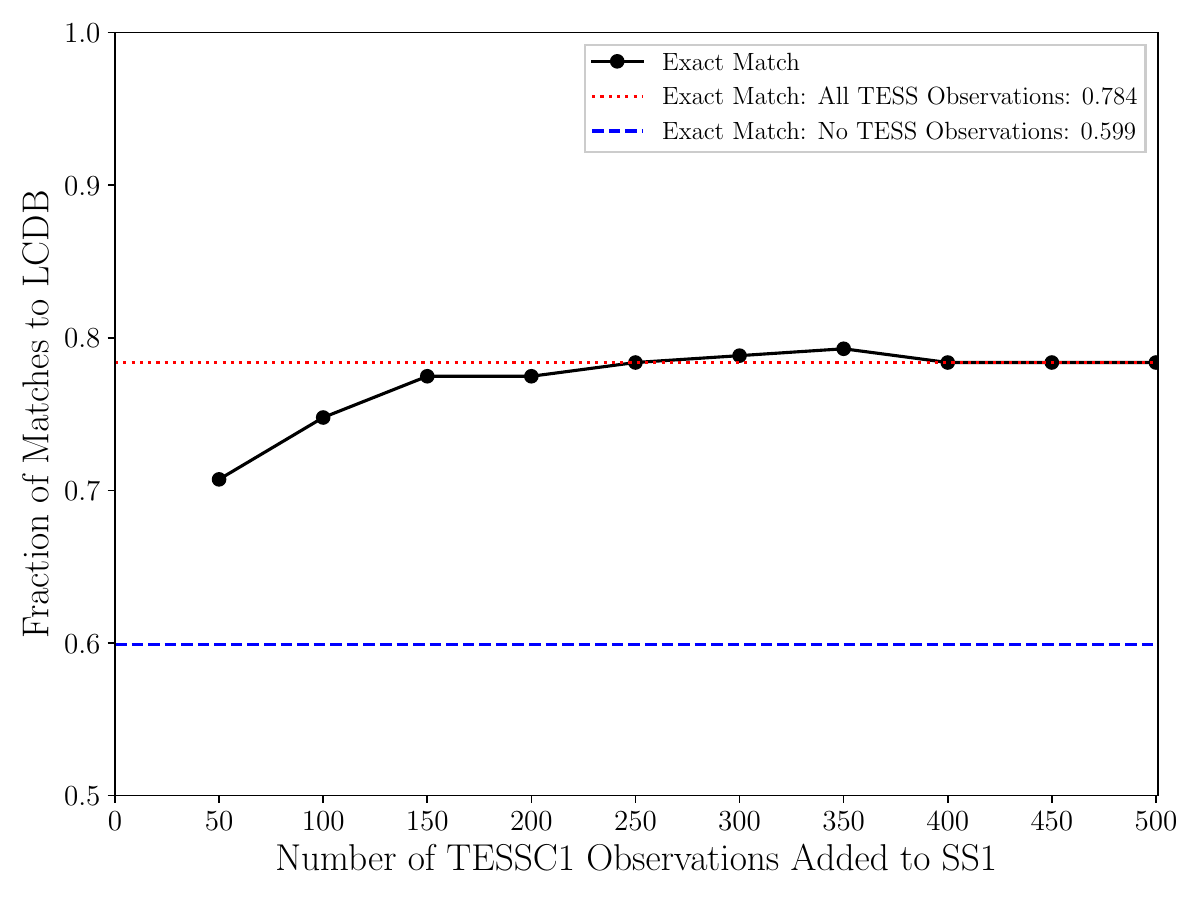}

 \caption{The exact match fraction to \lcdbdata as a function of the number of \tessdata observations added to \snapshotdata. The top horizontal line refers to the exact match fraction when all \tessdata observations are added to \snapshotdata and the bottom horizontal line refers to when there are no \tessdata observations added to \snapshotdata. These exact match fractions are 0.784 and 0.599, respectively (Table~\ref{tab:matchfraclcdb}).}
   \label{fig:incremental_TESS_to_ZTF}
\end{figure}

Figure~\ref{fig:incremental_TESS_to_ZTF} plots the exact match fraction to \lcdbdata as a function of the number of observations added per object from \tessdata to \snapshotdata. The upper and lower bound match fractions are given by the two horizontal lines, which refer to when all (or none) of the \tessdata observations are added to \snapshotdata. These fractions are reported in Table~\ref{tab:matchfraclcdb} where we find match fractions of 0.784 and 0.599, respectively.

We find that after adding roughly 150  \tessdata  observations to \snapshotdata, we obtain a match fraction approaching the upper bound of 0.784. Given that TESS uses a 30~minute full-frame image cadence, 150 observations/full-frames would require roughly 3 days of TESS observations.

As a targeted companion facility that is synchronized to LSST observations, 
this is an impractical solution, as LSST covers most of the available southern sky during a 3~day period, whereas TESS only covers a single field of view --- around 10\% of one hemisphere --- during this same time period. However, over the course of a year, TESS covers an entire hemisphere with $>$3~days of continuous coverage, and LSST also covers an entire hemisphere of the sky. The TESS-like data need not be contemporaneous with LSST-like observations; all that matters is the uninterrupted cadence that complements the LSST-like observations. Therefore, if TESS were observing during LSST's survey, the alias solutions would be much reduced and the overall match fraction increased as a natural consequence of combining these data sets with two different cadences.
(If this future TESS uses the 30~minute cadence then we still have no new power to address potential periods less than 3~hours.)

The above estimates may have under-predicted the delivered performance, though.
Consider that the \snapshotdata dataset derived from ZTF is roughly 10\% of the scale of LSST~\citep{trilling2023solar}. Thus, if the same experiment were to be carried out with the Rubin Observatory, we would expect that the match fraction without any companion telescope data would be higher than the 78.4\% reported here because each object would be observed more with the Rubin Observatory than ZTF over the same time period. Also, recall from Section~\ref{sec:low_amplitude_object_exclusion} that it is challenging to derive the rotation periods of asteroids with low lightcurve amplitudes. The Rubin Observatory will have lower photometric error \citep[$\leq$0.1~mag,][]{navarro2021asteroid} 
than ZTF and so we expect that  it will yield many more low amplitude objects with lightcurve amplitudes $<$0.1~mag that have correct rotation period solutions. Given that we expect many of the objects for which we cannot derive a correct rotation period solution to be those with low amplitudes, we believe that this capability will dramatically improve derived rotation period fidelity.

We therefore conclude that if TESS is still operational in 2025 and after, combining TESS data with LSST data will be extremely useful for increasing the fidelity of period solutions.



ESA's Euclid mission~\citep{euclid,racca} complements both TESS and LSST observations, and the joint Rubin-Euclid working group has already offered some recommendations about how the combined datasets may enhance the science return of both projects~\citep{rubineuclid}. The footprint of Euclid's Wide Survey avoids the ecliptic, but there are plenty of asteroids that will still be detected in Euclid data.
The imaging sequence of Euclid will provide several images over one hour~\citep{scaramella}. Thus, Euclid data may be useful to identify asteroids with short rotation periods, which neither TESS nor LSST will be very successful at identifying.
However, in general Euclid will not offer enough complementary data to significantly improve a large number of LSST period solutions.

\subsection{Augmenting LSST Data with Ground-based Data}

Additional data from the Rubin Observatory itself might be useful to resolve degeneracies through data from the Deep Drilling Fields (DDF). 
Not all DDF observations have been defined at this point, but it is likely that at least some of these DDF campaigns will be a continuous all-night stare at a single LSST pointing. This is a TESS-like (i.e., continuous) set of perhaps $\sim$1000~observations obtained over around 10~hours. This may help resolve degeneracies at the shortest periods, but is not likely to address diurnal aliases. In a future investigation we will model the performance of LSST period solutions in the presence of both the standard ``wide-fast-deep'' cadence (which is ZTF-like) and DDF observations.

It is possible that other ground-based observatories or networks that have broad longitudinal coverage, such as the
ATLAS network \citep{atlas}, which presently has four operational sites (one site in South Africa, one in Chile, and two in Hawai`i),
could collectively observe the same objects without a break in day-night observing. This would help reduce the number of rotation period solutions at aliases and would be less expensive than pursing a companion space-based telescope for LSST.
However, ATLAS' limiting magnitude of perhaps~20 is not well-matched to LSST's limiting magnitude of~24.5; the vast majority of sources observed by LSST will be too faint for ATLAS. 

In summary, there are no obvious existing or near-term ground-based assets that will provide the data necessary to improve period solutions on a large scale, but both LSST and ATLAS (and perhaps others) may provide limited capacity to improve solutions derived only from LSST wide-fast-deep data.

\section{Conclusion}\label{sec:conclusion}
This paper examined improving the derived asteroid rotation period fidelity when combining data from ground- and space-based facilities. In particular, we used the \snapshotdata~\citep{trilling2023solar} and \tessdata~\citep{mcneill2023tess} datasets which were collected with ZTF and TESS, respectively. We found that our baseline approach of combining the two datasets had the greatest benefit on improving our rotation period match fraction to \lcdbdata. Combining several optimizations further improved the match fraction in several instances; however, the impact of the optimizations have diminishing returns compared to simply combining the two datasets together. This is because many of the objects that had incorrect rotation periods were those that are well-known to be challenging to derive, such as those with low lightcurve amplitudes which often have rotation periods that are reported at aliases. Consequently, we find that the $\approx$85\% match fraction may be close to the upper limit for this dataset because many of the remaining $\approx$15\% of asteroids may have a spherical morphology or pole-on orientation that makes lightcurve derivation intractable. 

The \snapshotdata dataset is limited by 0.1~mag photometric error, and so we are not sensitive to amplitudes less than this value. The \tessdata dataset has even greater photometric error than \snapshotdata so it is even less sensitive to low amplitude (spherical) asteroids. The low amplitude problem described above will improve when the Rubin Observatory is online as it will have lower photometric error than ZTF or TESS. 

We also discussed Rubin Observatory operations and the prospects of augmenting the LSST data stream with data from space- or ground-based observatories to reduce the fraction of aliases. While none of the solutions for eliminating aliases are straightforward (or necessarily likely such as using TESS in 2025 and thereafter), the analysis points to several possible avenues that can be pursued over the 10 year lifespan of Rubin Observatory operations.

Recall that the $\gtrsim 0.1$~mag photometric error yielded by ZTF and TESS yields inaccurate period solutions when lightcurve amplitudes are $<$0.1~mag. An interesting future work direction that we will address is testing the ``sphere-abundant'' and ``sphere-limited'' hypotheses illustrated in Figure~\ref{fig:amplitude_histograms_model} (Strauss et al., in prep.).  This future work direction will be conducted as part of the DECam Ecliptic Exploration Project (DEEP)~\citep{trilling2023decam}, which will yield observations with a mean photometric error of roughly 0.03~mag. This is significantly lower than the $\sim$0.1~mag photometric error provided by ZTF and will allow us to determine which of the two hypotheses above are plausible.

Other future work includes incorporating data from other sources, such as observations of asteroids from Kepler~\citep{sergeyevspin} to create a unified observational record for analysis by our team and the community. Another direction is to examine accuracy as a function of precision for the binary classification approach to deriving rotation periods when using data from numerous sources. This direction may yield insights into the fundamental parameters needed to correctly derive (with very high probability) the rotation periods of asteroids using heterogeneous data sources. We will also incorporate these analyses into the Solar System Notification Alert Processing System (\snaps) alert broker infrastructure to further enhance our derived data products for the LSST era.

\section*{Acknowledgments}

This material is based upon work supported by the National Science Foundation under Grant No. 2042155 and 2206796.  We acknowledge support by the Arizona Board of Regents' Regents Innovation Fund and the Technology Research Initiative Research Fund (TRIF) Small Research Equipment Acquisition Program (SREAP) at Northern Arizona University. We thank Siegfried Eggl for fruitful discussions regarding the combination of data from space- and ground-based observatories. We thank Ryder Strauss for providing the estimated photmetric error for the DEEP project.

ZTF is a public-private partnership, with equal support from the ZTF Partnership and from the U.S. National Science Foundation through the Mid-Scale Innovations Program (MSIP).
The ZTF partnership is a consortium of the following universities and institutions (listed in descending longitude): TANGO Consortium of Taiwan; Weizmann Institute of Sciences, Israel; Oskar Klein Center, Stockholm University, Sweden; Deutsches Elektronen-Synchrotron \& Humboldt University, Germany; Ruhr University, Germany; Institut national de physique nucl\'eaire et de physique des particules, France; University of Warwick, UK; Trinity College, Dublin, Ireland; University of Maryland, College Park, USA; Northwestern University, Evanston, USA; University of Wisconsin, Milwaukee, USA; Lawrence Livermore National Laboratory, USA; IPAC, Caltech, USA; Caltech, USA.

\software{Astropy~\citep{robitaille2013astropy}, SciPy~\citep{virtanen2020scipy}, NumPy~\citep{harris2020array}, and pandas~\citep{reback2020pandas}.}

\bibliographystyle{aasjournal}

\begin{thebibliography}{}
\expandafter\ifx\csname natexlab\endcsname\relax\def\natexlab#1{#1}\fi
\providecommand{\url}[1]{\href{#1}{#1}}
\providecommand{\dodoi}[1]{doi:~\href{http://doi.org/#1}{\nolinkurl{#1}}}
\providecommand{\doeprint}[1]{\href{http://ascl.net/#1}{\nolinkurl{http://ascl.net/#1}}}
\providecommand{\doarXiv}[1]{\href{https://arxiv.org/abs/#1}{\nolinkurl{https://arxiv.org/abs/#1}}}

\bibitem[{{Bellm} {et~al.}(2019){Bellm}, {Kulkarni}, {Graham}, {Dekany},
  {Smith}, {Riddle}, {Masci}, {Helou}, {Prince}, {Adams}, {Barbarino},
  {Barlow}, {Bauer}, {Beck}, {Belicki}, {Biswas}, {Blagorodnova}, {Bodewits},
  {Bolin}, {Brinnel}, {Brooke}, {Bue}, {Bulla}, {Burruss}, {Cenko}, {Chang},
  {Connolly}, {Coughlin}, {Cromer}, {Cunningham}, {De}, {Delacroix}, {Desai},
  {Duev}, {Eadie}, {Farnham}, {Feeney}, {Feindt}, {Flynn}, {Franckowiak},
  {Frederick}, {Fremling}, {Gal-Yam}, {Gezari}, {Giomi}, {Goldstein},
  {Golkhou}, {Goobar}, {Groom}, {Hacopians}, {Hale}, {Henning}, {Ho}, {Hover},
  {Howell}, {Hung}, {Huppenkothen}, {Imel}, {Ip}, {Ivezi{\'c}}, {Jackson},
  {Jones}, {Juric}, {Kasliwal}, {Kaspi}, {Kaye}, {Kelley}, {Kowalski},
  {Kramer}, {Kupfer}, {Landry}, {Laher}, {Lee}, {Lin}, {Lin}, {Lunnan},
  {Giomi}, {Mahabal}, {Mao}, {Miller}, {Monkewitz}, {Murphy}, {Ngeow},
  {Nordin}, {Nugent}, {Ofek}, {Patterson}, {Penprase}, {Porter}, {Rauch},
  {Rebbapragada}, {Reiley}, {Rigault}, {Rodriguez}, {van Roestel}, {Rusholme},
  {van Santen}, {Schulze}, {Shupe}, {Singer}, {Soumagnac}, {Stein}, {Surace},
  {Sollerman}, {Szkody}, {Taddia}, {Terek}, {Van Sistine}, {van Velzen},
  {Vestrand}, {Walters}, {Ward}, {Ye}, {Yu}, {Yan}, \&
  {Zolkower}}]{2019PASP..131a8002B}
{Bellm}, E.~C., {Kulkarni}, S.~R., {Graham}, M.~J., {et~al.} 2019, \pasp, 131,
  018002, \dodoi{10.1088/1538-3873/aaecbe}

\bibitem[{{Bernardinelli} {et~al.}(2023){Bernardinelli}, {Bernstein}, {Jindal},
  {Abbott}, {Aguena}, {Andrade-Oliveira}, {Annis}, {Bacon}, {Bertin}, {Brooks},
  {Burke}, {Carnero Rosell}, {Carrasco Kind}, {Carretero}, {da Costa},
  {Pereira}, {Davis}, {Desai}, {Diehl}, {Doel}, {Everett}, {Ferrero},
  {Friedel}, {Frieman}, {Garc{\'\i}a-Bellido}, {Giannini}, {Gruen}, {Herner},
  {Hinton}, {Hollowood}, {Honscheid}, {James}, {Kuehn}, {Mena-Fern{\'a}ndez},
  {Menanteau}, {Miquel}, {Ogando}, {Pieres}, {Plazas Malag{\'o}n}, {Raveri},
  {Sanchez}, {Sevilla-Noarbe}, {Smith}, {Suchyta}, {Swanson}, {Tarle}, {To},
  {Walker}, {Wiseman}, \& {Zhang}}]{BernardinelliTNO2023}
{Bernardinelli}, P.~H., {Bernstein}, G.~M., {Jindal}, N., {et~al.} 2023, arXiv
  e-prints, arXiv:2304.03017, \dodoi{10.48550/arXiv.2304.03017}

\bibitem[{Cabrera-Vives {et~al.}(2017)Cabrera-Vives, Reyes, F{\"o}rster,
  Est{\'e}vez, \& Maureira}]{cabrera2017deep}
Cabrera-Vives, G., Reyes, I., F{\"o}rster, F., Est{\'e}vez, P.~A., \& Maureira,
  J.-C. 2017, The Astrophysical Journal, 836, 97

\bibitem[{Coughlin {et~al.}(2021)Coughlin, Burdge, Duev, Katz, van Roestel,
  Drake, Graham, Hillenbrand, Mahabal, Masci, Mróz, Prince, Yao, Bellm,
  Burruss, Dekany, Jaodand, Kaplan, Kupfer, Laher, Riddle, Rigault, Rodriguez,
  Rusholme, \& Zolkower}]{Coughlin2021}
Coughlin, M.~W., Burdge, K., Duev, D.~A., {et~al.} 2021, Monthly Notices of the
  Royal Astronomical Society, 505, 2954, \dodoi{10.1093/mnras/stab1502}

\bibitem[{Drake {et~al.}(2013)Drake, Catelan, Djorgovski, Torrealba, Graham,
  Belokurov, Koposov, Mahabal, Prieto, Donalek, {et~al.}}]{drake2013probing}
Drake, A., Catelan, M., Djorgovski, S., {et~al.} 2013, The Astrophysical
  Journal, 763, 32

\bibitem[{Duev {et~al.}(2019)Duev, Mahabal, Masci, Graham, Rusholme, Walters,
  Karmarkar, Frederick, Kasliwal, Rebbapragada, \& Ward}]{Duev2019}
Duev, D.~A., Mahabal, A., Masci, F.~J., {et~al.} 2019, Monthly Notices of the
  Royal Astronomical Society, 489, 3582, \dodoi{10.1093/mnras/stz2357}

\bibitem[{Erasmus {et~al.}(2021)Erasmus, Kramer, McNeill, Trilling,
  Janse van Rensburg, van Belle, Tonry, Denneau, Heinze, \&
  Weiland}]{Erasmus2021}
Erasmus, N., Kramer, D., McNeill, A., {et~al.} 2021, Monthly Notices of the
  Royal Astronomical Society, 506, 3872, \dodoi{10.1093/mnras/stab1888}

\bibitem[{{Euclid Collaboration} {et~al.}(2022){Euclid Collaboration},
  {Scaramella}, {Amiaux}, {Mellier}, {Burigana}, {Carvalho}, {Cuillandre}, {Da
  Silva}, {Derosa}, {Dinis}, {Maiorano}, {Maris}, {Tereno}, {Laureijs},
  {Boenke}, {Buenadicha}, {Dupac}, {Gaspar Venancio}, {G{\'o}mez-{\'A}lvarez},
  {Hoar}, {Lorenzo Alvarez}, {Racca}, {Saavedra-Criado}, {Schwartz}, {Vavrek},
  {Schirmer}, {Aussel}, {Azzollini}, {Cardone}, {Cropper}, {Ealet}, {Garilli},
  {Gillard}, {Granett}, {Guzzo}, {Hoekstra}, {Jahnke}, {Kitching}, {Maciaszek},
  {Meneghetti}, {Miller}, {Nakajima}, {Niemi}, {Pasian}, {Percival},
  {Pottinger}, {Sauvage}, {Scodeggio}, {Wachter}, {Zacchei}, {Aghanim},
  {Amara}, {Auphan}, {Auricchio}, {Awan}, {Balestra}, {Bender}, {Bodendorf},
  {Bonino}, {Branchini}, {Brau-Nogue}, {Brescia}, {Candini}, {Capobianco},
  {Carbone}, {Carlberg}, {Carretero}, {Casas}, {Castander}, {Castellano},
  {Cavuoti}, {Cimatti}, {Cledassou}, {Congedo}, {Conselice}, {Conversi},
  {Copin}, {Corcione}, {Costille}, {Courbin}, {Degaudenzi}, {Douspis},
  {Dubath}, {Duncan}, {Dusini}, {Farrens}, {Ferriol}, {Fosalba}, {Fourmanoit},
  {Frailis}, {Franceschi}, {Franzetti}, {Fumana}, {Gillis}, {Giocoli},
  {Grazian}, {Grupp}, {Haugan}, {Holmes}, {Hormuth}, {Hudelot}, {Kermiche},
  {Kiessling}, {Kilbinger}, {Kohley}, {Kubik}, {K{\"u}mmel}, {Kunz},
  {Kurki-Suonio}, {Lahav}, {Ligori}, {Lilje}, {Lloro}, {Mansutti}, {Marggraf},
  {Markovic}, {Marulli}, {Massey}, {Maurogordato}, {Melchior}, {Merlin},
  {Meylan}, {Mohr}, {Moresco}, {Morin}, {Moscardini}, {Munari}, {Nichol},
  {Padilla}, {Paltani}, {Peacock}, {Pedersen}, {Pettorino}, {Pires}, {Poncet},
  {Popa}, {Pozzetti}, {Raison}, {Rebolo}, {Rhodes}, {Rix}, {Roncarelli},
  {Rossetti}, {Saglia}, {Schneider}, {Schrabback}, {Secroun}, {Seidel},
  {Serrano}, {Sirignano}, {Sirri}, {Skottfelt}, {Stanco}, {Starck},
  {Tallada-Cresp{\'\i}}, {Tavagnacco}, {Taylor}, {Teplitz}, {Toledo-Moreo},
  {Torradeflot}, {Trifoglio}, {Valentijn}, {Valenziano}, {Verdoes Kleijn},
  {Wang}, {Welikala}, {Weller}, {Wetzstein}, {Zamorani}, {Zoubian}, {Andreon},
  {Baldi}, {Bardelli}, {Boucaud}, {Camera}, {Di Ferdinando}, {Fabbian},
  {Farinelli}, {Galeotta}, {Graci{\'a}-Carpio}, {Maino}, {Medinaceli}, {Mei},
  {Neissner}, {Polenta}, {Renzi}, {Romelli}, {Rosset}, {Sureau}, {Tenti},
  {Vassallo}, {Zucca}, {Baccigalupi}, {Balaguera-Antol{\'\i}nez}, {Battaglia},
  {Biviano}, {Borgani}, {Bozzo}, {Cabanac}, {Cappi}, {Casas}, {Castignani},
  {Colodro-Conde}, {Coupon}, {Courtois}, {Cuby}, {de la Torre}, {Desai},
  {Dole}, {Fabricius}, {Farina}, {Ferreira}, {Finelli}, {Flose-Reimberg},
  {Fotopoulou}, {Ganga}, {Gozaliasl}, {Hook}, {Keihanen}, {Kirkpatrick},
  {Liebing}, {Lindholm}, {Mainetti}, {Martinelli}, {Martinet}, {Maturi},
  {McCracken}, {Metcalf}, {Morgante}, {Nightingale}, {Nucita}, {Patrizii},
  {Potter}, {Riccio}, {S{\'a}nchez}, {Sapone}, {Schewtschenko}, {Schultheis},
  {Scottez}, {Teyssier}, {Tutusaus}, {Valiviita}, {Viel}, {Vriend}, \&
  {Whittaker}}]{scaramella}
{Euclid Collaboration}, {Scaramella}, R., {Amiaux}, J., {et~al.} 2022, \aap,
  662, A112, \dodoi{10.1051/0004-6361/202141938}

\bibitem[{Friedman(1984)}]{friedman1984variable}
Friedman, J.~H. 1984.
\newblock \url{http://www.slac.stanford.edu/cgi-wrap/getdoc/slac-pub-3477.pdf}

\bibitem[{Gowanlock {et~al.}(2022)Gowanlock, Butler, Trilling, \&
  McNeill}]{gowanlock2022gpu}
Gowanlock, M., Butler, N., Trilling, D., \& McNeill, A. 2022, Astronomy and
  Computing, 38, 100511, \dodoi{https://doi.org/10.1016/j.ascom.2021.100511}

\bibitem[{Gowanlock {et~al.}(2021)Gowanlock, Kramer, Trilling, Butler, \&
  Donnelly}]{gowanlock2021fast}
Gowanlock, M., Kramer, D., Trilling, D., Butler, N., \& Donnelly, B. 2021,
  Astronomy and Computing, 36, 100472,
  \dodoi{https://doi.org/10.1016/j.ascom.2021.100472}

\bibitem[{Graham {et~al.}(2013)Graham, Drake, Djorgovski, Mahabal, Donalek,
  Duan, \& Maker}]{graham2013comparison}
Graham, M.~J., Drake, A.~J., Djorgovski, S.~G., {et~al.} 2013, Monthly Notices
  of the Royal Astronomical Society, 434, 3423, \dodoi{10.1093/mnras/stt1264}

\bibitem[{{Guy} {et~al.}(2022){Guy}, {Cuillandre}, {Bachelet}, {Banerji},
  {Bauer}, {Collett}, {Conselice}, {Eggl}, {Ferguson}, {Fontana}, {Heymans},
  {Hook}, {Aubourg}, {Aussel}, {Bosch}, {Carry}, {Hoekstra}, {Kuijken},
  {Lanusse}, {Melchior}, {Mohr}, {Moresco}, {Nakajima}, {Paltani}, {Troxel},
  {Allevato}, {Amara}, {Andreon}, {Anguita}, {Bardelli}, {Bechtol}, {Birrer},
  {Bisigello}, {Bolzonella}, {Botticella}, {Bouy}, {Brinchmann}, {Brough},
  {Camera}, {Cantiello}, {Cappellaro}, {Carlin}, {Castander}, {Castellano},
  {Chari}, {Chisari}, {Collins}, {Courbin}, {Cuby}, {Cucciati}, {Daylan},
  {Diego}, {Duc}, {Fotopoulou}, {Fouchez}, {Gavazzi}, {Gruen}, {Hatfield},
  {Hildebrandt}, {Landt}, {Hunt}, {Ibata}, {Ilbert}, {Jasche}, {Joachimi},
  {Joseph}, {Knight}, {Kotak}, {Laigle}, {Lan{\c{c}}on}, {Larsen}, {Lavaux},
  {Leclercq}, {Leonard}, {von der Linden}, {Liu}, {Longo}, {Magliocchetti},
  {Maraston}, {Marshall}, {Mart{\'\i}n}, {Mattila}, {Maturi}, {McCracken},
  {Metcalf}, {Montes}, {Mortlock}, {Moscardini}, {Narayan}, {Paolillo},
  {Papaderos}, {Pello}, {Pozzetti}, {Radovich}, {Rejkuba}, {Rom{\'a}n},
  {S{\'a}nchez-Janssen}, {Sarpa}, {Sartoris}, {Schrabback}, {Sluse}, {Smartt},
  {Smith}, {Snodgrass}, {Talia}, {Tao}, {Toft}, {Tortora}, {Tutusaus}, {Usher},
  {van Velzen}, {Verma}, {Vernardos}, {Voggel}, {Wandelt}, {Watkins}, {Weller},
  {Wright}, {Yoachim}, {Yoon}, \& {Zucca}}]{rubineuclid}
{Guy}, L.~P., {Cuillandre}, J.-C., {Bachelet}, E., {et~al.} 2022, in Zenodo id.
  5836022, Vol.~58, 5836022, \dodoi{10.5281/zenodo.5836022}

\bibitem[{Harris {et~al.}(2020)Harris, Millman, van~der Walt, Gommers,
  Virtanen, Cournapeau, Wieser, Taylor, Berg, Smith, Kern, Picus, Hoyer, van
  Kerkwijk, Brett, Haldane, del R{\'{i}}o, Wiebe, Peterson,
  G{\'{e}}rard-Marchant, Sheppard, Reddy, Weckesser, Abbasi, Gohlke, \&
  Oliphant}]{harris2020array}
Harris, C.~R., Millman, K.~J., van~der Walt, S.~J., {et~al.} 2020, Nature, 585,
  357, \dodoi{10.1038/s41586-020-2649-2}

\bibitem[{Juric {et~al.}(2021)Juric, Eggl, Jones, Moeyens, Bellm, Ivezic, Lust,
  Stetzler, Cornwall, Berres, Chernyavskaya, \& Project}]{SSPDB2021}
Juric, M., Eggl, S., Jones, L., {et~al.} 2021, Bulletin of the AAS, 53

\bibitem[{Kramer {et~al.}(2023{\natexlab{a}})Kramer, Gowanlock, \&
  Trilling}]{kramer2023submitted}
Kramer, D., Gowanlock, M., \& Trilling, D. 2023{\natexlab{a}}, The Astronomical
  Journal, submitted

\bibitem[{Kramer {et~al.}(2023{\natexlab{b}})Kramer, Gowanlock, Trilling,
  McNeill, \& Erasmus}]{KRAMER2023}
Kramer, D., Gowanlock, M., Trilling, D., McNeill, A., \& Erasmus, N.
  2023{\natexlab{b}}, Astronomy and Computing, 44, 100711,
  \dodoi{https://doi.org/10.1016/j.ascom.2023.100711}

\bibitem[{{Laureijs} {et~al.}(2011){Laureijs}, {Amiaux}, {Arduini},
  {Augu{\`e}res}, {Brinchmann}, {Cole}, {Cropper}, {Dabin}, {Duvet}, {Ealet},
  {Garilli}, {Gondoin}, {Guzzo}, {Hoar}, {Hoekstra}, {Holmes}, {Kitching},
  {Maciaszek}, {Mellier}, {Pasian}, {Percival}, {Rhodes}, {Saavedra Criado},
  {Sauvage}, {Scaramella}, {Valenziano}, {Warren}, {Bender}, {Castander},
  {Cimatti}, {Le F{\`e}vre}, {Kurki-Suonio}, {Levi}, {Lilje}, {Meylan},
  {Nichol}, {Pedersen}, {Popa}, {Rebolo Lopez}, {Rix}, {Rottgering},
  {Zeilinger}, {Grupp}, {Hudelot}, {Massey}, {Meneghetti}, {Miller}, {Paltani},
  {Paulin-Henriksson}, {Pires}, {Saxton}, {Schrabback}, {Seidel}, {Walsh},
  {Aghanim}, {Amendola}, {Bartlett}, {Baccigalupi}, {Beaulieu}, {Benabed},
  {Cuby}, {Elbaz}, {Fosalba}, {Gavazzi}, {Helmi}, {Hook}, {Irwin}, {Kneib},
  {Kunz}, {Mannucci}, {Moscardini}, {Tao}, {Teyssier}, {Weller}, {Zamorani},
  {Zapatero Osorio}, {Boulade}, {Foumond}, {Di Giorgio}, {Guttridge}, {James},
  {Kemp}, {Martignac}, {Spencer}, {Walton}, {Bl{\"u}mchen}, {Bonoli},
  {Bortoletto}, {Cerna}, {Corcione}, {Fabron}, {Jahnke}, {Ligori}, {Madrid},
  {Martin}, {Morgante}, {Pamplona}, {Prieto}, {Riva}, {Toledo}, {Trifoglio},
  {Zerbi}, {Abdalla}, {Douspis}, {Grenet}, {Borgani}, {Bouwens}, {Courbin},
  {Delouis}, {Dubath}, {Fontana}, {Frailis}, {Grazian}, {Koppenh{\"o}fer},
  {Mansutti}, {Melchior}, {Mignoli}, {Mohr}, {Neissner}, {Noddle}, {Poncet},
  {Scodeggio}, {Serrano}, {Shane}, {Starck}, {Surace}, {Taylor},
  {Verdoes-Kleijn}, {Vuerli}, {Williams}, {Zacchei}, {Altieri}, {Escudero
  Sanz}, {Kohley}, {Oosterbroek}, {Astier}, {Bacon}, {Bardelli}, {Baugh},
  {Bellagamba}, {Benoist}, {Bianchi}, {Biviano}, {Branchini}, {Carbone},
  {Cardone}, {Clements}, {Colombi}, {Conselice}, {Cresci}, {Deacon}, {Dunlop},
  {Fedeli}, {Fontanot}, {Franzetti}, {Giocoli}, {Garcia-Bellido}, {Gow},
  {Heavens}, {Hewett}, {Heymans}, {Holland}, {Huang}, {Ilbert}, {Joachimi},
  {Jennins}, {Kerins}, {Kiessling}, {Kirk}, {Kotak}, {Krause}, {Lahav}, {van
  Leeuwen}, {Lesgourgues}, {Lombardi}, {Magliocchetti}, {Maguire}, {Majerotto},
  {Maoli}, {Marulli}, {Maurogordato}, {McCracken}, {McLure}, {Melchiorri},
  {Merson}, {Moresco}, {Nonino}, {Norberg}, {Peacock}, {Pello}, {Penny},
  {Pettorino}, {Di Porto}, {Pozzetti}, {Quercellini}, {Radovich}, {Rassat},
  {Roche}, {Ronayette}, {Rossetti}, {Sartoris}, {Schneider}, {Semboloni},
  {Serjeant}, {Simpson}, {Skordis}, {Smadja}, {Smartt}, {Spano}, {Spiro},
  {Sullivan}, {Tilquin}, {Trotta}, {Verde}, {Wang}, {Williger}, {Zhao},
  {Zoubian}, \& {Zucca}}]{euclid}
{Laureijs}, R., {Amiaux}, J., {Arduini}, S., {et~al.} 2011, arXiv e-prints,
  arXiv:1110.3193, \dodoi{10.48550/arXiv.1110.3193}

\bibitem[{Lomb(1976)}]{lomb1976least}
Lomb, N.~R. 1976, Astrophysics and space science, 39, 447,
  \dodoi{10.1007/BF00648343}

\bibitem[{{Matheson} {et~al.}(2021){Matheson}, {Stubens}, {Wolf}, {Lee},
  {Narayan}, {Saha}, {Scott}, {Soraisam}, {Bolton}, {Hauger}, {Silva},
  {Kececioglu}, {Scheidegger}, {Snodgrass}, {Aleo}, {Evans-Jacquez}, {Singh},
  {Wang}, {Yang}, \& {Zhao}}]{2021AJ....161..107M}
{Matheson}, T., {Stubens}, C., {Wolf}, N., {et~al.} 2021, \aj, 161, 107,
  \dodoi{10.3847/1538-3881/abd703}

\bibitem[{McNeill {et~al.}(2023)McNeill, Gowanlock, Mommert, Trilling, Llama,
  \& Paddock}]{mcneill2023tess}
McNeill, A., Gowanlock, M., Mommert, M., {et~al.} 2023, The Astronomical
  Journal, 166, 152, \dodoi{10.3847/1538-3881/acf194}

\bibitem[{Navarro-Meza {et~al.}(2021)Navarro-Meza, Aadland, \&
  Trilling}]{navarro2021asteroid}
Navarro-Meza, S., Aadland, E., \& Trilling, D. 2021, Research Notes of the AAS,
  5, 111

\bibitem[{Oelkers {et~al.}(2017)Oelkers, Rodriguez, Stassun, Pepper, Somers,
  Kafka, Stevens, Beatty, Siverd, Lund, Kuhn, James, \&
  Gaudi}]{oelkers2017variability}
Oelkers, R.~J., Rodriguez, J.~E., Stassun, K.~G., {et~al.} 2017, The
  Astronomical Journal, 155, 39, \dodoi{10.3847/1538-3881/aa9bf4}

\bibitem[{{P{\'a}l} {et~al.}(2020){P{\'a}l}, {Szak{\'a}ts}, {Kiss}, {B{\'o}di},
  {Bogn{\'a}r}, {Kalup}, {Kiss}, {Marton}, {Moln{\'a}r}, {Plachy},
  {S{\'a}rneczky}, {Szab{\'o}}, \& {Szab{\'o}}}]{pal2020}
{P{\'a}l}, A., {Szak{\'a}ts}, R., {Kiss}, C., {et~al.} 2020, \apjs, 247, 26,
  \dodoi{10.3847/1538-4365/ab64f0}

\bibitem[{pandas~development team(2020)}]{reback2020pandas}
pandas~development team, T. 2020, pandas-dev/pandas: Pandas, latest,  Zenodo,
  \dodoi{10.5281/zenodo.3509134}

\bibitem[{{Racca} {et~al.}(2016){Racca}, {Laureijs}, {Stagnaro}, {Salvignol},
  {Lorenzo Alvarez}, {Saavedra Criado}, {Gaspar Venancio}, {Short}, {Strada},
  {B{\"o}nke}, {Colombo}, {Calvi}, {Maiorano}, {Piersanti}, {Prezelus},
  {Rosato}, {Pinel}, {Rozemeijer}, {Lesna}, {Musi}, {Sias}, {Anselmi},
  {Cazaubiel}, {Vaillon}, {Mellier}, {Amiaux}, {Berth{\'e}}, {Sauvage},
  {Azzollini}, {Cropper}, {Pottinger}, {Jahnke}, {Ealet}, {Maciaszek},
  {Pasian}, {Zacchei}, {Scaramella}, {Hoar}, {Kohley}, {Vavrek}, {Rudolph}, \&
  {Schmidt}}]{racca}
{Racca}, G.~D., {Laureijs}, R., {Stagnaro}, L., {et~al.} 2016, in Society of
  Photo-Optical Instrumentation Engineers (SPIE) Conference Series, Vol. 9904,
  Space Telescopes and Instrumentation 2016: Optical, Infrared, and Millimeter
  Wave, ed. H.~A. {MacEwen}, G.~G. {Fazio}, M.~{Lystrup}, N.~{Batalha},
  N.~{Siegler}, \& E.~C. {Tong}, 99040O, \dodoi{10.1117/12.2230762}

\bibitem[{Ricker {et~al.}(2015)Ricker, Winn, Vanderspek, Latham, Bakos, Bean,
  Berta-Thompson, Brown, Buchhave, Butler, {et~al.}}]{ricker2015transiting}
Ricker, G.~R., Winn, J.~N., Vanderspek, R., {et~al.} 2015, Journal of
  Astronomical Telescopes, Instruments, and Systems, 1, 014003,
  \dodoi{10.1117/1.JATIS.1.1.014003}

\bibitem[{Robitaille {et~al.}(2013)Robitaille, Tollerud, Greenfield,
  Droettboom, Bray, Aldcroft, Davis, Ginsburg, Price-Whelan, Kerzendorf,
  {et~al.}}]{robitaille2013astropy}
Robitaille, T.~P., Tollerud, E.~J., Greenfield, P., {et~al.} 2013, Astronomy \&
  Astrophysics, 558, A33, \dodoi{10.1051/0004-6361/201322068}

\bibitem[{{Saha} {et~al.}(2016){Saha}, {Wang}, {Matheson}, {Narayan},
  {Snodgrass}, {Kececioglu}, {Scheidegger}, {Axelrod}, {Jenness}, {Ridgway},
  {Seaman}, {Taylor}, {Toeniskoetter}, {Welch}, {Yang}, \& {Zaidi}}]{ANTARES}
{Saha}, A., {Wang}, Z., {Matheson}, T., {et~al.} 2016, in Society of
  Photo-Optical Instrumentation Engineers (SPIE) Conference Series, Vol. 9910,
  Observatory Operations: Strategies, Processes, and Systems VI, ed. A.~B.
  {Peck}, R.~L. {Seaman}, \& C.~R. {Benn}, 99100F, \dodoi{10.1117/12.2232095}

\bibitem[{{Scargle}(1982)}]{1982ApJ...263..835S}
{Scargle}, J.~D. 1982, \apj, 263, 835, \dodoi{10.1086/160554}

\bibitem[{Sergeyev {et~al.}(2023)Sergeyev, Carry, Eggl, Berthier, Vachier, \&
  Santerne}]{sergeyevspin}
Sergeyev, A., Carry, B., Eggl, S., {et~al.} 2023, in ACM Conference 2023,
  Asteroids, Comets, Meteors (ACM) Conference.
\newblock \url{https://www.hou.usra.edu/meetings/acm2023/pdf/2175.pdf}

\bibitem[{S{\"u}veges {et~al.}(2015)S{\"u}veges, Guy, Eyer, Cuypers, Holl,
  Lecoeur-Ta{\"\i}bi, Mowlavi, Nienartowicz, Blanco, Rimoldini,
  {et~al.}}]{suveges2015comparative}
S{\"u}veges, M., Guy, L.~P., Eyer, L., {et~al.} 2015, Monthly Notices of the
  Royal Astronomical Society, 450, 2052, \dodoi{10.1093/mnras/stv719}

\bibitem[{{Tonry} {et~al.}(2018){Tonry}, {Denneau}, {Heinze}, {Stalder},
  {Smith}, {Smartt}, {Stubbs}, {Weiland}, \& {Rest}}]{atlas}
{Tonry}, J.~L., {Denneau}, L., {Heinze}, A.~N., {et~al.} 2018, \pasp, 130,
  064505, \dodoi{10.1088/1538-3873/aabadf}

\bibitem[{Trilling {et~al.}(2023{\natexlab{a}})Trilling, Gowanlock, Kramer,
  McNeill, Donnelly, Butler, \& Kececioglu}]{trilling2023solar}
Trilling, D.~E., Gowanlock, M., Kramer, D., {et~al.} 2023{\natexlab{a}}, The
  Astronomical Journal, 165, 111, \dodoi{10.3847/1538-3881/acac7f}

\bibitem[{Trilling {et~al.}(2023{\natexlab{b}})Trilling, Gerdes, Juric,
  Trujillo, Bernardinelli, Napier, Smotherman, Strauss, Fuentes, Holman,
  {et~al.}}]{trilling2023decam}
Trilling, D.~E., Gerdes, D.~W., Juric, M., {et~al.} 2023{\natexlab{b}}, arXiv
  preprint arXiv:2309.03417

\bibitem[{VanderPlas(2018)}]{VanderPlas2018}
VanderPlas, J.~T. 2018, The Astrophysical Journal Supplement Series, 236, 16,
  \dodoi{10.3847/1538-4365/aab766}

\bibitem[{{{\v{D}}urech} {et~al.}(2022){{\v{D}}urech}, {V{\'a}vra},
  {Van{\v{c}}o}, \& {Erasmus}}]{2022FrASS...909771D}
{{\v{D}}urech}, J., {V{\'a}vra}, M., {Van{\v{c}}o}, R., \& {Erasmus}, N. 2022,
  Frontiers in Astronomy and Space Sciences, 9, 809771,
  \dodoi{10.3389/fspas.2022.809771}

\bibitem[{Virtanen {et~al.}(2020)Virtanen, Gommers, Oliphant, Haberland, Reddy,
  Cournapeau, Burovski, Peterson, Weckesser, Bright,
  {et~al.}}]{virtanen2020scipy}
Virtanen, P., Gommers, R., Oliphant, T.~E., {et~al.} 2020, Nature methods, 17,
  261, \dodoi{10.1038/s41592-019-0686-2}

\bibitem[{{Warner} {et~al.}(2009){Warner}, {Harris}, \& {Pravec}}]{warner2009}
{Warner}, B.~D., {Harris}, A.~W., \& {Pravec}, P. 2009, \icarus, 202, 134,
  \dodoi{10.1016/j.icarus.2009.02.003}

\end{thebibliography}


\appendix

\section{Derived Rotation Periods}\label{sec:appendix_table}
\startlongtable
\begin{deluxetable*}{l|r|r|r|r|r|r|r}
\tablecaption{\intersectsnapstess rotation periods (in hours) and amplitudes (mag) for $n=1623$ objects with high confidence ($\sim$85\%). We also show the rotation periods for \snapshotdata, \tessdata, and \lcdbdata (if applicable), and whether the rotation period was the result of a direct replacement using \tessdata. Column ``Class'' refers to the result of the conservative binary classifier where ``I'' and ``C'' refer to incorrect and correct class labels, respectively.  (This table is available in its entirety in machine-readable form.)}\label{tab:appendix_rotationperiods}
\tablewidth{\textwidth}
\tabletypesize{\footnotesize}
\tablehead{
\colhead{MPC Desig.}&
\colhead{Rot. Per.}&
\colhead{Amplitude (mag)}& 
\colhead{Rot. Per. \tessdata}&
\colhead{Rot. Per. \snapshotdata}&
\colhead{TESSC1 Replacement}&
\colhead{Class}&
\colhead{Rot. Per. \lcdbdata}
}
\startdata
100052 & 4.682 & 0.482 & 34.946 & 4.682 & N & C &  \\
10026 & 9.977 & 0.199 & 33.726 & 8.257 & N & I &  \\
10041 & 47.834 & 0.255 & 816.327 & 47.963 & N & C & 2.564 \\
10065 & 20.050 & 0.535 & 20.047 & 20.050 & N & C &  \\
100683 & 18.261 & 0.499 & 232.480 & 18.259 & N & C &  \\
100727 & 3.175 & 0.368 & 36.920 & 3.175 & N & I &  \\
10080 & 7.051 & 0.177 & 7.052 & 6.146 & N & C &  \\
10117 & 6.880 & 0.467 & 6.881 & 6.880 & N & C &  \\
10128 & 9.573 & 0.142 & 9.571 & 7.978 & N & C &  \\
101632 & 121.458 & 1.702 & 121.458 & 684.825 & Y & C &  \\
\enddata
\end{deluxetable*}




\end{document}